\documentclass[12pt]{article}
\pdfoutput=1
\usepackage{jheppub}

\usepackage{subfigure}
\usepackage{amsmath,amssymb,amsthm,amsfonts,array,bbm}
\usepackage{graphicx}
\usepackage{color}
\usepackage[american]{babel}

\usepackage[mathscr]{euscript}
\usepackage{MnSymbol}

\usepackage{tikz}
\usetikzlibrary{automata}
\usetikzlibrary{arrows}
\usetikzlibrary{calc}
\usetikzlibrary{decorations.markings}
\usetikzlibrary{decorations.pathreplacing}
\usetikzlibrary{intersections}
\usetikzlibrary{positioning}
\usetikzlibrary{topaths}
\usetikzlibrary{shapes.geometric}
\usetikzlibrary{shapes.misc}
\tikzset{->-/.style = {
    decoration = {markings, mark = at position #1 with {\arrow{>}}},
    postaction = {decorate}}}
\tikzset{color-group/.style = {
    shape = circle,
    minimum size = 2.5ex,
    inner sep = .5ex,
    draw}}
\tikzset{flavor-group/.style = {
    shape = rectangle,
    minimum size = 2.5ex,
    inner sep = .5ex,
    draw}}
\tikzset{cf-group/.style = {
    shape = rounded rectangle,
    rounded rectangle right arc = none,
    draw}}
\tikzset{fc-group/.style = {
    shape = rounded rectangle,
    rounded rectangle left arc = none,
    draw}}
\tikzset{cross/.style={minimum width=1pt, path picture={
      \draw[black, very thick]
               (path picture bounding box.south east)
            -- (path picture bounding box.north west)
               (path picture bounding box.south west)
            -- (path picture bounding box.north east);
          }}}

\newcommand{\lp}{\left(}
\newcommand{\rp}{\right)}
\newcommand{\ti}{\widetilde}
\newcommand{\tr}{\textrm{Tr} \,}

\newcommand{\p}{\partial}

\def\Nf2{\lfloor N_f/2 \rfloor}
\def\Nfo2{\lfloor \frac{N_f}{2} \rfloor}

\def\CP1{\bC\bP^1}

\numberwithin{equation}{section}

\newcommand{\be}{\begin{equation}} \newcommand{\ee}{\end{equation}}
\newcommand{\bea}{\begin{equation} \begin{aligned}} \newcommand{\eea}{\end{aligned} \end{equation}}

\newcommand{\vev}[1]{{\langle {#1} \rangle}}

\newcommand{\cA}{\mathcal{A}}

\newcommand{\cC}{\mathcal{C}}

\newcommand{\cF}{\mathcal{F}}

\newcommand{\cN}{\mathcal{N}}

\newcommand{\cP}{\mathcal{P}}

\newcommand{\cR}{\mathcal{R}}
\newcommand{\cS}{\mathcal{S}}

\newcommand{\bC}{\mathbb{C}}

\newcommand{\bP}{\mathbb{P}}

\newcommand{\bR}{\mathbb{R}}
\newcommand{\bZ}{\mathbb{Z}}

\def\repa{\raise4pt\hbox{$\square$}\mkern-14mu\raise-4pt\hbox{$\square$}}
\def\repab{\overline{\raise4pt\hbox{$\square$}\mkern-14mu\raise-4pt\hbox{$\square$}\mkern-1mu}}

\numberwithin{equation}{section}       

\begin{document}

\begin{titlepage}

\vspace*{-2cm} 
\begin{flushright}
	{\tt KIAS-P18051 \\ CERN-TH-2018-140} 
\end{flushright}
	
	\begin{center}

		\vskip .5in 
		\noindent

		{\Large \bf{Wilson loops in 5d $\cN=1$ theories and S-duality}}
		
		\bigskip\medskip
		
		Benjamin Assel$^1$ and Antonio Sciarappa$^2$\\
		
		\bigskip\medskip
		{\small 
			$^1$ Theory Department, CERN, CH-1211, Geneva 23, Switzerland \\
			$^2$ School of Physics, Korea Institute for Advanced Study \\
85 Hoegiro, Dongdaemun-gu, Seoul 130-722, Republic of Korea			
		}
		
		\vskip .5cm 
		{\small \tt benjamin.assel@gmail.com, asciara@kias.re.kr}
		\vskip .9cm 
		{\bf Abstract }
		\vskip .1in
\end{center}
	
\noindent
We study the action of S-duality on half-BPS Wilson loop operators in 5d $\cN=1$ theories. The duality is the statement that different massive deformations of a single 5d SCFT are described by different gauge theories, or equivalently that the SCFT points in parameter space of two gauge theories coincide. The pairs of dual theories that we study are realized by brane webs in type IIB string theory that are S-dual to each other. We focus on $SU(2)$ SQCD theories with $N_f \le 4$ flavors, which are self-dual, and on $SU(3)$ SQCD theories, which are dual to $SU(2)^2$ quiver theories. From string theory engineering we predict that Wilson loops are mapped to dual Wilson loops under S-duality. We confirm the predictions with exact computations of Wilson loop VEVs, which we extract from the 5d half-index in the presence of auxiliary loop operators (also known as higher qq-characters) sourced by D3 branes placed in the brane webs. A special role is played by Wilson loops in tensor products of the (anti)fundamental representation, which provide a natural basis to express the S-duality action. The exact computations also reveal the presence of additional multiplicative factors in the duality map, in the form of background Wilson loops. 


\vfill

\end{titlepage}

\setcounter{page}{1}

\noindent\hrulefill
\tableofcontents

\noindent\hrulefill

\bigskip



\section{Introduction and summary of results}
\label{sec:Intro}

Five-dimensional SCFTs deformed by relevant operators often admit a low energy description in terms of 5d $\cN=1$ SYM gauge theories with matter. The corresponding massive parameters are interpreted in the gauge theory as the Yang-Mills couplings $t = g_{\rm YM}^{-2}$ for the simple factors in the gauge group and the masses of the matter hypermultiplets. The SCFTs are then viewed as the limit of infinite coupling $g_{\rm YM} \to \infty$ ($t \to 0$) and massless matter of the gauge theories. This was described first in the seminal paper of Seiberg \cite{Seiberg:1996bd} for $SU(2)$ gauge theories with $N_f \le 7$ flavor hypermultiplets, where it was argued from their string theory engineering that the SCFTs have enhanced $E_{N_f+1}$ global symmetry. Many more SCFTs have been contructed from 5d $\cN=1$ quiver gauge theories and related to brane systems and geometric engineering in string theory (see \cite{Douglas:1996xp,Morrison:1996xf,Intriligator:1997pq,Aharony:1997ju,Aharony:1997bh} for some early papers).\footnote{Recently there were some attempts to classify low rank 5d $\cN=1$ SCFTs based on their Coulomb branch or their engineering in M-theory \cite{Jefferson:2017ahm,Jefferson:2018irk}. See also \cite{Chang:2017cdx} for an analysis of low rank 5d SCFTs based on numerical bootstrap techniques.}

It can happen that different massive deformations of a SCFT lead to different gauge theory descriptions. Typically a deformation with parameter $t>0$ or $t<0$ can lead to two different gauge theories with couplings $g_{\rm YM}^{-2} \sim |t|$. Thus deformations in different ``chambers" of the parameter space may be described by different gauge theories.\footnote{However this is not a generic phenomenon. Often some regions of parameter space simply do not admit a gauge theory description.} This can be phrased as a duality between the gauge theories which are obtained from deformations of the same SCFT. 

One such duality is realized by S-duality in the type IIB brane setup realizing the 5d theories and we will thus call it S-duality. An important class of S-dual theories are $SU(N)^{M-1}$ linear quivers and their dual $SU(M)^{N-1}$ linear quivers. They can be realized as the low-energy theories of webs of 5-branes in type IIB string theory as described in \cite{Aharony:1997ju,Aharony:1997bh} and the action of S-duality exchanges the brane webs of the dual theories. 
This duality can be tested by computing observables in the dual theories and matching them with appropriate identification of parameters. This assumes that the gauge theory observables in question can be analytically continued in the deformation parameters to the full parameter space of the SCFT.
Such tests have been performed with exact results from topological strings \cite{Bao:2011rc,Mitev:2014jza} and  from supersymmetric localization at the level of the partition function (or supersymmetric index) of the gauge theories \cite{Bergman:2013aca}.

An important challenge is to understand how S-duality acts on loop operators. In this paper we answer this question for half-BPS Wilson loop operators,
\be
W = \tr_{\cR} \,  \cP \exp\big(\int i A + \sigma dx \big) \,,
\ee
where $A$ is the one-form gauge potential, $\sigma$ the adjoint scalar in the vector multiplet and $\cR$ is a representation of the gauge group.
 We find that S-duality acts as an automorphism on the space of Wilson loops, namely that Wilson loops are mapped to Wilson loops. This differs from 4d S-duality where Wilson loops are mapped to 't Hooft loops (or in general to dyonic loops), and from 3d mirror symmetry where they are mapped to vortex loops \cite{Assel:2015oxa}. 
Our findings are guided by the type IIB brane realization of half-BPS loop operator insertions. We relate Wilson loops to configurations with specific arrays of strings stretched between D5 branes and auxiliary D3 branes. Through standard brane manipulations we identify these configurations across S-duality and deduce a prediction for the S-duality map between Wilson loops.

We consider two classes of theories, which are those with lowest gauge algebra rank. In Section \ref{sec:SU2} and \ref{sec:SU2Nf} we consider $SU(2)$ theories with $N_f \le 4$ flavor hypermultiplets. These theories are self-dual under S-duality. For instance the pure $SU(2)$ theory is dual to another pure $SU(2)$ theory with the gauge couplings related by $t = - \ti t$ (namely the region of negative $t$ is described by the dual $SU(2)_{\ti t=-t}$ theory).
In Section \ref{sec:SU3dualities} we consider examples of $SU(3)$ theories and their $SU(2)\times SU(2)$ quiver duals.

 An important prediction of the brane analysis is that there is a privileged basis of Wilson loops in which to express the S-duality map: these are Wilson loops in tensor products of fundamental and anti-fundamental representations for each gauge node.\footnote{For higher rank theories, the relevant representations are tensor products of anti-symmetric representations for each gauge node.} They are naturally realized in the brane setup. For these loops we predict a one-to-one S-duality map. For Wilson loops in other representations (which are linear combinations of loops in the privileged basis), each individual loop is mapped to a linear combination of loops in the dual theory. We therefore focus on Wilson loops of the former kind.
 
To test the proposed duality map we compute the exact VEVs of the Wilson loops wrapping the circle at the origin of the 5d Omega background $S^1\times \bR^4_{\epsilon_1,\epsilon_2}$, or `half-index' in the presence of Wilson loop insertions, using supersymmetric localization.
This happens to be a challenging computation because the modifications of the instanton corrections (in particular to the moduli space of singular instantons at the origin) in the presence of a Wilson loop are not yet completely understood (to our knowledge). To side-step this problem we follow the approach advocated in \cite{Kim:2016qqs} (see also \cite{Agarwal:2018tso}) and compute instead the VEVs of certain $\cN=(0,4)$ SQM loop operators, which are roughly speaking generating functions for some Wilson loops. They are defined by an array of 1d fermions with a subgroup of the flavor symmetry gauged with 5d fields in an $\cN=(0,4)$ supersymmetry preserving manner (they preserve the same supersymmetry as the half-BPS Wilson loops). The relevant SQM loops are those sourced by stacks of $n$ D3 branes placed in the brane web.\footnote{These SQM loop operators are also known under the name of fundamental (for $n = 1$ D3) or higher (for $n > 1$) qq-characters in the language of \cite{Nekrasov:2015wsu,Bourgine:2016vsq,Kimura:2015rgi,Bourgine:2017jsi}, although the relation to Wilson loops is not discussed in those works.} The Wilson loop VEVs can then be identified as certain residues of the SQM loop VEVs in the SQM flavor fugacities. This property is inferred from string considerations.
The virtue of the SQM loops is that one can use their brane realization as a guide to find the appropriate modification of the ADHM quiver quantum mechanics computing instanton corrections. Our results confirm the validity of the procedure by correctly reproducing the classical contributions to the Wilson loop VEVs and by confirming the conjectured S-duality map. 

We find however a somewhat surprising feature: Wilson loops in the appropriate tensor product representations do not transform exactly into their dual Wilson loop, but rather come with an extra multiplicative factor which can be interpreted as a background Wilson loop. We say that they transform covariantly under S-duality. Let us summarize our results:
\begin{itemize}
\item For the self-S-dual $SU(2)$ theories with $N_f\le 4$ flavors we consider Wilson loops in the representations $\mathbf{2}^{\otimes n}$ and find that they transform under S-duality as
\be
S. W_{\mathbf{2}^{\otimes n}}(t,m_k) =  Y^{-n} \, \ti W_{\mathbf{2}^{\otimes n}} (\ti t, \ti m_k) \,,
\label{introStransfoSU2}
\ee
with $t,m_k, \ti t, \ti m_k$ the gauge coupling and mass parameters in the dual theories respectively (see Section \ref{sec:SU2Nf} and Appendix \ref{app:Nf34} for the precise maps), and 
$Y= e^{\frac t2 + \frac 14 \sum_{k=1}^{N_f} (-1)^k m_k} = \ti Y^{-1}$.  We also find that at each order in the appropriate expansion parameter the contributions to the Wilson loop VEVs are organized into characters of the $E_{N_f+1}$ symmetry, confirming the symmetry enhancement. S-duality is then a transformation in the Weyl group of $E_{N_f+1}$ \cite{Mitev:2014jza}. The parameter $Y$ can be understood as a charge one background Wilson loop for a $U(1)$ subgroup of $E_{N_f+1}$. We strongly believe that these results hold for $N_f=5,6,7$ (but we were not able to test it).
\item For $SU(3)$ theories we consider Wilson loops in representations $\mathbf{3}^{\otimes n_1} \otimes \overline{\mathbf 3}{}^{\otimes n_2}$ and in the dual $SU(2)\times SU(2)$ quiver Wilson loops in representations $(\mathbf{2}^{\otimes n_1},\mathbf{2}^{\otimes n_2})$. We find the exact map
\bea
& S.W_{\mathbf{3}^{\otimes n_1} \otimes \overline{\mathbf 3}{}^{\otimes n_2}} = Y_1^{-n_1} Y_2^{-n_2} \ti W_{(\mathbf{2}^{\otimes n_1},\mathbf{2}^{\otimes n_2})} \,,
\label{introStransfoSU3}
\eea
with background Wilson loops $Y_1,Y_2$ which are given, for instance in the duality relating the $SU(3)$ $N_f=2$ to the $SU(2)\times SU(2)$ quiver without fundamental hypermultiplet (see Section \ref{ssec:SU3Nf2duality}) by $Y_1 = e^{-\frac{2\ti t_1+ \ti t_2}{3}} = e^{\frac t2 + \frac{m_1+m_2}{4}}$ and $Y_2= e^{-\frac{\ti t_1+ 2\ti t_2}{3}}= e^{\frac t2 - \frac{m_1+m_2}{4}}$. 
\end{itemize}
These results are based on exact computations up to 2 or 3-instanton corrections and for the Wilson loops in the lowest rank representations, namely with $n=1,2$ (sometimes $n=3$) and $n_1+n_2 \le 2$, which is as far as we could reasonably go technically (using Mathematica).
We conjecture a generalization of the S-duality map of Wilson loops with the relation \eqref{SmapGen} for the duality relating $SU(M)$ SQCD theories to $SU(2)^{M-1}$ quivers.

Before moving to the bulk of the discussion it is worth mentioning some related work. Analogous dualities of 5d $\cN=1$ theories for $SU(N)$ theory with $N_f$ flavors and Chern-Simons level $N- \frac{N_f}{2}$ were studied in \cite{Gaiotto:2015una} (with generalization to quiver theories). In that case the theories are self-dual with a map of massive parameters which reverses the sign of the (squared) gauge coupling. The paper describes duality interface theories for this duality, but also study the action of the duality on Wilson loop operators. This involves a dressing factor in the form of a background Wilson loop as well.\footnote{The map proposed in \cite{Gaiotto:2015una} (Equation (4.2)) is not quite analogous to what we find for S-duality, because it acts covariantly on Wilson loops in irreducible representations, instead of tensor product representations. We observe that the proposal does not seem consistent with the fact that Wilson loops in rank $N$ antisymmetric representations are trivial. The duality studied is not S-duality in general, but it should be S-duality for $SU(2)$ theories (up to $E_{N_f+1}$ Weyl transformations). 
Moreover the method proposed to compute the half-index in the presence of a Wilson loop differs from the one proposed in this paper and might explain the slightly different results. We believe that the method we present provides a more robust framework to carry out such computations.}
The enhancement to $E_{N_f+1}$ global symmetry seen from the computation of the superconformal index in $SU(2)$ SQCD theories was also found in \cite{Kim:2012gu,Hwang:2014uwa} and with Wilson ray insertions in \cite{Chang:2016iji}, using closely related computational methods.

\medskip

The rest of the paper is organized as follows. In section \ref{sec:BranesLoops} we discuss the brane realization of Wilson loops and SQM loops in IIB string theory, their relations and the action of S-duality inferred from type IIB S-duality. In Section \ref{sec:SU2} we explain the computation of the half-index with Wilson loops in detail and derive the exact S-duality action for the pure $SU(2)$ theory. Section \ref{sec:SU2Nf} contains the computation and the results for the $SU(2)$ theory with $N_f=1$ and $N_f=2$ flavors. We relegate to Appendix \ref{app:Nf34} the study of $SU(2)$ with $N_f=3$ and $4$. In Section \ref{sec:SU3dualities} we study the duality relating $SU(3)$ theories to $SU(2)$ quivers and we generalize it in Section \ref{sec:Generalization}. The remaining appendices contain details about the ADHM instanton computations (Appendix \ref{app:ADHMcomputations}) and some exact results which were too voluminous to fit in the main text (Appendix \ref{app:longresults}).

\section{Branes and Loops}
\label{sec:BranesLoops}

In this section we give a brief review of the brane realization of 5d $\cN=1$ theories following \cite{Aharony:1997bh} and we explain how the insertion of half-BPS loop operators can be achieved by adding extra branes or strings to the construction.

\subsection{Brane setup}
\label{ssec:Branes}

The 5d $\cN=1$ theories that we will study are engineered with a 5-brane web in type IIB string theory, with the orientations described in the first entries of Table \ref{tab:braneorientations}.
\begin{table}[h]
\begin{center}
\begin{tabular}{|c||c|c|c|c|c|c|c|c|c|c|}
  \hline
      & 0 & 1 & 2 & 3 & 4 & 5 & 6 & 7 & 8 & 9 \\ \hline
  D5  & X & X & X & X &  X & X  &   &   &   &   \\
  NS5  & X & X & X & X  & X &  & X &   &   &   \\
  5$_{(p,q)}$ & X & X & X & X  & X  & $\theta$  & $\theta$  &  &  &  \\  \hline
  F1 & X &  &  &   &   &   & X  &  &  &  \\ 
  D1 & X &  &  &   &   & X  &   &  &  &  \\ 
  D3 & X &  &  &   &   &   &   & X & X & X \\  \hline
\end{tabular}
\caption{\footnotesize Brane array for 5d $\cN=1$ theories and half-BPS loop operators. ($\tan\theta=\frac{q}{p}$)}
\label{tab:braneorientations}
\end{center}
\end{table}
A 5$_{(p,q)}$ brane spans a line in the $x^{56}$ plane defined by $\cos(\theta) x^5 + \sin(\theta) x^6 =0$ with $\tan\theta=\frac{q}{p}$. In this convention we have D5 = 5$_{(0,1)}$ and NS5 = 5$_{(1,0)}$. In pictures we stick to the usual convention that D5 branes are horizontal lines, while NS5 branes are vertical lines, which means we draw pictures in the $x^{56}$ plane.

\begin{figure}[th]
\centering
\includegraphics[scale=0.35]{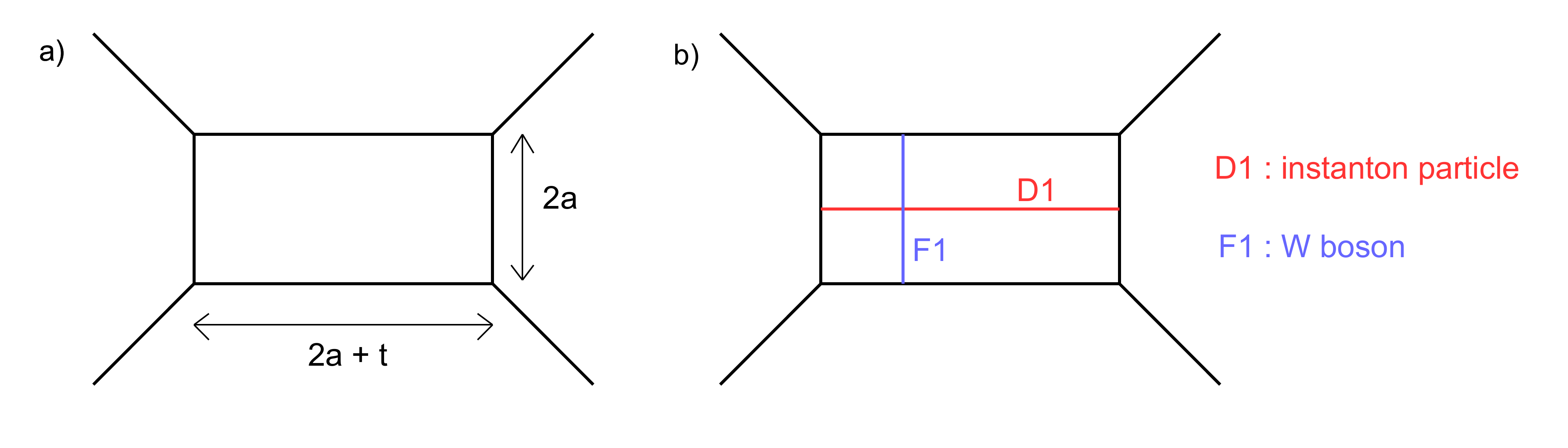}
\vspace{-0.5cm}
\caption{\footnotesize{a) Brane realization of the pure $SU(2)$ theory. b) Half-BPS strings excitations for W-bosons and instanton particles.}}
\label{SU2}
\end{figure}

The brane setups have parallel D5 branes spanning an interval along $x^5$ and supporting a 5d Yang-Mills gauge theory at energies lower than the size of the interval. The simplest example is that of Figure \ref{SU2}-a with two parallel D5s supporting an $SU(2)$ gauge theory.\footnote{For $N$ D5 brane segments, it is believed that the diagonal $U(1) \subset U(N)$ subgroup of the gauge group living on the D5s is massive, so that the theory at energies sufficiently small is an $SU(N)$ gauge theory.}
There are two distances in this configuration: the distance $2a$ between the D5s corresponding to the VEV of the real scalar in the vector multiplet 
 $\vev{\phi} = 
\lp 
\begin{array}{cc}
a & 0 \cr
0 & -a \cr
\end{array}
\rp$,
and the distance $t_{\rm eff} := \frac{1}{g^2_{\rm eff}}= t + 2a$ between the NS5s corresponding to the effective abelian coupling on either of the D5 branes, where we denoted $t:= \frac{1}{g^2}$ the bare Yang-Mills coupling of the $SU(2)$ theory. 
The brane setup thus describes the gauge theory at finite coupling $t$ and on the Coulomb branch of vacua. The SCFT is obtained as the configuration where these two sizes are set to zero, namely when the D5s and NS5s are shrunk to zero size and the configuration looks like two intersecting 5$_{(1,1)}$ and 5$_{(1,-1)}$ branes.

\medskip

In this picture one can add strings stretched between 5-branes associated to particle excitations of the 5d $\cN=1$ theory, as shown in Figure \ref{SU2}-b. F1 strings stretched between D5s are W-bosons excitations with mass $2a$, while D1 strings stretched between NS5s are instanton particles with mass $t +2a$.

\medskip

To add flavor hypermultiplets to the 5d theory one should add external (semi-infinite) D5 branes to the construction. To increase the rank of the gauge group one should add D5 segments to the construction. We will look at these more elaborate brane setups in later sections.

\subsection{Half-BPS loop operators}
\label{ssec:Loops}

Half-BPS loop operators are realized by adding semi-infinite F1 strings, D1 strings and/or D3 branes to the setup with the orientations given in the second entries of Table \ref{tab:braneorientations}. The F1 strings, D1 strings and D3 branes all preserve the same four supercharges, so we can consider configurations with all of them together if we wish. 
The presence of the strings and/or D3 branes break the supersymmetry to a 1d $\cN=(0,4)$ subalgebra. 

\begin{figure}[th]
\centering
\includegraphics[scale=0.3]{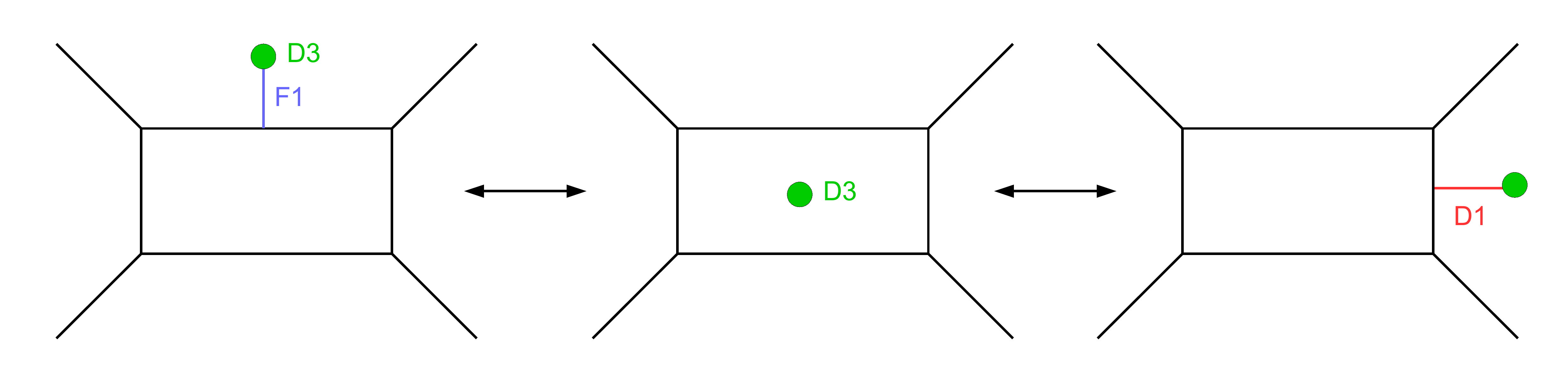}
\vspace{-0.5cm}
\caption{\footnotesize{Hanany-Witten string creation/annihilation effects as we move a D3 brane (green dot) in the $x^{56}$ plane.}}
\label{SU2Loops}
\end{figure}

Importantly the D3 and D5 branes are in a Hanany-Witten orientation relative to each other, with a F1 string creation effect, which means that as the D3 brane crosses the D5 brane a F1 string is created stretched between them. 
Similarly (and remarkably) the D3 and NS5 branes are also in a Hanany-Witten orientation relative to each other, but with a D1 string creation effect: as a D3 brane crosses an NS5 brane a D1 string is created. We illustrate these effects in Figure \ref{SU2Loops}.
This will be important since, according to \cite{Hanany:1996ie}, the low energy physics is not affected by moving the D3 brane along $x^{56}$ as long as one takes into account these string creation effects. 

This also comes with an important property, usually called {\it s-rule} : at low energies there can be at most one F1 string stretched between a D3 and a D5, and similarly at most a D1 string stretched between a D3 and an NS5. This is because the lowest energy mode on such a string is fermionic.
\medskip

The interpretation of the semi-infinite F1 and D1 string as operator insertions in the $SU(2)$ gauge theory is the following. 
A semi-infinite F1 string stretched from infinity (along $x^5$) to the D5s inserts a half-BPS Wilson loop in the fundamental representation of $SU(2)$. There are two configurations -- the string ending on one or the other D5 -- corresponding to the two states traced over in the fundamental representation. 
A semi-infinite D1 string stretched from infinity (along $x^6$) to the NS5s inserts a loop operator which should be a 1d defect related to instantons in the 5d theory, however we do not know of any description of these loops in terms of a singularity prescription for the 5d fields.
We will not try to characterize them further in this paper, however we observe from Figure \ref{SU2Loops} that such loops are related to standard Wilson loops through Hanany-Witten moves, therefore it is enough in principle to study Wilson loops.
\smallskip

The interpretation of a D3 brane placed in the middle of the 5-brane array is not strictly speaking as the insertion of a loop operator since the D3 brane support a 4d $\cN=4$ $U(1)$ SYM theory at low-energies. However the 4d theory is coupled to 5d theory along a line, through charged localized 1d fields, and the whole 5d-4d-1d setup preserves the same supersymmetry as a half-BPS loop operator in the 5d theory, namely 1d $\cN=(0,4)$ supersymmetry. Moreover the 4d theory will not play a role in our computations and we can consider it as non-dynamical.\footnote{It would be interesting to study the full 5d-4d theories interacting along a line defect and to understand the duality properties of such systems.}
Therefore we interpret this setup as inserting a loop operator described by coupling a (0,4) SQM to the 5d theory.

\begin{figure}[th]
\centering
\includegraphics[scale=0.4]{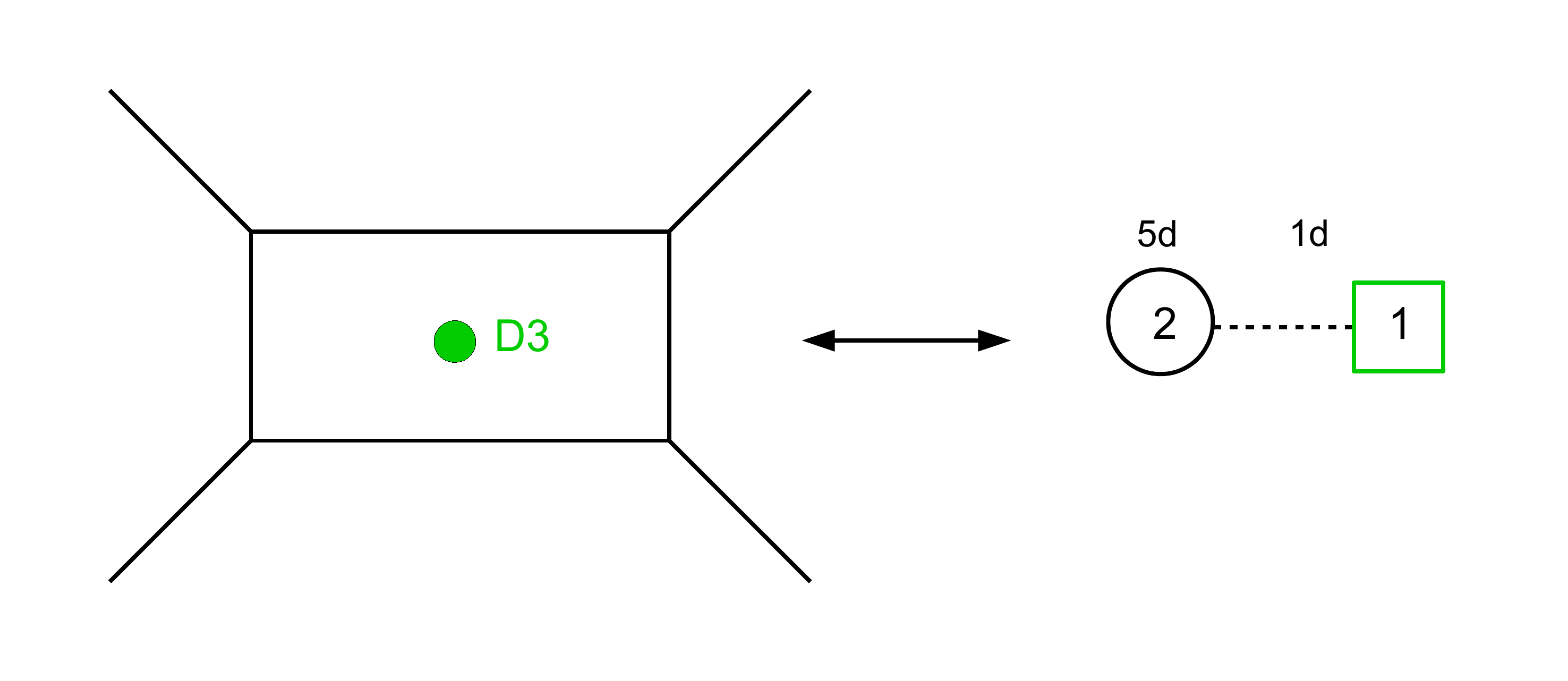}
\vspace{-0.5cm}
\caption{\footnotesize{5d-1d quiver theory corresponding to the addition of a D3 brane in the center. The circle indicates the 1d flavor symmetry gauged with 5d fields. The dashed line 
indicates a bifundamental Fermi multiplet (two fermions).}}
\label{D3SQM}
\end{figure}

At low energies the localized 1d modes are two complex fermions $\chi_{a=1,2}$ (two (0,4) Fermi multiplets), which arise as the lowest excitation of strings stretched between the D3 and D5 branes. They form a doublet of $SU(2)$ which is identified with the 5d gauge group.  The 1d fermions $\chi_a$ are coupled to the 5d `bulk' theory via gauging the $SU(2)$ symmetries by the 5d vector multiplet at the location of the line defect.\footnote{The 5d $\cN=1$ vector multiplet can be decomposed into 1d $\cN=(0,4)$ multiplets. This provides a 1d vector multiplet which gauges the $SU(2)$ flavor symmetry of the defect theory.} This leaves a $U(1)_f$ flavor symmetry acting on both fermions with the same charge.
The corresponding mixed 5d-1d quiver theory is shown in Figure \ref{D3SQM}. 
The 1d action is (in implicit notation)
\be
S_{\chi} = \int dt \, \bar\chi^a (\p_t - i A^{(5d)}{}_{a}{}^b +  \phi^{(5d)}{}_{a}{}^b - M \, \delta_a^b ) \chi_b \,.
\ee
The VEVs of the vector multiplet scalar $\phi^{(5d)}$ and the real mass $M$ can be identified with the positions of the D5s and of the D3 along $x^5$ respectively. Denoting $M$ the position of the D3 brane and $a_1,a_2$ the positions of the D5s along $x^5$, the fermions have mass $a_1 - M$ and $a_2 - M$.

\medskip

It will be central in our discussion to understand the relation between such `D3-loop operators' or `SQM loop operators' and the Wilson loop operators that we wish to study. This is because the exact computation of Wilson loop VEVs is at the moment not completely understood, therefore in order to 
evaluate them we will have to make use of certain relations between the VEVs of SQM operators and the VEVs of Wilson loops on a given manifold.
To this purpose we make the following heuristic argument.

\begin{figure}[th]
\centering
\includegraphics[scale=0.3]{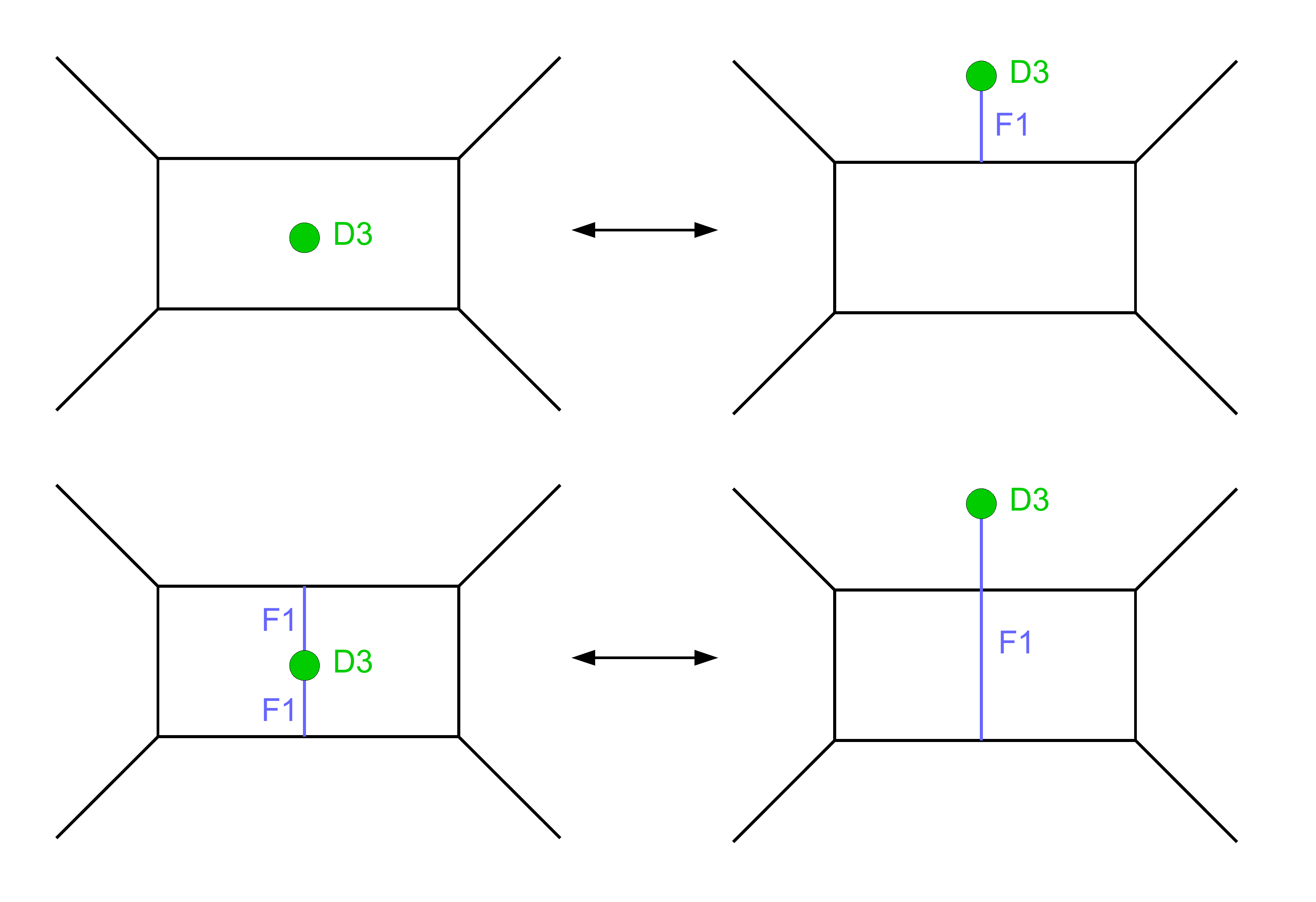}
\vspace{-0.5cm}
\caption{\footnotesize{The (0,0) and (1,1) string configurations (on the left) are related to the two configurations for the fundamental Wilson loop insertion (on the right) by Hanany-Witten moves.}}
\label{SU2Wcontrib}
\end{figure}

We consider the supersymmetric partition function on some manifold of the SQM theory associated to the presence of the D3 brane, by which we mean the partition function of the 5d-1d theory, and we normalize it by the partition function of the 5d theory alone $Z_{\text{5d-1d}}/Z_{\text{5d}}$. We define this as the normalized VEV of the SQM loop.
It receives contributions from the degrees of freedom sourced by (fundamental) strings stretched between the D3 and D5 branes.
Since there can be at most one F1 string stretched between the D3 and a D5, there are four possible configurations with F1 strings contributing: $(0,0), (0,1),(1,0),(1,1)$, where $(n,m)$ stands for $n$ strings stretched to the top D5 and $m$ strings stretched to the bottom D5. 
In the configurations $(1,0)$ and $(0,1)$, with a single string, one can move the D3 brane to the top or to the bottom of the brane setup so that no string ends on it anymore (taking into account the string annihilation effect). Such configurations carry almost trivial contributions to the SQM loop VEV \footnote{Because of the flux induced by the D3 brane on the D5 worldvolumes (and vice-versa) \cite{Hanany:1996ie}, such a contribution is not 1, but rather a simple classical factor, as we will see in later sections.} since the D3 brane is decoupled from the brane web.
In the two other configurations, (0,0) and (1,1), by moving the D3 vertically to the top of the brane configuration we obtain a brane configuration with a string stretched between the D3 and one of the D5s, corresponding of the two setups of the fundamental Wilson loop insertion. This is illustrated in Figure \ref{SU2Wcontrib}. Therefore the (normalized) fundamental Wilson loop VEV corresponds to a sector of the SQM loop, which is associated to the two configurations (0,0) and (1,1).
These configurations are those with zero net number of strings ending on the D3 (when placed in the middle of the web)\footnote{The net number of strings ending on the D3 is the number of strings ending on the D3 counted with a sign according to their orientation, namely the difference between the number of strings ending on the top and on the bottom of the D3 brane in the picture.} and correspond to states with no charge under the $U(1)_f$ symmetry. The same considerations apply in the presence of D1 strings stretched between the two NS5s, corresponding to instanton sectors of the gauge theory, and in the presence of extra F1 strings stretched between the two D5s, corresponding to sectors with W-boson excitations. Therefore we arrive at the proposal for the pure $SU(2)$ theory,
\be
\langle\text{fundamental Wilson loop} \rangle \  = \  \langle \text{SQM loop} \rangle \Big|_{U(1)_f \, \text{neutral sector}} \,. 
\label{SU2RelFund}
\ee
In explicit computations this means that the Wilson loop will be obtained by taking a residue in the $U(1)_f$ flavor fugacity. Of course the heuristic argument that we gave is not precise enough to predict the overall coefficient in the above relation and we will find in later sections that it holds up  to a sign.

\medskip

\begin{figure}[th]
\centering
\includegraphics[scale=0.4]{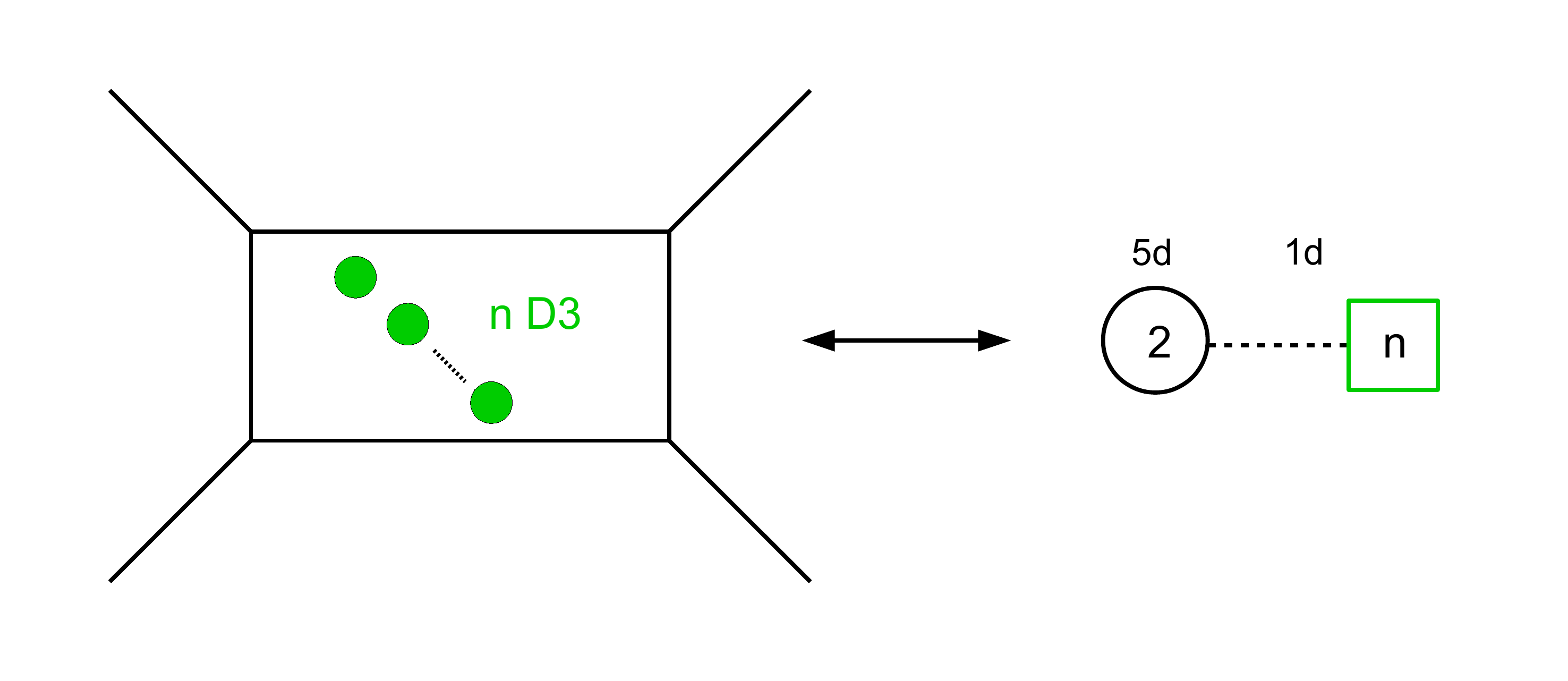}
\vspace{-0.5cm}
\caption{\footnotesize{Adding $n$ D3 branes realizes a quiver theory with a Fermi multiplet in the bifundamental of the $SU(2)\times U(n)_f$ flavor symmetry.}}
\label{D3SQM2}
\end{figure}

To access Wilson loops in higher representations we need to consider more D3 branes. Let us place $n$ D3 branes in the middle of the 5-brane web, as in Figure \ref{D3SQM2}. 
The SQM theory has now Fermi multiplets transforming in the bifundamental representation of $SU(2)\times U(n)_f$, with $U(n)_f$ flavor symmetry associated to the stack of D3s. 
Once again we can think of configurations with strings stretched between the D3s and the D5s and try to isolate those corresponding to Wilson loops insertions. We take the D3s separated, namely we give generic masses to the $n$ fundamental Fermi multiplets. Each D3 brane of type (0,0) or (1,1) (zero or two strings attached) contributes as the insertion of a fundamental Wilson loop. The sum over configurations with D3s of type (0,0) or (1,1) only can be mapped to the trace over states in the tensor product representation $\mathbf{2}^{\otimes n} := \mathbf{2} \otimes \mathbf{2} \otimes \cdots \otimes \mathbf{2}$ ($n$ times). It corresponds to the sector of the SQM theory neutral under a $U(1)_f^n$ maximal torus of $U(n)_f$. We thus arrive at the proposal 
\be
\langle \text{Wilson loop in} \  \mathbf{2}^{\otimes n} \rangle \ = \ \langle \text{SQM loop} \rangle \Big|_{U(1)_f^n \, \text{neutral sector}} \,. 
\label{SU2RelTensProd}
\ee
Finally we may think about identifying configurations related by D3 permutations, which correspond to averaging over $U(n)_f$ Weyl transformations. The resulting reduced set of configurations reproduces the states in the symmetric representation of rank $n$ of $SU(2)$, or spin $n$ representation, and corresponds to projecting to the $U(n)_f$ invariant sector in the SQM, 
\be
\langle \text{Wilson loop in spin $n$} \rangle \ = \ \langle \text{SQM loop} \rangle \Big|_{U(n)_f \, \text{neutral sector}}  \,. 
\label{SU2RelSym}
\ee
These are the predictions we can make from the analysis of the brane setup realizing half-BPS loop insertions. As we will see in the next sections, some more refined prescription will be needed to extract the Wilson loop VEVs from the SQM loop VEVs, in the form of a precise residue integration. We will now try to make these claims more precise, to confirm them by exact computations and use the results to understand the S-duality map of Wilson loops.

\bigskip

Before proceeding we should make a comment.
In the description of the SQM defect theory there is no excitation corresponding to D1 strings stretched between the D3 and the NS5 branes, although these are present in the brane setup. These should correspond to 't Hooft loops in the 4d SYM theory living on the D3 branes (that we consider as frozen). This means that in our field theory description we are restricting to a sector of the full system which excludes these excitations. One consequence of this is that when applying S-duality to the brane setup we will not be able to map the full SQM operator to a dual SQM operator, but we will only map the Wilson loops which are sectors of the SQM loop.

\subsection{S-duality}
\label{ssec:Sduality}

A type IIB brane configuration realizing a 5d gauge theory can be transformed by S-duality, namely the element 
$S= \lp 
\begin{array}{cc}
0 & -1 \cr
1 & 0
\end{array} 
\rp $
of the $SL(2,\bZ)$ symmetry of IIB string theory, to a dual  brane configuration, which may realize a different 5d gauge theory. S-duality in type IIB thus implies a duality or equivalence of the two 5d gauge theories and in particular the identification of their infinite coupling SCFT limit. We will refer to the duality of 5d theories as S-duality again.

In the brane picture S-duality transforms a 5$_{(p,q)}$ brane into a 5$_{(-q,p)}$ brane. For convenience we combine it with a reflection $x^5 \leftrightarrow x^6$ so that NS5 and D5 branes are still horizontal and vertical respectively in the brane picture.\footnote{The reflection can be seen as a $\frac{\pi}{2}$ rotation in $x^{56}$, followed by a parity $x^5 \to - x^5$ reversing the orientation of one type of 5-branes. Combined with IIB S-duality, it ensures that NS5s and D5s are exchanged. This convention is different from part of the literature on the topic where S is only combined with the $\frac{\pi}{2}$ rotation.}
Therefore under S-duality the brane picture is simply flipped around the $x^5=x^6$ diagonal. 

In many situations the dual brane configuration has no D5 branes and we cannot read a dual field theory. We will only discuss situations where there is a dual 5d field theory.
When this is the case, in general the dual 5d gauge theories have different gauge gauge groups and hypermultiplet representations. The Coulomb parameters are exchanged with the effective abelian gauge couplings. 
 
 In the simplest cases, and in particular for the pure $SU(2)$ theory, S-duality brings back the brane configuration to itself with the Coulomb parameter and effective coupling exchanged $2a \leftrightarrow t + 2a$. This means that the theory is mapped to itself under this map of parameters. We say that the pure $SU(2)$ theory is self-dual. 
 We will see that $SU(2)$ theories with $N_f$ flavors are also self-dual, while $SU(N)$ theories with $N>2$ are dual to $SU(2)$ quiver theories. We will study both situations in this paper.

 \bigskip
 
 The action of S-duality on loop operators can be understood from their realization in the IIB brane picture. 
 F1 strings and D1 strings are swapped under S-duality, which means that in general Wilson loops will be exchanged with the loops created by the D1 strings. However, brane manipulations like those in Figure \ref{SU2Loops} suggest that these two classes of loops are not independent, but rather form a single class of half-BPS loops which can all be realized with D3 branes placed in the middle of the brane web. One way to phrase this is that Wilson loops of one theory are mapped to Wilson loops of the dual theory under S-duality. This is the conjecture that we  wish to verify.
 \begin{center}
\begin{tikzpicture}
  \node (A) at (1,1.5) {Theory A};
  \node (C) at (1,0.5) {Wilson loop in $\cR_A$ \quad };
  \node (B) at (6,1.5) {Theory B};
   \node (D) at (6,0.5) {Wilson loop in $\cR_B$ \quad };
  \draw[<->] (C) -- (D);
\end{tikzpicture}
\end{center}
We will make this mapping more precise in examples by providing the map of representations labelling the Wilson loops $\cR_A \leftrightarrow \cR_B$. We will see that the mapping of Wilson loops is actually slightly more complicated in the presence of massive deformations because it involves some dressing factors corresponding to background Wilson loops. 
\medskip

In the case of a self-dual theory, the brane picture predicts that the set of all Wilson loops 
gets mapped to itself under the exchange of deformation parameters, with contributions from W-boson excitations exchanged with contributions from instanton excitations. We will see that Wilson loops in certain representations -- the tensor products of fundamental representations -- are directly mapped back to themselves under the duality, they transform {\it covariantly} under S, while loops in other $SU(2)$ representations are mapped to linear combinations of Wilson loops.

\section{Loops in pure $SU(2)$ theory}
\label{sec:SU2}

The simplest theory to analyse is the pure $SU(2)$ theory, whose brane web is shown in Figure \ref{SU2}. 
According to the discussion in the previous section we expect the set of all Wilson loops to be mapped to itself under S-duality. We now wish to find precisely how S-duality acts on each individual Wilson loop. 

To do so we propose to compute the exact half-index of the 5d theory in the presence of a Wilson loop, which is the VEV of a Wilson loop on
$S^1 \times \bR^4_{\epsilon_1,\epsilon_2}$, where the loop wraps $S^1$ and is placed at the origin in $\bR^4_{\epsilon_1,\epsilon_2}$.\footnote{The name half-index comes from the fact that the superconformal index is computed by the partition function on $S^1\times S^4$, which can be obtained by  a gluing procedure from two copies of $S^1 \times \bR^4_{\epsilon_1,\epsilon_2}$. It is sometimes called hemisphere index.} Here $\bR^4_{\epsilon_1,\epsilon_2}$ denotes the $\bR^4$ Omega background with equivariant parameters $\epsilon_1,\epsilon_2$. To be more precise we will be considering the VEVs of Wilson loops normalized by the partition function. 

Such supersymmetric observables can in principle be computed by equivariant localization techniques, as discussed for example in \cite{Gaiotto:2014ina,Bullimore:2014upa,Bullimore:2014awa,Gaiotto:2015una} following the seminal works \cite{Nekrasov:2002qd,Pestun:2007rz}. 
However, in practice one encounters difficulties because the computations reduce to an integration over the moduli spaces of singular instantons localized at the origin of $\bR^4_{\epsilon_1,\epsilon_2}$. The presence of a Wilson loop affects these moduli spaces in a way that is not completely understood to our knowledge. To circumvent this difficulty it has been proposed in particular cases \cite{Kim:2016qqs} (building on the analysis of \cite{Tong:2014yna,Tong:2014cha}) that Wilson loop observables can be identified as certain contributions in partition functions of 5d-1d coupled systems, namely contributions in SQM loop observables (aka qq-characters \cite{Nekrasov:2015wsu,Bourgine:2016vsq,Kimura:2015rgi,Bourgine:2017jsi}). To compute the SQM loop observables $\vev{L_{\rm SQM}}$ one then relies on the string theory realization of the defect theory. From the brane construction it is possible to understand how the loop affects each instanton sector, as we will see below. Explicit proposals and computations have been made in \cite{Kim:2016qqs} for Wilson loops in completely antisymmetric representations in 5d $\cN=1^\ast$ $U(N)$ theory and 5d $\cN=1$ pure $U(N)$ theory, as well as in \cite{Agarwal:2018tso} for  Wilson loops in more general representations in 5d $\cN=1^\ast$ $U(N)$ theory. Here we apply the same approach to study Wilson loops in all possible tensor product of antisymmetric representations for a larger class of 5d $\cN=1$ theories with unitary gauge groups. In section \ref{sec:BranesLoops} we have proposed a relation between Wilson loops and SQM loops.  Based on the brane realization of the SQM loops we will be able to carry out the computations and extract the exact results for the Wilson loops. The validity of the method will be ensured by consistency checks, including nice S-duality properties.

\subsection{Half-index computations from residues}
\label{ssec:ResiduesComputations}

In section \ref{ssec:Loops} we predicted the equality \eqref{SU2RelTensProd} between the (normalized) VEVs of the Wilson loop in the tensor product representation $\vev{W_{\mathbf{2}^{\otimes n} }}$ and the $U(1)_f^n$ neutral sector of the (normalized) SQM loop VEVs realized with $n$ D3 branes $\vev{L_{{\rm SQM}}}$. 

\begin{figure}[th]
\centering
\includegraphics[scale=0.4]{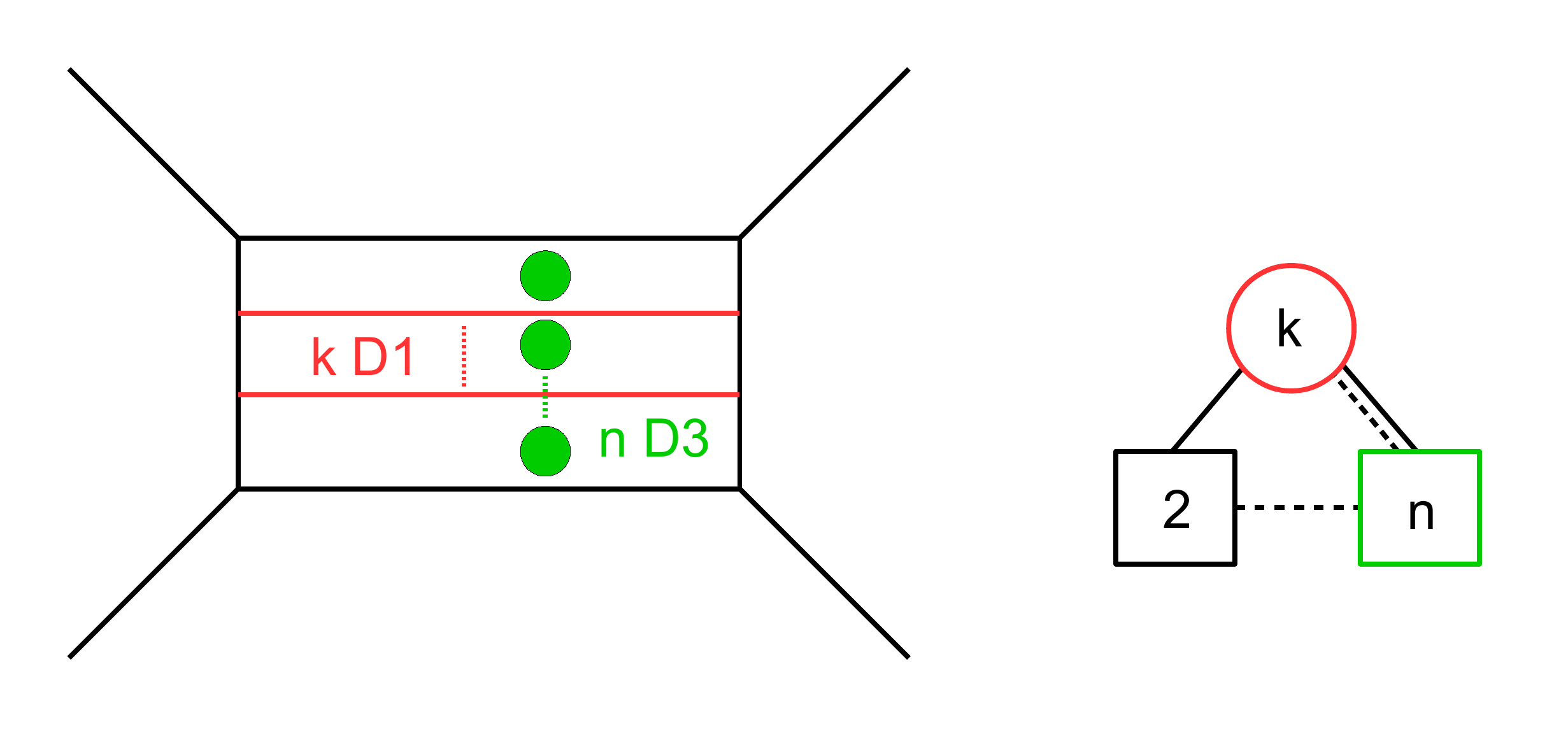}
\vspace{-0.5cm}
\caption{\footnotesize{Brane setup for the $k$-instanton sector in the presence of the SQM loop and associated $\cN=(0,4)$ ADHM quiver SQM. The circle denotes a $U(k)$ vector multiplet with an adjoint hypermultiplet, the continuous line a bifundamental hypermultiplet, the dashed line a  Fermi multiplet with single fermion and the mixed-doubled line a twisted hypermultiplet and a Fermi multiplet with two fermions. The $SU(2)$ flavor symmetry is gauged with 5d fields in the 5d-1d theory.}}
\label{SU2ADHM}
\end{figure}

The evaluation of the SQM loop on $S^1\times \bR^{4}_{\epsilon_1,\epsilon_2}$ is obtained from standard equivariant localization techniques. The exact result has the form of a supersymmetric index and depends on various fugacities:
\begin{itemize}
\item $q_1=e^{\epsilon_1}$ and $q_2=e^{\epsilon_2}$  are the fugacities associated to the symmetry generators $\frac 12 (j_1+j_2+R)$ and $\frac 12 (j_2-j_1+R)$ respectively, with $j_1,j_2$ the Cartans of the $SO(4)_{1234}\sim SU(2)_1\times SU(2)_2$ rotation symmetry on $\bR^4$ and $R$ the Cartan of the $SO(3)_{789} \sim SU(2)_R$ R-symmetry;
\item $\alpha=e^{a}$ is the fugacity associated to the Cartan generator of global $SU(2)$ gauge symmetries;
\item $Q = e^{-t}$ is the fugacity associated to the $U(1)_{\rm inst}$ symmetry (instanton counting parameter);
\item $x_i = e^{M_i}$ are the $U(n)_f$ flavor symmetry fugacities of the defect theory, with $M_i$ the masses of the SQM multiplets.
\end{itemize}
In the $\cN=(0,4)$ SQM, the R-symmetry is $SO(4) \sim SU(2)_2\times SU(2)_R$.
The result of the computation is organized in an expansion in instanton sectors weighted by $Q^{k}$, $k\ge 0$, multiplied by a common perturbative part. Since we normalize the SQM loop by the partition function in the absence of the defect, we have the following structure
\bea
Z_{\text{5d}} & = Z^{\rm pert}_{\rm 5d} (\alpha) \sum_{k\ge 0} Z^{\text{inst}, (k)}_{\rm 5d} (\alpha) \, Q^k \,, \cr
Z_{\text{5d-1d}} & = Z^{\rm pert}_{\rm 5d} (\alpha) \sum_{k\ge 0} Z^{\text{inst}, (k)}_{\text{5d-1d}}(\alpha , x) \, Q^k \,, \cr
\vev{L_{\rm SQM}} & = \frac{Z_{\text{5d-1d}}}{Z_{\text{5d}}} = \sum_{k\ge 0} c_{k}(\alpha,x) Q^k\,.
\eea
Since $Z^{\rm pert}_{\rm 5d}$ cancels in the normalization  there is no need to compute it.
The coefficient $Z^{\text{inst}, (k)}_{\rm 5d} (\alpha)$ is computed as the supersymmetric index of the ADHM quantum mechanics of the instanton sector $k$. The coefficient $Z^{\text{inst}, (k)}_{\text{5d-1d}}(\alpha , x)$ arises from a modified $\cN=(0,4)$ ADHM quantum mechanics\footnote{1d $\cN=(0,4)$ supermultiplets and Lagrangians can be constructed as the dimensional reduction of the 2d $\cN=(0,4)$ supermultiplets and Lagrangians. See \cite{Hwang:2014uwa,Kim:2016qqs} for a detailed presentation of the (0,4) ADHM quiver data. See e.g. \cite{Tong:2014yna,Putrov:2015jpa} for details on 2d (0,4) supersymmetry.\label{foot:1d04}} which can be read off from the brane realization of the SQM loop and which is shown in Figure \ref{SU2ADHM} for the SQM loop realized with $n$ D3 branes. The various (0,4) supermultiplets arise from the lowest modes of fundamental strings stretched between various D-branes. We have a $U(k)$ gauge theory with a vector multiplet and an adjoint hypermultiplet (both symbolized by a circle in the figure), $2$ fundamental hypermultiplets (continuous line), $n$ fundamental twisted hypermultiplets and $n$ Fermi multiplets with two complex fermions (doubled continuous-dashed lines), and $2n$ uncharged Fermi multiplets with a single fermion (dashed line). In addition there are potential terms (J and E terms)  required by (0,4) supersymmetry and other potentials coupling 1d and 5d fields\footnote{We did not study in detail the form of the $J$ and $E$ terms. Being Q-exact, they do not affect the computations, except for the identifications of 1d and 5d flavor symmetries which are implicit in Appendix \ref{app:ADHMcomputations}. The $J$ and $E$ terms ensuring $\cN=(0,4)$ supersymmetry can be found e.g. in \cite{Tong:2014yna}.}. 
The flavor symmetries of the ADHM theory are $SU(2)\times U(n)_f$ with fugacities $\alpha$ for $SU(2)$, identified with the global $SU(2)$ gauge transformations of the 5d theory, and $x_{i=1,\cdots, n}$ for $U(n)_f$. Closely related ADHM quantum mechanics were already considered in \cite{Tong:2014yna,Tong:2014cha,Kim:2016qqs,Agarwal:2018tso} in relation to Wilson loops in 5d $\cN=1^\ast$ theories.

We relegate the details of the computations to appendix \ref{app:ADHMcomputations}. It is still worth mentioning that we obtain our results by first considering the 5d $U(2)$ gauge theory with fugacities $\alpha_1=e^{a_1}, \alpha_2=e^{a_2}$, and then projecting onto the $SU(2)$ theory by imposing the traceless condition$a_1 = -a_2 =a$ with $\alpha=e^{a}$. There are additional subtleties to this procedure that arise when including matter hypermultiplets (see next sections and appendix \ref{app:ADHMcomputations}) and we follow \cite{Bergman:2013aca} for the precise method. 
To keep the formulas short we show some results only at the one instanton order, although we computed them up to three instanton order. 

For the $n=1$ SQM loop (single D3 brane), we find 
\be
\vev{L^{n=1}_{\rm SQM}} =  x - \left( \alpha + \alpha^{-1} - Q \frac{q_1 q_2(\alpha + \alpha^{-1})}{(1-\alpha^2 q_1 q_2)(1-\alpha^{-2} q_1 q_2)} + O(Q^2) \right)  + x^{-1} \,.
\label{Zn1}
\ee
It is a Laurent polynomial in the $U(1)_f$ fugacity $x$. We can easily relate the various terms in this polynomial to contributions from strings in the brane setup with a single D3 brane (Figure \ref{D3SQM}). In particular, the terms $x$ and $x^{-1}$ can be associated to the contributions with one string stretched from the D3 (placed in the middle) to the upper and to the lower D5 respectively; moving the D3 brane to the top, respectively to the bottom, of the brane web and taking into account the string annihilation effect we observe that the D3 brane decouples from the 5-brane array, explaining the almost trivial contribution to the SQM loop (no instanton correction). The counting parameter $x$ and $x^{-1}$ can be associated to the presence of fluxes induced by the D5 brane on the D3 worldvolume \cite{Hanany:1996ie}: with a D3 at (exponentiated) position $x$ and a D5 at (exponentiated) position $y$ we associate a classical contribution $\sqrt{x/y}$ or $\sqrt{y/x}$ if the D3 is above or below the D5. In addition, if a string is stretched from the D3 to the D5 we add a factor $y/x$ or $x/y$  if the D3 is above or below the D5. These rules ensure that the contribution of a given configuration of strings is invariant under Hanany-Witten moves of the D3 brane along $x^6$. Using these rules we understand the four classical contributions $x = \sqrt{\alpha/x}\sqrt{x\alpha}(x/\alpha)$, $x^{-1} = \sqrt{\alpha/x}\sqrt{x\alpha}(1/x\alpha)$, $\alpha = \sqrt{\alpha/x}\sqrt{x\alpha}$ and $\alpha^{-1} = \sqrt{\alpha/x}\sqrt{x\alpha}(x/\alpha)(1/x\alpha)$ as associated to the four possible string configurations (1,0), (0,1), (0,0) and (1,1) discussed in Section \ref{ssec:Loops} (the positions of the two D5s are $y=\alpha$ and $y=1/\alpha$). The other terms are instanton terms and only affect the sectors (0,0),(1,1), namely the sector neutral under $U(1)_f$, that we recognized as the fundamental Wilson loop.

Following our prescription \eqref{SU2RelFund} we can extract the fundamental Wilson loop $\vev{W_{\mathbf{2}}}$ by taking a residue over $x$, which selects the contributions from $U(1)_f$ neutral states, 
\be
\vev{W_{\mathbf{2}}} = - \oint \frac{dx}{2\pi i x} \vev{L^{n=1}_{\rm SQM}}(x) \,.
\ee
Here we have fixed the coefficient in the relation to $-1$, so that the classical contribution to the Wilson loop matches usual conventions.
This leads to 
\be
\vev{W_{\mathbf{2}}} = \alpha + \alpha^{-1} - Q \frac{q_1 q_2(\alpha + \alpha^{-1})}{(1-\alpha^2 q_1 q_2)(1-\alpha^{-2} q_1 q_2)} + O(Q^2) \,.
\label{W2}
\ee
\medskip

We can now look at larger values of $n$, where the SQM loop is defined by coupling $n$ fundamental fermions to the 5d $SU(2)$ theory (Figure \ref{D3SQM2}), with $n$ flavor fugacities $x_i$.
For $n=2$ (two D3 branes) the SQM loop evaluates to
\be 
\begin{split}
\vev{L^{n=2}_{\text{SQM}}} & = x_1 x_2 + x_1^{-1} x_2^{-1} + x_1 x_2^{-1} + x_1^{-1} x_2
- (x_1 + x_2 + x_1^{-1} + x_2^{-1}) \vev{W_{\mathbf{2}}} \\
& \quad + \vev{W_{\mathbf{2} \,\otimes \mathbf{2}}} - Q \dfrac{(1 - q_1)(1 - q_2)(1 + q_1 q_2) x_1 x_2}{(x_1 - q_1 q_2 x_2)(x_2 - q_1 q_2 x_1)} \,,
\label{Zn2}
\end{split}
\ee
where we have identified the contribution $\vev{W_{\mathbf{2}}}$, given by \eqref{W2}, and the contribution $\vev{W_{\mathbf{2} \,\otimes \mathbf{2}}}$ for the Wilson loop in the tensor product representation $\mathbf{2} \,\otimes \mathbf{2}$, with
\bea
\vev{W_{\mathbf{2} \,\otimes \mathbf{2}}} &= \oint_{\cC} \dfrac{dx_1}{2\pi i x_1} \dfrac{dx_2}{2\pi i x_2} \vev{L^{n=2}_{\rm SQM}}(x_1, x_2) \cr
&= \alpha^2 + 2 + \alpha^{-2} 
+ Q \dfrac{(1-q_1)(1-q_2)(1+q_1 q_2) - 2 q_1 q_2 (\alpha^2 + 2 + \alpha^{-2})}{(1-\alpha^2 q_1 q_2)(1-\alpha^{-2} q_1 q_2)} + O(Q^2) \,,
\label{W22}
\eea
with the contour $\cC$ for $x_1,x_2$ being circles around the origin with radii such that 
$|x_2| < q_1q_2 |x_1|$ and $|x_2| < (q_1q_2)^{-1}|x_1|$ (see explanation below).  
Here again the classical contributions to $\vev{L^{n=2}_{\text{SQM}}}$ (zero-instanton level) can be understood as associated to the possible configurations of strings stretched between the two D3s and the two D5s. The Wilson loop VEV $\vev{W_{\mathbf{2} \,\otimes \mathbf{2}}}$ corresponds, according to our prescription \eqref{SU2RelTensProd}, to the $U(1)_f^2$ invariant sector, which can be isolated by taking the residue over the two fugacities $x_1,x_2$ \eqref{W22}. Indeed we recognize the classical contribution as that of the $\mathbf{2} \,\otimes \mathbf{2}$ $SU(2)$ character. 

The appearance of the fundamental Wilson loop $\vev{W_{\mathbf{2}}}$ can be understood as the contribution from string configurations where one D3 brane has a single string attached. We can move and decouple such a D3 brane from the brane web, leaving a single D3 in the middle of the web, sourcing a fundamental Wilson loop. There are four such configurations (with one D3 in the middle and one D3 moved outside) corresponding to the four factors $\vev{W_{\mathbf{2}}}$ appearing in \eqref{Zn2}. 

In addition to the classical and Wilson loop factors there is a extra contribution in $\vev{L^{n=2}_{\text{SQM}}}$ at one-instanton level (but not at higher instanton levels) in the form of a rational function of $x_1,x_2$. We notice that this term has poles at $x_2/x_1 = q_1q_2$ and $x_2/x_1 = (q_1q_2)^{-1}$. We interpret this term in the string/brane language as arising from the motion of a D1 segment stretched between the two D3 branes. Indeed, such modes have (exponentiated) mass parameters $(x_2/x_1)^{\pm 1}$ when the D3 branes are in flat space (corresponding to $q_1q_2=1$), explaining the presence of the poles. They contribute to the VEV of a line operator in the D3 brane theory\footnote{By taking a residue over $\alpha$ we can  isolate this extra factor and recognize it as a monopole bubbling contribution for an 't Hooft loop of minimal magnetic charge in the 4d $U(2)$ SYM theory living on the D3 branes, with $x_1,x_2$ identified with the 4d Coulomb branch parameters (see \cite{Ito:2011ea,Brennan:2018yuj,Brennan:2018moe}).}
and should a priori not contribute to the Wilson loop VEV of the 5d theory that we would like to compute.

\begin{figure}[th]
\centering
\includegraphics[scale=0.7]{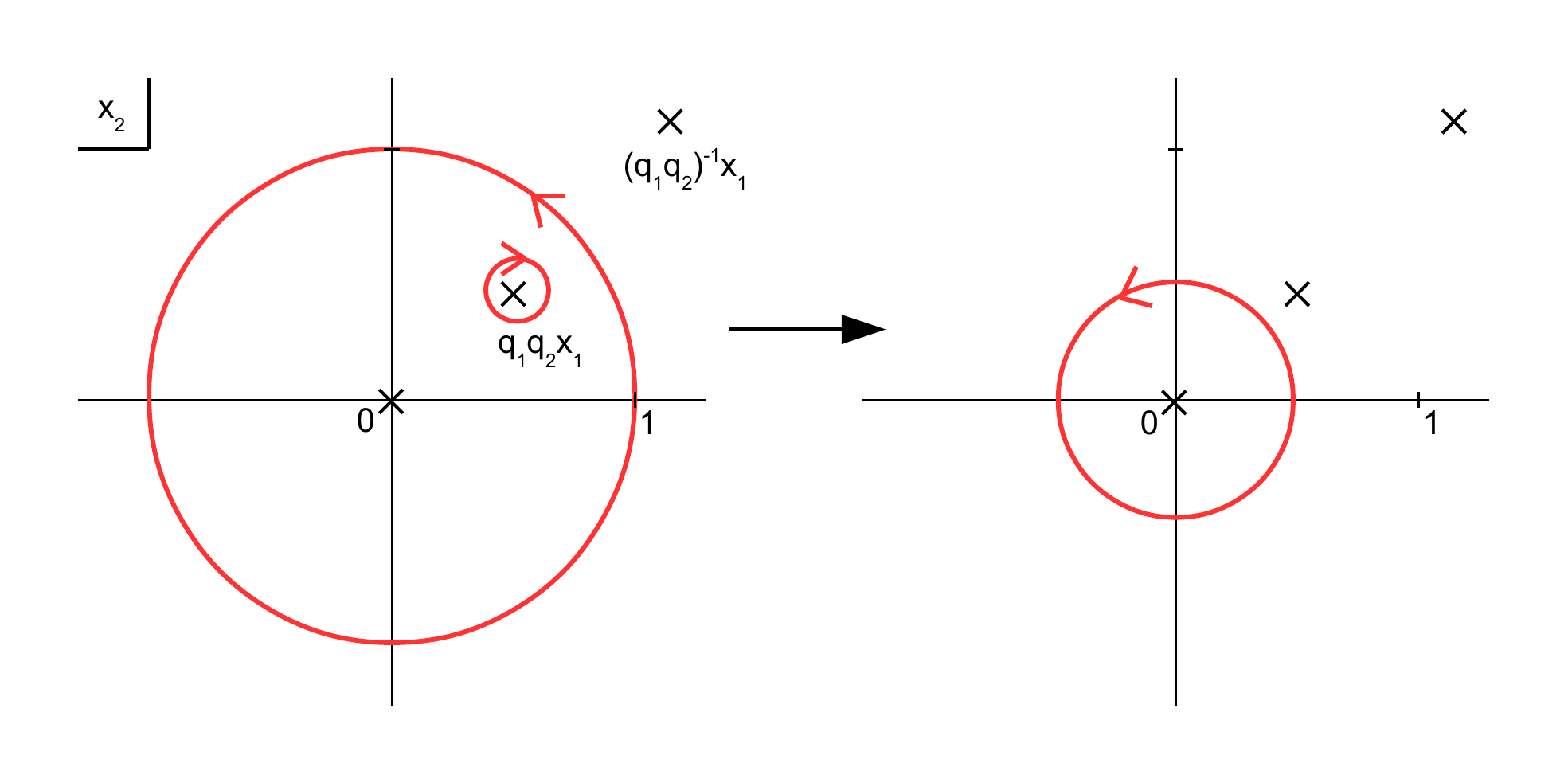}
\vspace{-0.5cm}
\caption{\footnotesize{Contour $\cC$ (red) for the $x_2$ integration, assuming $x_1$ is integrated (after $x_2$) on the unit circle. On the left: we exclude the residue at $x_2 = q_1q_2 x_1$ (keeping only the residue at zero). This can be achieved by choosing the integration circle on the right.}}
\label{contour}
\end{figure}

If we take a naive contour of integration $\cC$ as two unit circles, we would pick a residue contribution from these terms at $x_2 = (q_1q_2)^{\pm 1}x_1$. Based on the above discussion, we believe that these residues should be excluded. One way to achieve this is to define the contour $\cC$ as described above. We illustrate this in Figure \ref{contour}. This choice provides a consistent picture in the study of S-duality in the later sections. 

\medskip

The method generalizes to any $n$. The Wilson loop in the tensor product representation $\mathbf{2}^{\otimes n}$ is extracted from the SQM loop by the residue computation
\be
\vev{W_{\mathbf{2}^{\otimes n}}}  = (-1)^n \oint_{\cC} \prod_{i=1}^n \frac{dx_i}{2\pi i x_i} \vev{L^{n}_{\text{SQM}}}(x_1, \ldots, x_n) \,,
\label{residueformula}
\ee
with a contour $\cC$ around the origin such that poles at $x_i = (q_1q_2)^{\pm 1} x_j$ are excluded. For instance one can pick contours as circles around zero radii such that $|x_{i+1}| < (q_1q_2)^{\pm 1} |x_i|$, $i=1,\cdots, n-1$. This reproduces the prediction from the heuristic brane argument \eqref{SU2RelTensProd}.

Let us give one more explicit results for $n=3$,
\bea
& \vev{W_{\mathbf{2} \,\otimes \mathbf{2} \, \otimes \mathbf{2}}}  = - \oint_{\cC} \dfrac{dx_1}{2\pi i x_1} \dfrac{dx_2}{2\pi i x_2} \dfrac{dx_3}{2\pi i x_3} \vev{L_{\text{SQM}}^{n = 3}}(x_1, x_2, x_3) \cr
 & \quad  = \alpha^{3} + 3 \alpha + 3 \alpha^{-1} + \alpha^{-3}  \cr
& \quad \ + Q (\alpha + \alpha^{-1}) \dfrac{(1 - q_1)(1 - q_2)(2 + q_1 + q_2 + 2 q_1 q_2) - 3 q_1 q_2 (\alpha^2 + 2 + \alpha^{-2})}{(1 - \alpha^2 q_1 q_2)(1 - \alpha^{-2} q_1 q_2)} + O(Q^2) \,.
\eea
The fact that we always  recover the correct classical part for the Wilson loop VEVs is a confirmation of the validity of our residue procedure.
\medskip

From the evaluation of the Wilson loops in the tensor product representation $\mathbf{2}^{\otimes n}$, one can compute Wilson loops in any representation. For instance the Wilson loop in the rank two symmetric representation (spin 1/2) is given by $W_{\cS_2} = W_{\mathbf{2} \,\otimes \mathbf{2}} - W_{\cA_2} = W_{\mathbf{2} \,\otimes \mathbf{2}} -1$, where we used the fact that the rank two antisymmetric representation $\cA_2$ is trivial. 
\medskip

Although we will focus only on Wilson loops in tensor product representations in this paper, we can also compute directly the VEV of Wilson loops in rank $n$ symmetric representations $\cS_n$, which are simply the irreducible spin $n$ representations of $SU(2)$, by a different residue prescription. Following the logic of section \ref{ssec:Loops} we expect that such Wilson loops can be extracted from the SQM loop $\vev{L^{n}_{\rm SQM}}$ by projecting onto the $U(n)$ invariant sector. This is achieved by computing the residue in $x_1,x_2,\cdots, x_n$ with the $U(n)$ Haar measure,
\be
\vev{W_{\cS_n}}  = (-1)^n \oint_{\cC} \prod_{i=1}^n \dfrac{dx_i}{2\pi i x_i} \frac{1}{n!}\prod_{i\neq j} \big( 1 - \frac{x_i}{x_j} \big)\,  \vev{L^{n}_{\text{SQM}}}(x_1, \ldots, x_n) \,.
\label{residueformula2}
\ee
Once again we define the contour $\cC$ as unit circles with residues at $x_i = (q_1q_2)^{\pm 1} x_j$ removed. In explicit computations we recover, for instance, the identity $W_{\mathbf{2} \,\otimes \mathbf{2}} = 1 + W_{\cS_2}$.

\begin{figure}[th]
\centering
\includegraphics[scale=0.4]{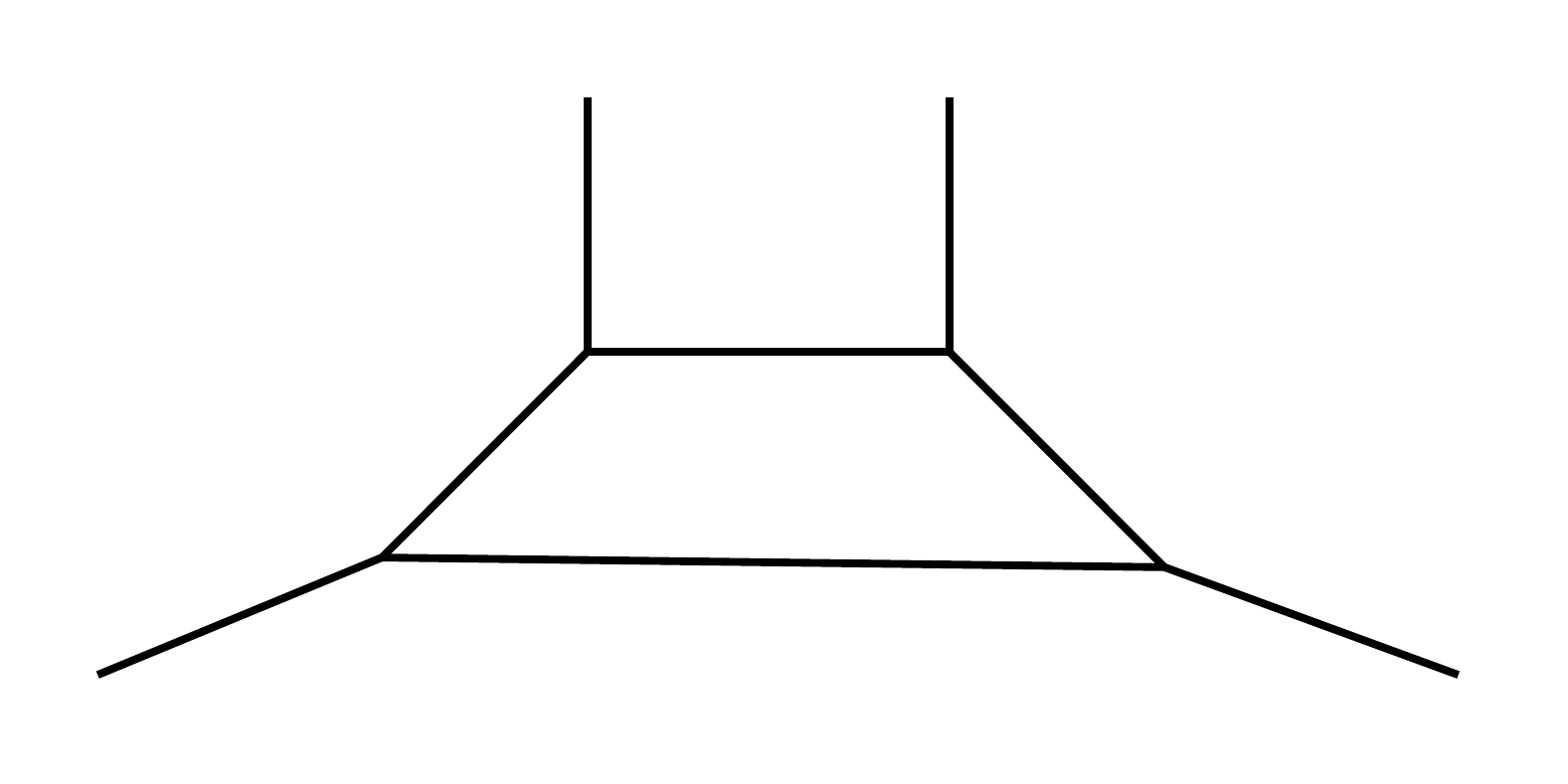}
\vspace{-0.5cm}
\caption{\footnotesize{A possible alternative brane realization of the 5d $\mathcal{N} = 1$ pure $SU(2)$ theory.}}
\label{SU2alt}
\end{figure}

Before concluding this section, for the sake of completeness, we should also mention that different string theory realizations of the 5d $\mathcal{N} = 1$ pure $SU(2)$ theory appear to have different SQM loop operators, but same Wilson loop observables. Consider for example the brane configuration in Figure \ref{SU2alt}: as argued in \cite{Bergman:2013ala,Bergman:2013aca,Bao:2013pwa,Hayashi:2013qwa} this describes the same pure $SU(2)$ theory as Figure \ref{SU2}, after removing the contribution of extra decoupled states associated to the parallel external NS5-branes\footnote{In computations, the difference arises because one starts from $U(2)$ with Chern-Simons level $\kappa=-2$ rather than $\kappa=0$, before projecting to $SU(2)$.} The partition functions of the two brane configurations coincide, modulo a factor which is independent of the $SU(2)$ gauge fugacity $\alpha$ but only depends on the instanton fugacity $Q$ \cite{Bergman:2013aca,Kim:2012gu,Hwang:2014uwa}. The situation is somehow similar, although slightly more complicated, for our SQM loop operator. For example, when adding one D3 brane the configuration in Figure \ref{SU2alt} gives
\be
\vev{L_{\text{SQM}}^{n=1}} = x (1 + Q) - \vev{W_{\mathbf{2}}} + x^{-1} \,,
\ee
with $\vev{W_{\mathbf{2}}}$ as in \eqref{W2}. Comparing with \eqref{Zn1} we see that the only difference appears in the $x$ sector, which receives a single instanton correction (due to the interaction between D1 stretched along the parallel external NS5 branes and the D3 inside of them), while the fundamental $SU(2)$ Wilson loop is the same. With two D3 branes we find instead
\bea 
\vev{L_{\text{SQM}}^{n=2}} & =  x_1 x_2 (1 + Q)^2 + x_1^{-1} x_2^{-1} + (x_1 x_2^{-1} + x_1^{-1} x_2)(1 + Q)
- (x_1 + x_2)(1 + Q) \vev{W_{\mathbf{2}}} \\
& - (x_1^{-1} + x_2^{-1}) \vev{W_{\mathbf{2}}} + \vev{W_{\mathbf{2} \,\otimes \mathbf{2}}} - Q \dfrac{(1 - q_1)(1 - q_2)(1 + q_1 q_2) x_1^2 x_2^2}{(x_1 - q_1 q_2 x_2)(x_2 - q_1 q_2 x_1)}  \,,
\eea
with $\vev{W_{\mathbf{2}}}$, $\vev{W_{\mathbf{2} \,\otimes \mathbf{2}}}$ as in \eqref{W2}, \eqref{W22} respectively. Comparing with \eqref{Zn2} we again notice that although the sectors involving positive powers of $x_1$, $x_2$ receive $Q$ corrections (and the extra rational function is also slightly modified), the Wilson loops still coincide. A similar pattern can be observed at higher number of D3 branes, as well as in more complicated theories. It is however not clear to us whether only one SQM loop VEV is the correct result, or whether the different options correspond to several SQM loops in the $SU(2)$ theory.

\subsection{S-duality of Wilson loops}
\label{SdualitySU2loops}

As we explained in the previous section the pure $SU(2)$ theory is self-dual under S-duality with the exchange of massive parameters $2a \leftrightarrow 2a+ t$, which is the map $(2a,t) \to (2a + t, -t)$. We see here that S-duality relates the theory at coupling $t$ to the theory at coupling $-t$, i.e. at negative $\frac{1}{g^2}$. It is not obvious how to make sense of the 5d theory at negative $t$. One needs to analytically continue the theory to negative $t$, assuming that observables are holomorphic in $t$. This may be possible, however we only need to assume something weaker, which is that the theory is well-defined as long as the effective coupling $t+2a$ is positive, which can be seen as a constraint on the space of vacua ($a > -t/2$). This condition ensures for instance that instantons on the Coulomb branch have positive mass.

It is convenient to introduce the exponentiated parameters, or ``fugacities",
\be
Q_F = e^{-2a} \,, \quad Q_B = e^{-t - 2a} \,;
\ee
in terms of these variables the S-duality map is 
\be
\underline{\text{S-duality map}}: \quad  (Q_F,Q_B) \to (Q_B, Q_F) \,.
\ee
The terminology $Q_F,Q_B$ refer to the fiber-base duality of toric Calabi-Yau three-folds, realizing the 5d SCFTs in M-theory, studied in \cite{Mitev:2014jza}. The M-theory realization is dual to the type IIB brane realization and the fiber-base duality of the Calabi-Yaus is the S-duality that we want to study.
\medskip

In the previous section we evaluated the Wilson loop VEVs in a small $Q=Q_B/Q_F$ expansion. To check S-duality we should further expand in small $Q_F$ and write the result as a double expansion in $Q_F,Q_B$. We find\footnote{We use the evaluation of the Wilson loop up to three instantons, which we did not explicitly write in the previous section.}
\be 
\begin{split}
Q_F^{1/2} \vev{W_{\mathbf{2}}} & = 1 + Q_F + Q_B + \chi^{A_1}_{\mathbf{3}}(q_+) Q_F Q_B
+ \chi^{A_1}_{\mathbf{5}}(q_+) Q_F Q_B (Q_F + Q_B) \\
& + Q_F Q_B (Q_F^2 + Q_F Q_B + Q_B^2) \chi^{A_1}_{\mathbf{7}}(q_+) \\
& + Q_F^2 Q_B^2 \Big( \chi^{A_1}_{\mathbf{7}}(q_+) + \chi^{A_1}_{\mathbf{5}}(q_+) + \chi^{A_1}_{\mathbf{2}}(q_-) \chi^{A_1}_{\mathbf{8}}(q_+) \Big) + \ldots ,
\end{split}
\ee
\be
\begin{split}
 Q_F \vev{W_{\mathbf{2} \,\otimes \mathbf{2}}}  &= 1 + 2(Q_F + Q_B) + \left( Q_F^2 + Q_F Q_B + Q_B^2 \right)
+ \Big( \chi^{A_1}_{\mathbf{3}}(q_+) + \chi^{A_1}_{\mathbf{2}}(q_+) \chi^{A_1}_{\mathbf{2}}(q_-) \Big) Q_F Q_B \\
&  + Q_F Q_B (Q_F + Q_B) \Big( \chi^{A_1}_{\mathbf{5}}(q_+) + \chi^{A_1}_{\mathbf{3}}(q_+) + \chi^{A_1}_{\mathbf{4}}(q_+) \chi^{A_1}_{\mathbf{2}}(q_-) \Big) \\
&  + Q_F Q_B (Q_F^2 + Q_F Q_B + Q_B^2) \Big( \chi^{A_1}_{\mathbf{7}}(q_+) + \chi^{A_1}_{\mathbf{5}}(q_+) + \chi^{A_1}_{\mathbf{6}}(q_+) \chi^{A_1}_{\mathbf{2}}(q_-) \Big) \\
&  + Q_F^2 Q_B^2 \Big( \chi^{A_1}_{\mathbf{8}}(q_+) \chi^{A_1}_{\mathbf{2}}(q_-) + \chi^{A_1}_{\mathbf{6}}(q_+) \chi^{A_1}_{\mathbf{2}}(q_-) + \chi^{A_1}_{\mathbf{4}}(q_+) \chi^{A_1}_{\mathbf{2}}(q_-) 
\\
&  \hspace{1.8 cm} + \chi^{A_1}_{\mathbf{7}}(q_+) \chi^{A_1}_{\mathbf{3}}(q_-)  + \chi^{A_1}_{\mathbf{7}}(q_+) + 2 \chi^{A_1}_{\mathbf{5}}(q_+) + 1 \Big) + \ldots,
\end{split}
\ee
\be
\begin{split}
Q_F^{3/2} \vev{W_{\mathbf{2} \,\otimes \mathbf{2} \, \otimes \mathbf{2}}} & = 1 + 3(Q_F + Q_B) + 3 (Q_F^2 + Q_F Q_B + Q_B^2) \\
& + \Big( \chi^{A_1}_{\mathbf{3}}(q_+) + \chi^{A_1}_{\mathbf{3}}(q_-)  + \chi^{A_1}_{\mathbf{2}}(q_+) \chi^{A_1}_{\mathbf{2}}(q_-) + 2 \Big) Q_F Q_B \\
& + \Big( \chi^{A_1}_{\mathbf{5}}(q_+) + 3 \chi^{A_1}_{\mathbf{3}}(q_+) + \chi^{A_1}_{\mathbf{3}}(q_+) \chi^{A_1}_{\mathbf{3}}(q_-) + \chi^{A_1}_{\mathbf{4}}(q_+) \chi^{A_1}_{\mathbf{2}}(q_-) \\ 
& \;\;\;\;\;\; + \chi^{A_1}_{\mathbf{2}}(q_+) \chi^{A_1}_{\mathbf{2}}(q_-) \Big)  
Q_F Q_B (Q_F + Q_B) + \Big(Q_F^3 + Q_F^2 Q_B + Q_F Q_B^2 + Q_B^3\Big) \\
& + \ldots.
\end{split}
\ee$SU(2) \, (\sim A_1)$ characters for various representations. Indeed $q_+$ and $q_-$ are fugacities for two $SU(2)$ symmetries of the theory: $SU(2)_{\rm diag}=$diag$(SU(2)_2\times SU(2)_R)$ and $SU(2)_1$ respectively.
\medskip

Every term in the above expansions is invariant under the S-duality map $Q_F \leftrightarrow Q_B$. However we had to multiply each Wilson loop by a factor $Q_F^{n/2}$ to obtain this result. We therefore have the identity
\be
\vev{W_{\mathbf{2}^{\otimes n}}} (Q_F,Q_B) = \lp\frac{Q_B}{Q_F}\rp^{n/2} \vev{W_{\mathbf{2}^{\otimes n}}} (Q_B,Q_F) \,.
\ee
This means that the Wilson loops are not invariant under S-duality, but rather covariant with the transformation
\be
S. W_{\mathbf{2}^{\otimes n}}(a,t) =  e^{-\frac{nt}{2}} \, W_{\mathbf{2}^{\otimes n}}(a+t/2,-t) \,,
\label{StransfoNf0}
\ee
with ``$S.$"  denoting the action of S-duality.  In the CFT limit $t\to 0$, the Wilson loops become invariant under S-duality.
The multiplicative factor $e^{-\frac{nt}{2}}$ can be interpreted as background Wilson loop of charge $-n$ for the a $U(1)_{\rm inst}$ global symmetry associated with the instanton charge.

\medskip

We thus find that Wilson loops in tensor product representations $\mathbf{2}^{\otimes n}$ transform covariantly under S-duality. From here we can deduce the transformation of Wilson loops in any representation. What we find is that in general Wilson loops do not transform covariantly, but rather pick up an inhomogeneous part in the transformation. In particular all the Wilson loops in spin $n$ representations are mapped to combinations of Wilson loops involving various representations with different multiplicative factors.

\medskip

\noindent{\bf $E_1$ symmetry}

\medskip

This is not the whole story since the pure $SU(2)$ theory is conjectured to have an  $E_1 = SU(2)_I$ global symmetry in the CFT limit ($t=0$), enhanced from the $U(1)_{\rm inst}$ symmetry, and
S-duality should correspond to the $\bZ_2$ Weyl transformation in $SU(2)_I$. To make the $SU(2)_I$ symmetry manifest
one should introduce a different set of variables,
\be
A = e^{-\frac{t}{4} - a} \,, \quad y = e^{\frac{t}{2}} = Q^{-\frac 12} \,.
\ee
The parameters $Q_F, Q_B$ are re-expressed as 
$Q_F = A^2 y$ and $Q_B = \frac{A^2}{y}$. The S-duality (or Weyl transformation) then corresponds to 
\be
\underline{\text{S-duality map}}: \quad  (A,y) \to (A, y^{-1}) \,.
\ee
The parameter $y$ is the $SU(2)_I$ fugacity.
Expanding observables in powers of $A^2$, one expects coefficients $f_n(y)$ which are $SU(2)$ characters.
This was checked at the level of the $S^1 \times \bR^4_{\epsilon_1,\epsilon_2}$ partition function or ``half-index" in \cite{Mitev:2014jza} at the first few orders in $A$, using the topological vertex formalism.

Expanding the Wilson loops in this new set of parameters we find
\be 
\begin{split}
A y^{1/2} \vev{W_{\mathbf{2}}} & = 1 + \chi^{A_1}_{\mathbf{2}}(y) A^2 
+ \chi^{A_1}_{\mathbf{3}}(q_+) A^4 + \chi^{A_1}_{\mathbf{2}}(y) 
\chi^{A_1}_{\mathbf{5}}(q_+) A^6 \\[6 pt]
& +  \Big( \chi^{A_1}_{\mathbf{3}}(y) \chi^{A_1}_{\mathbf{7}}(q_+) + \chi^{A_1}_{\mathbf{7}}(q_+) + \chi^{A_1}_{\mathbf{5}}(q_+) + \chi^{A_1}_{\mathbf{2}}(q_-) \chi^{A_1}_{\mathbf{8}}(q_+) \Big) A^8 + \ldots,
\end{split}
\ee
\be
\begin{split}
A^2 y \vev{W_{\mathbf{2} \,\otimes \mathbf{2}}} & = 1 + 2 \chi^{A_1}_{\mathbf{2}}(y) A^2 +  \Big( \chi^{A_1}_{\mathbf{3}}(y) + \chi^{A_1}_{\mathbf{3}}(q_+) + \chi^{A_1}_{\mathbf{2}}(q_+) \chi^{A_1}_{\mathbf{2}}(q_-) \Big) A^4 \\
& + \chi^{A_1}_{\mathbf{2}}(y) \Big( \chi^{A_1}_{\mathbf{5}}(q_+) + \chi^{A_1}_{\mathbf{3}}(q_+) + \chi^{A_1}_{\mathbf{4}}(q_+) \chi^{A_1}_{\mathbf{2}}(q_-) \Big) A^6 
\\
& + \chi^{A_1}_{\mathbf{3}}(y) \Big( \chi^{A_1}_{\mathbf{7}}(q_+) + \chi^{A_1}_{\mathbf{5}}(q_+) + \chi^{A_1}_{\mathbf{6}}(q_+) \chi^{A_1}_{\mathbf{2}}(q_-) \Big) A^8 
\\ 
& + \Big( \chi^{A_1}_{\mathbf{8}}(q_+) \chi^{A_1}_{\mathbf{2}}(q_-) + \chi^{A_1}_{\mathbf{6}}(q_+) \chi^{A_1}_{\mathbf{2}}(q_-) + \chi^{A_1}_{\mathbf{4}}(q_+) \chi^{A_1}_{\mathbf{2}}(q_-) + \chi^{A_1}_{\mathbf{7}}(q_+) \chi^{A_1}_{\mathbf{3}}(q_-) \\
& \;\;\;\;\;\; + \chi^{A_1}_{\mathbf{7}}(q_+) + 2 \chi^{A_1}_{\mathbf{5}}(q_+) + 1 \Big) A^8 + \ldots,
\end{split}
\ee
\be
\begin{split}
A^3 y^{3/2} \vev{W_{\mathbf{2} \,\otimes \mathbf{2} \, \otimes \mathbf{2}}} & = 
1 + 3 \chi^{A_1}_{\mathbf{2}}(y) A^2 + \Big( 3 \chi^{A_1}_{\mathbf{3}}(y) + 
\chi^{A_1}_{\mathbf{3}}(q_+) + \chi^{A_1}_{\mathbf{3}}(q_-)  + \chi^{A_1}_{\mathbf{2}}(q_+) \chi^{A_1}_{\mathbf{2}}(q_-) + 2  \Big) A^4 \\
& + \chi^{A_1}_{\mathbf{2}}(y) \Big( \chi^{A_1}_{\mathbf{5}}(q_+) + 3 \chi^{A_1}_{\mathbf{3}}(q_+) + \chi^{A_1}_{\mathbf{3}}(q_+) \chi^{A_1}_{\mathbf{3}}(q_-) + \chi^{A_1}_{\mathbf{4}}(q_+) \chi^{A_1}_{\mathbf{2}}(q_-)  \\
& \hspace{1.8 cm} + \chi^{A_1}_{\mathbf{2}}(q_+) \chi^{A_1}_{\mathbf{2}}(q_-) \Big) A^6 
+ \chi^{A_1}_{\mathbf{4}}(y) A^6 + \ldots \,.
\end{split}
\ee 
The $SU(2)_I$ characters do appear, but only after multiplying the Wilson loop by a factor $(A^2 y)^{n/2}$.

\section{Loops in $SU(2)$ theories with matter}
\label{sec:SU2Nf}

The discussion of Wilson loops in the pure $SU(2)$ theory generalizes to $SU(2)$ theories with $N_f$ fundamental flavors. These are realized via 5-brane webs with extra external D5 and NS5 branes. They are again self-dual under S-duality and we will show that the Wilson loops in the $\mathbf{2}^{\otimes n}$ representations transform covariantly under S-duality. It is well-known that the $SU(2)$ theories with $N_f$ flavors enjoy a conjectured symmetry enhancement $U(1)_{\rm inst}\times SO(2N_f) \to E_{N_f+1}$ at the CFT locus. The S-duality is again a Weyl transformation in $E_{N_f+1}$ \cite{Mitev:2014jza}.  We check this remarkable conjecture by showing that the Wilson loop VEVs on $S^1\times \bR^4_{\epsilon_1,\epsilon_2}$ admit an expansion in $E_{N_f+1}$ characters.

Because of technical limitations we studied only the cases $N_f = 1,2,3,4$, however we strongly believe that the Wilson loops in the remaining theories with $N_f =5,6,7$ have qualitatively identical properties.
In this section we provide the results for $N_f=1$ and $N_f=2$ flavors, while the theories with $N_f=3,4$ are discussed in Appendix \ref{app:Nf34} to shorten the presentation. Our results strongly support the general relation \eqref{StransfoNfgeneral} for the action of S-duality on Wilson loops in tensor product representations $\mathbf{2}^{\otimes n}$ at finite massive deformations.

\subsection{$N_f = 1$}
\label{ssec:Nf1}

\begin{figure}[th]
\centering
\includegraphics[scale=0.4]{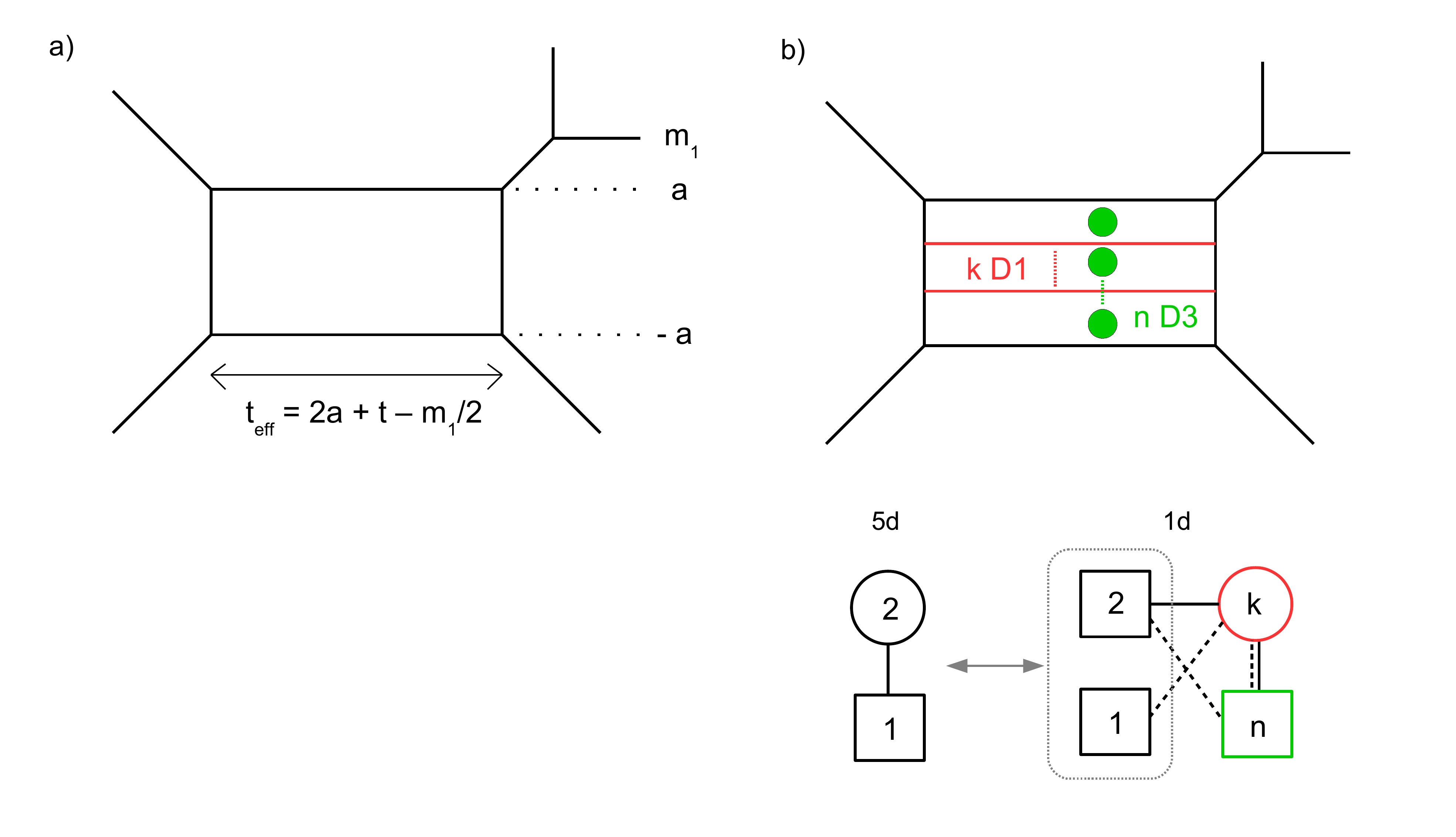}
\vspace{-1cm}
\caption{\footnotesize{a) Brane realization of the $SU(2)$ theory with $N_f=1$, with $a_1=-a_2=a$. b) Brane setup and ADHM quiver SQM for the $k$-instanton sector in the presence of an $n$ D3 branes SQM loop.}}
\label{SU2Nf1}
\end{figure}

We start by considering the $SU(2)$ gauge theory with one fundamental hypermultiplet. The brane web realizing the theory is shown in Figure \ref{SU2Nf1}-a.  It is useful to see it as arising from the $U(2)_{-1/2}$ theory with $N_f=1$, by ungauging the diagonal $U(1)$. The index $-\frac 12$ indicates a Chern-Simons at level $-\frac 12$ for the diagonal $U(1)$.\footnote{This `parent' $U(2)$ theory is also  in principle the theory realized by the brane setup \ref{SU2Nf1}-a. However the diagonal $U(1)$ subgroup of the gauge group is massive, since there is only one Coulomb branch deformation of the brane web (i.e. preserving the positions of the exterior 5-branes), corresponding to the $SU(2)$ Coulomb parameter.} 
This $U(2)$ theory is used to facilitate explicit half-index computations (see appendix \ref{app:ADHMcomputations}).

The vertical positions of the internal D5 branes are $a_1, a_2$ for the Coulomb parameters, and the vertical position of the external D5 brane is $m_1$ for the mass parameter of the hypermultiplet. 
The horizontal distance between the two NS5 branes is the effective gauge coupling $t_{\rm eff}$ of the abelian theory on a single D5. At a generic point on the Coulomb branch the adjoint real scalar is $\phi = $diag$(a_1,a_2)$, with say $a_1 > a_2$, and the prepotential evaluates to \cite{Intriligator:1997pq}
\be
\cF = \frac t2 (a_1^2 + a_2^2) + \frac{1}{6} (a_1-a_2)^3 -\frac{1}{12} \big((m_1-a_1)^3 +(m_1-a_2)^3 \big) - \frac{1}{12} (a_1^3 + a_2^3) \,,
\ee
where we assumed $m_1 > a_i$ for $i=1,2$, as in the figure.
The effective coupling on a D5 brane is 
\be
t_{\rm eff} = \frac{\p^2\cF}{\p a_1{}^2} = t + (a_1-a_2) -\frac{m_1}{2} \,.
\ee
We now impose the traceless condition $a_1 = -a_2 = a$ and define the fugacities
\be
\alpha = e^a \,, \quad \mu_1 = e^{m_1} \,.
\ee

The half-index in the presence of a Wilson loop in the tensor product representation $\mathbf{2}^{\otimes n}$ is computed using the same technology as for the pure $SU(2)$ theory. We identify the Wilson loop VEVs with sectors of the SQM loop realized by the addition of $n$ D3 branes in the center of the brane web. This SQM loop $L_{\rm SQM}^{n}$ is described, as for the pure $SU(2)$ SYM theory, by a (0,4) SQM theory with $2n$ Fermi multiplets with flavor symmetry $SU(2)\times U(n)_f$ and the $SU(2)$ flavor gauged with 5d fields.

The SQM loop VEV $\vev{L^n_{\text{SQM}}}$ is computed with the modified ADHM quiver for the $k$-instanton sector shown in Figure \ref{SU2Nf1}-b, deduced from the brane setup with $n$ D3 branes and $k$ D1 branes. This ADHM quiver is not the same as in the pure $SU(2)$ theory (there are (0,4) Fermi multiplets from strings stretched between the D1s and the external D5 and superpotential terms identifying 1d and 5d flavor symmetries). Finally the Wilson loop in $\mathbf{2}^{\otimes n}$ is extracted by the residue computation (the same as \eqref{residueformula})
\be
\vev{W_{\mathbf{2}^{\otimes n}}} =  (-1)^n \oint_{\cC} \prod_{i=1}^n \dfrac{dx_i}{2\pi i x_i} \vev{L^{n}_{\text{SQM}}}(x_1, \ldots, x_n) \,,
\ee
where $x_1,\cdots,x_n$ are the fugacities for the $U(n)_f$ SQM flavor symmetry and the contour $\cC$ is chosen such that $|x_{i+1}| < (q_1q_2)^{\pm 1} |x_i|$, for $i=1,\cdots, n-1$.

\noindent We find for $n=1$,
\bea
& \vev{L^{n=1}_{\text{SQM}}} = x_1 - \vev{W_{\mathbf{2}}} + x_1^{-1}  , \\[5 pt]
& \vev{W_{\mathbf{2}}} = \alpha + \alpha^{-1} + Q q_1 q_2 \dfrac{\mu_1^{-1/2}(q_1^{1/2} q_2^{1/2} + q_1^{-1/2} q_2^{-1/2}) - \mu_1^{1/2}(\alpha + \alpha^{-1})}{(1 - \alpha^2 q_1 q_2)(1 - \alpha^{-2} q_1 q_2)} + O(Q^2) \,.
\eea
For $n=2$,
\bea
& \vev{L^{n=2}_{\text{SQM}}}  = x_1 x_2 + x_1^{-1} x_2^{-1} + x_1 x_2^{-1} + x_1^{-1} x_2
- (x_1 + x_2 + x_1^{-1} + x_2^{-1}) \vev{W_{\mathbf{2}}} + \vev{W_{\mathbf{2} \,\otimes \mathbf{2}}} \cr
& - Q \mu_1^{1/2} \dfrac{(1 - q_1)(1 - q_2)(1 + q_1 q_2) x_1 x_2}{(x_1 - q_1 q_2 x_2)(x_2 - q_1 q_2 x_1)} 
+ Q q_1^{1/2} q_2^{1/2} \mu_1^{-1/2} \dfrac{(1 - q_1)(1 - q_2) x_1 x_2 (x_1 + x_2)}{(x_1 - q_1 q_2 x_2)(x_2 - q_1 q_2 x_1)} , \\[8 pt]
& \vev{W_{\mathbf{2} \,\otimes \mathbf{2}}} = \alpha^2 + 2 + \alpha^{-2} 
+ Q \mu_1^{1/2} \dfrac{(1-q_1)(1-q_2)(1+q_1 q_2) - 2 q_1 q_2 (\alpha^2 + 2 + \alpha^{-2})}{(1-\alpha^2 q_1 q_2)(1-\alpha^{-2} q_1 q_2)} \cr
& - Q q_1^{1/2} q_2^{1/2} \mu_1^{-1/2} \dfrac{(\alpha + \alpha^{-1})(1 + q_1)(1 + q_2)}{(1-\alpha^2 q_1 q_2)(1-\alpha^{-2} q_1 q_2)} + O(Q^2) \,.
\eea

Acting with S-duality in the brane setup  ($x^5 \leftrightarrow x^6$ reflection) we find that $2a$ is exchanged with $t_{\rm eff}= t + 2a - m_1/2$ and $m_1$ becomes $m_1-a + t_{\rm eff}/2 = 3m_1/4 +t/2$. 
The S-symmetry is the Weyl transformation in the full $E_2 = SU(2) \times U(1)$ global symmetry (enhanced from $SO(2) \times U(1)$). To make this symmetry apparent, we define 
\be
A = e^{-a -\frac t4 + \frac{m_1}{8}} \,, \quad y = e^{\frac{t}{2}-\frac{m_1}{4}} \,, \quad v = e^{-\frac{t}{4}-\frac{7m_1}{8}} \,,
\ee
giving the map of fugacities
\be
\alpha =  A^{-1} y^{-1/2}  \,, \quad \mu_1 = y^{-1/2}v^{-1} \,.
\ee
The parameter $A$ captures the Coulomb branch moduli, while $y$ and $v$ are fugacities for the $SU(2)$ and $U(1)$ global symmetries respectively.
S-duality corresponds to the action $y \to y^{-1}$, with $A$ and $v$ invariant.

Expanding further the above results (at 3-instanton order) at small $A$, we find 
\bea
A y^{1/2} \vev{W_{\mathbf{2}}} & = 1 + \chi^{A_1}_{\mathbf{2}}(y) A^2 - \chi^{A_1}_{\mathbf{2}}(q_+) v A^3 + \chi^{A_1}_{\mathbf{3}}(q_+) A^4 \cr
& \ - \chi^{A_1}_{\mathbf{2}}(y) \chi^{A_1}_{\mathbf{4}}(q_+) v A^5
 + \chi^{A_1}_{\mathbf{2}}(y) \chi^{A_1}_{\mathbf{5}}(q_+)  A^6 + \chi^{A_1}_{\mathbf{5}}(q_+) v^2 A^6 + \ldots , \cr
A^2 y \vev{W_{\mathbf{2} \,\otimes \mathbf{2}}}  & = 1 + 2 \chi^{A_1}_{\textbf{2}}(y) A^2 - \Big( \chi^{A_1}_{\textbf{2}}(q_+) +  \chi^{A_1}_{\textbf{2}}(q_-) \Big) v A^3 \cr
& \ + \Big( \chi^{A_1}_{\textbf{3}}(y) + \chi^{A_1}_{\textbf{3}}(q_+) + \chi^{A_1}_{\textbf{2}}(q_+) \chi^{A_1}_{\textbf{2}}(q_-) \Big)  A^4 \cr
& \ - \chi^{A_1}_{\textbf{2}}(y) \chi^{A_1}_{\textbf{3}}(q_+) \Big( \chi^{A_1}_{\textbf{2}}(q_+) + \chi^{A_1}_{\textbf{2}}(q_-) \Big) v A^5 \cr
& \ + \chi^{A_1}_{\textbf{2}}(y) \Big( \chi^{A_1}_{\textbf{5}}(q_+) + \chi^{A_1}_{\textbf{3}}(q_+) + \chi^{A_1}_{\textbf{4}}(q_+) \chi^{A_1}_{\textbf{2}}(q_-) \Big) A^6 \cr
& \ + \Big( \chi^{A_1}_{\textbf{5}}(q_+) + \chi^{A_1}_{\textbf{4}}(q_+) \chi^{A_1}_{\textbf{2}}(q_-) + 1 \Big) v^2 A^6 + \ldots \,.
\eea
The coefficients are expressed as characters of $SU(2)$ as in the previous section. 

Here again the characters of the $SU(2)\subset E_2$ global symmetry arise only after multiplying the Wilson loops by a factor $(A^2 y)^{n/2}$. We deduce that under S-duality the Wilson loops transform covariantly, with the S action
\be
S. W_{\mathbf{2}^{\otimes n}}(A,y,v) =  y^{-n} \, W_{\mathbf{2}^{\otimes n}}(A,y^{-1},v) \,.
\label{StransfoNf1}
\ee
This is the same transformation as in the pure $SU(2)$ theory \eqref{StransfoNf0}, except that now the parameter $y$ is $y= e^{\frac{t}{2}-\frac{m_1}{4}}$.

\subsection{$N_f = 2$}
\label{ssec:Nf2}

\begin{figure}[th]
\centering
\includegraphics[scale=0.38]{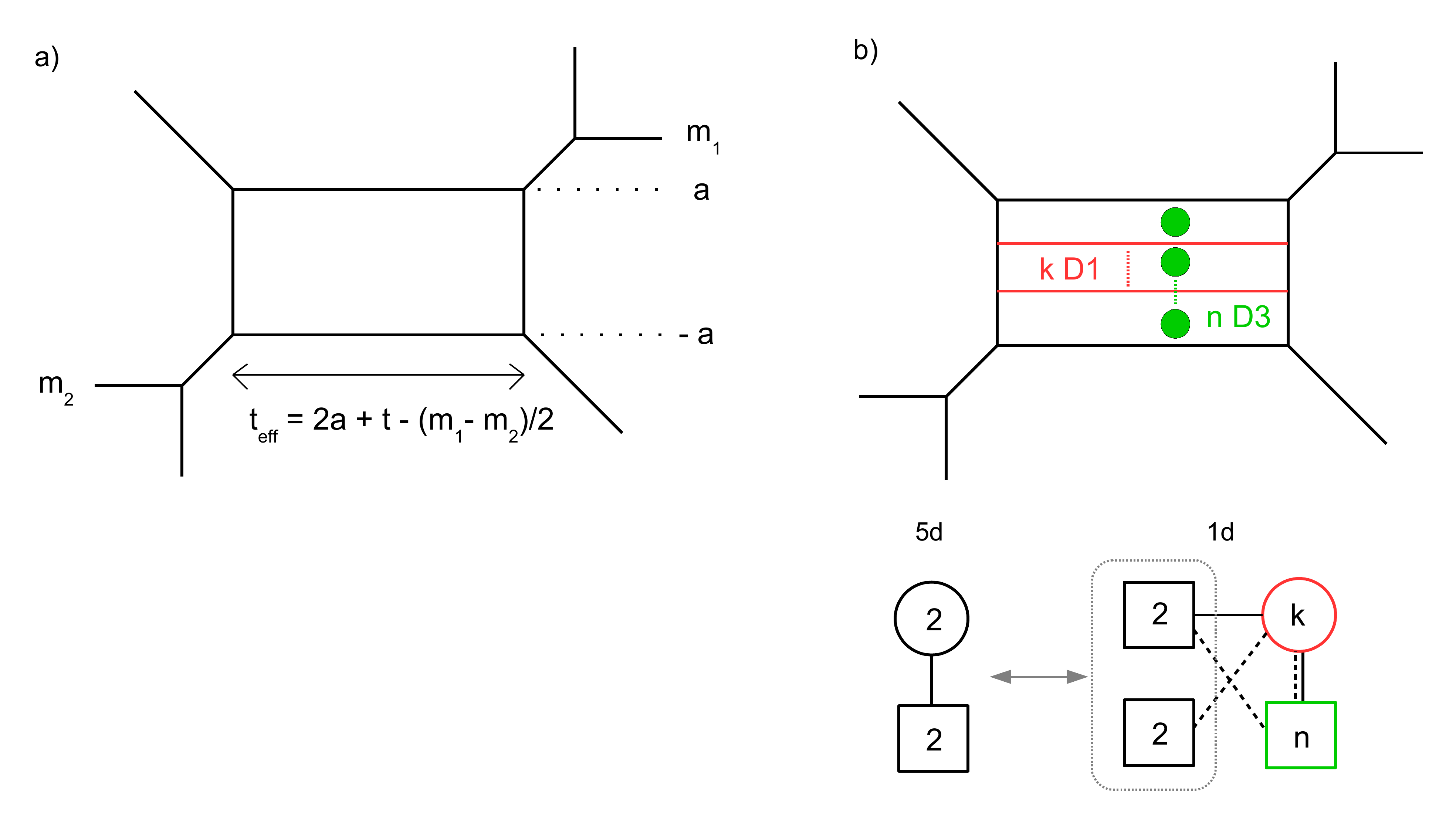}
\vspace{-0.5cm}
\caption{\footnotesize{a) Brane realization of the $SU(2)$ theory with $N_f=2$ (with $a_1=-a_2=a$). b) Brane setup and ADHM quiver SQM for the $k$-instanton sector in the presence of an $n$ D3 branes SQM loop.}}
\label{SU2Nf2}
\end{figure}

The brane realization of the $SU(2)$ theory with $N_f=2$ fundamental hypermultiplets is shown in Figure \ref{SU2Nf2}.
We can regard the theory as arising from the $U(2)$ theory with $N_f=2$ (without Chern-Simons term), by ungauging the diagonal $U(1)$. 
We denote $m_1,m_2$ the masses of the fundamental hypermultiplets.

The prepotential of the theory on the Coulomb branch, with parameter ranges $m_2 < a_1,a_2 < m_1$ (corresponding to the brane configuration of Figure \ref{SU2Nf2}), is
\be
\cF = \frac t2 (a_1^2 + a_2^2) + \frac{1}{6} (a_1-a_2)^3 -\frac{1}{12} \sum_{i=1,2}\big[ (m_1-a_i)^3 + (a_i-m_2)^3  \big] \,,
\ee
and the effective abelian coupling is
\be
t_{\rm eff} = \frac{\p^2\cF}{\p a_1{}^2} = t + (a_1-a_2) -\frac{m_1-m_2}{2} = t + 2a -\frac{m_1-m_2}{2}  \,,
\ee
corresponding to the distance between the NS5 branes in the brane configuration. In the last equality we imposed the traceless condition $a_1 = -a_2 =a$. We define the fugacities
\be
\alpha = e^a \,, \quad \mu_1 = e^{m_1} \,, \quad \mu_2 = e^{m_2} \,.
\ee

The Wilson loops $W_{\mathbf{2}^{\otimes n}}$ are evaluated from the residue formula \eqref{residueformula} from the SQM loop $L^{n}_{\rm SQM}$ defined as before, but with the modified $k$-instanton ADHM SQM shown in Figure \ref{SU2Nf2}-b. 
We find for $n = 1$
\bea
\vev{L^{n=1}_{\text{SQM}}} &= x_1 - \vev{W_{\mathbf{2}}} + x_1^{-1} \,, \\[5 pt]
\vev{W_{\mathbf{2}}} &= \alpha + \alpha^{-1} + Q q_1 q_2 \dfrac{(\mu_1^{1/2} \mu_2^{1/2} + \mu_1^{-1/2} \mu_2^{-1/2})(q_1^{1/2} q_2^{1/2} + q_1^{-1/2} q_2^{-1/2})}{(1 - \alpha^2 q_1 q_2)(1 - \alpha^{-2} q_1 q_2)}  \\
& - Q q_1 q_2 \dfrac{(\mu_1^{1/2} \mu_2^{-1/2} + \mu_1^{-1/2} \mu_2^{1/2})(\alpha + \alpha^{-1})}{(1 - \alpha^2 q_1 q_2)(1 - \alpha^{-2} q_1 q_2)} + O(Q^2)  \,, 
\eea
while for $n = 2$
\bea
\vev{L^{n=2}_{\text{SQM}}} &= x_1 x_2 + x_1^{-1} x_2^{-1} + x_1 x_2^{-1} + x_1^{-1} x_2
- (x_1 + x_2 + x_1^{-1} + x_2^{-1}) \vev{W_{\mathbf{2}}} + \vev{W_{\mathbf{2} \,\otimes \mathbf{2}}}  \cr
& + Q \frac{(1 - q_1)(1 - q_2) }{\mu_1^{1/2}\mu_2^{1/2}} 
\dfrac{ q_1^{1/2} q_2^{1/2} (x_1 x_2 + \mu_1 \mu_2) (x_1 + x_2) - x_1 x_2 (1 + q_1 q_2) (\mu_1 + \mu_2)}{(x_1 - q_1 q_2 x_2)(x_2 - q_1 q_2 x_1)} \,, \\[8 pt]
\vev{W_{\mathbf{2} \,\otimes \mathbf{2}}}  &= \alpha^2 + 2 + \alpha^{-2} 
+ Q \dfrac{\mu_1 + \mu_2}{\sqrt{\mu_1 \mu_2}} \dfrac{(1 - q_1)(1 - q_2)(1 + q_1 q_2) - 2 q_1 q_2 (\alpha^2 + 2 + \alpha^{-2})}{(1-\alpha^2 q_1 q_2)(1-\alpha^{-2} q_1 q_2)} \cr
& + Q \dfrac{1 + \mu_1 \mu_2}{\sqrt{\mu_1 \mu_2}} \dfrac{\sqrt{q_1 q_2}(1 + q_1)(1 + q_2) (\alpha + \alpha^{-1})}{(1-\alpha^2 q_1 q_2)(1-\alpha^{-2} q_1 q_2)} + O(Q^2) \,.
\eea

\medskip

\noindent S-duality, implemented by the $x^5 \leftrightarrow x^6$ reflection, acts on the parameters as follows:
\bea
\underline{\text{S action}}: \quad & a \to \frac t2 + a - \frac{m_1-m_2}{4} \,, \quad t \to - \frac t2 + \frac 34 (m_1-m_2) \,, \cr
&  m_1 \to \frac t2 +\frac{3m_1+m_2}{4} \,, \quad m_2 \to - \frac t2 +\frac{m_1+3m_2}{4} \,.
\eea
To make the $E_3 = SU(2) \times SU(3)$ global symmetries (enhanced from $SO(4) \times U(1)$) appear, we define the new set of fugacities 
\bea
& A = e^{-\frac{t}{3} - a} \,, \quad u = e^{-\frac{m_1+m_2}{2}} \,, \cr
& y_1 = e^{\frac{2t}{3}} \,, \quad y_2 = e^{-\frac{t}{3} + \frac{m_1-m_2}{2}} \,, \quad  y_3 = e^{-\frac{t}{3} - \frac{m_1-m_2}{2}} \,,
\eea
satisfying $y_1 y_2 y_3 =1$. The $y_i$ are the $SU(3)$ fugacities and $u$ is the $SU(2)$ fugacity.
In terms of the new parameters, the S action is simply $y_1 \leftrightarrow y_2$ (with the other parameters invariant) and corresponds to a Weyl transformation in $SU(3)$. In particular it does not commute with the flavor symmetry $m_1 \leftrightarrow m_2$, which is the Weyl transformation $y_2 \leftrightarrow y_3$.
The full group of Weyl symmetries of $SU(2)\times SU(3)$ corresponds to the action $u \to u^{-1}$ for $SU(2)$ and the permutations of $y_1,y_2,y_3$ for $SU(3)$. 

Expanding further the above results (at 3-instanton order) at small $A$, we find 
\bea
A y_1^{1/2} \vev{W_{\mathbf{2}}} & = 1 + \chi^{A_2}_{\textbf{3}}(\vec{y}) A^2 
- \chi^{A_1}_{\textbf{2}}(q_+) \chi^{A_1}_{\textbf{2}}(u) A^3 
+ \chi^{A_1}_{\textbf{3}}(q_+) \chi^{A_2}_{\overline{\textbf{3}}}(\vec{y}) A^4 \cr
& \ \ - \chi^{A_1}_{\textbf{4}}(q_+) \chi^{A_2}_{\textbf{3}}(\vec{y}) \chi^{A_1}_{\textbf{2}}(u) A^5 + \chi^{A_1}_{\mathbf{5}}(q_+) \chi^{A_1}_{\mathbf{3}}(u) A^6 + \chi^{A_1}_{\mathbf{5}}(q_+) \chi^{A_2}_{\textbf{8}}(\vec{y}) A^6 \cr
& \ \ + \Big( \chi^{A_1}_{\mathbf{6}}(q_+) \chi^{A_1}_{\mathbf{2}}(q_-) + \chi^{A_1}_{\mathbf{5}}(q_+) + \chi^{A_1}_{\mathbf{3}}(q_+) \Big) A^6 + \ldots \,, \cr
A^2 y_1 \vev{W_{\mathbf{2} \,\otimes \mathbf{2}}} & =
1 + 2 \chi^{A_2}_{\mathbf{3}}(\vec{y}) A^2 - \Big( \chi^{A_1}_{\mathbf{2}}(q_+) + \chi^{A_1}_{\mathbf{2}}(q_-) \Big) \chi^{A_1}_{\mathbf{2}}(u) A^3 \cr
& \ \ + \chi^{A_2}_{\mathbf{6}}(\vec{y}) A^4  
+ \Big( \chi^{A_1}_{\mathbf{3}}(q_+) + \chi^{A_1}_{\mathbf{2}}(q_+) \chi^{A_1}_{\mathbf{2}}(q_-) \Big) \chi^{A_2}_{\overline{\mathbf{3}}}(\vec{y}) A^4 \cr
& \ \ - \Big( \chi^{A_1}_{\mathbf{4}}(q_+) + \chi^{A_1}_{\mathbf{2}}(q_+) + \chi^{A_1}_{\mathbf{3}}(q_+) \chi^{A_1}_{\mathbf{2}}(q_-) \Big) \chi^{A_1}_{\mathbf{2}}(v) \chi^{A_2}_{\mathbf{3}}(\vec{y}) A^5 \cr
& \ \ + \Big( \chi^{A_1}_{\mathbf{5}}(q_+) + \chi^{A_1}_{\mathbf{3}}(q_+) + \chi^{A_1}_{\mathbf{4}}(q_+) \chi^{A_1}_{\mathbf{2}}(q_-) \Big) \chi^{A_2}_{\mathbf{8}}(\vec{y}) A^6 \cr
& \ \ + \Big( \chi^{A_1}_{\mathbf{5}}(q_+) + \chi^{A_1}_{\mathbf{4}}(q_+) \chi^{A_1}_{\mathbf{2}}(q_-) + 1 \Big) \chi^{A_1}_{\mathbf{3}}(u) A^6 + \ldots \,.
\eea
We find the expansions in characters of the $E_3$ global symmetry after multiplying the Wilson loops by a factor $(A^2 y_1)^{n/2}$. We deduce that under S-duality the Wilson loops transform covariantly, with the S action
\be
S. W_{\mathbf{2}^{\otimes n}}(A,y_1,y_2,y_3, u) =  \lp\frac{y_1}{y_2}\rp^{-\frac n2} \, W_{\mathbf{2}^{\otimes n}}(A,y_2,y_1,y_3,u) \,,
\label{StransfoNf2}
\ee
with the multiplicative parameter $(y_1/y_2)^{1/2} = e^{\frac t2 - \frac{m_1-m_2}{4}}$. Since it does not commute with the flavor Weyl symmetry $F$ exchanging $m_1$ and $m_2$ ($y_2 \leftrightarrow y_3$), we can define a second S-duality action $S' = F^{-1}.S.F$ which implements $y_1 \leftrightarrow y_3$ and transform the Wilson loop with a multiplicative parameter $(y_1/y_3)^{1/2} = e^{\frac t2 + \frac{m_1-m_2}{4}}$. The flavor symmetry transformation $F$ exchanges $S$ and $S'$.

\bigskip

We study similarly the $SU(2)$ theories with $N_f=3$ and $N_f=4$ flavors in Appendix \ref{app:Nf34}. We find again that the Wilson loop VEVs $\vev{ W_{\mathbf{2}^{\otimes n}}}$ are computed from the residue formula \eqref{residueformula}, with appropriate SQM loop $L^n_{\rm SQM}$ derived from the brane configurations with $n$ D3 branes. The results for $n=1,2$ are again consistent with the enhanced $E_{N_f+1}$ flavor symmetry at the CFT point. 

Under S-duality we find that the Wilson loops $W_{\mathbf{2}^{\otimes n}}$ transform covariantly,
\be
S. W_{\mathbf{2}^{\otimes n}}(\vec y) =  Y^{-n} \, W_{\mathbf{2}^{\otimes n}}(\vec y') \,,
\label{StransfoNfgeneral}
\ee
with $\vec y$ the fugacities, $\vec y '$ their S-transform, and  $Y = e^{\frac t2 + \frac 14 \sum_{k=1}^{N_f} (-1)^k m_k}$.\footnote{This one S-duality. Other S-duality transformations are obtained by conjugating $S$ by flavor Weyl symmetries (permutations of the $m_k$).} The parameter $Y$ can be understood as a charge one background Wilson loop for a $U(1)$ subgroup of $E_{N_f+1}$.
This is our main result for $SU(2)$ theories with $N_f \le 4$ flavor hypermultiplets. We conjecture that this will hold for $N_f=5,6,7$ and $n\ge 3$ as well.

\section{$SU(3)$-$SU(2)^2$ dualities}
\label{sec:SU3dualities}

We now explore the action of S-duality in theories which are not self-dual. The lowest rank examples relate $SU(3)$ theories with flavor hypermultiplets to $SU(2)\times SU(2)$ quiver theories. They are part of a larger group of dualities relating $SU(N)^{M-1}$ quivers to $SU(M)^{N-1}$ quivers, proposed in \cite{Aharony:1997ju,Aharony:1997bh,Katz:1997eq} and studied e.g. in \cite{Bao:2011rc,Mitev:2014jza}. We will discuss two instances of such dualities and find how the Wilson loops of one theory are mapped to the Wilson loops of the dual theory.

\subsection{$SU(3)$ $N_f = 2$ and $SU(2)_{\pi} \times SU(2)_\pi$}
\label{ssec:SU3Nf2duality}

First we consider the $SU(3)$ theory with $N_f=2$ fundamental hypermultiplets. Its brane realization is shown in Figure \ref{SU3Nf2}-a. Acting with S-duality on the brane configuration we obtain the web diagram of Figure \ref{SU2xSU2}-a, which realizes the quiver theory  $SU(2)_{\pi} \times SU(2)_{\pi}$, which has one bifundamental hypermultiplet. The index $\pi$ indicates that the $SU(2)$ gauge nodes have a non-trivial theta angle.\footnote{In the previous sections the theta angle was always vanishing.} Indeed in five dimensions an $SU(2)$ gauge theory admits a $\bZ_2$ valued deformation, parametrized by $\theta =0,\pi$, which affects the weight of instanton contributions in the path integral. We refer to \cite{Douglas:1996xp} for a more detailed discussion on the theta angle deformation and to \cite{Bergman:2013aca} for the determination of the theta angles from the brane configuration.

We will see that the exact computations of the half index with Wilson loop insertions support the S-duality map between loops that one can read from the brane picture. We start by computing the Wilson loop VEVs in the two dual theories from residues of SQM loops.

\subsubsection{$SU(3), N_f=2$ loops}
\label{sssec:SU3Nf2}

\begin{figure}[th]
\centering
\includegraphics[scale=0.3]{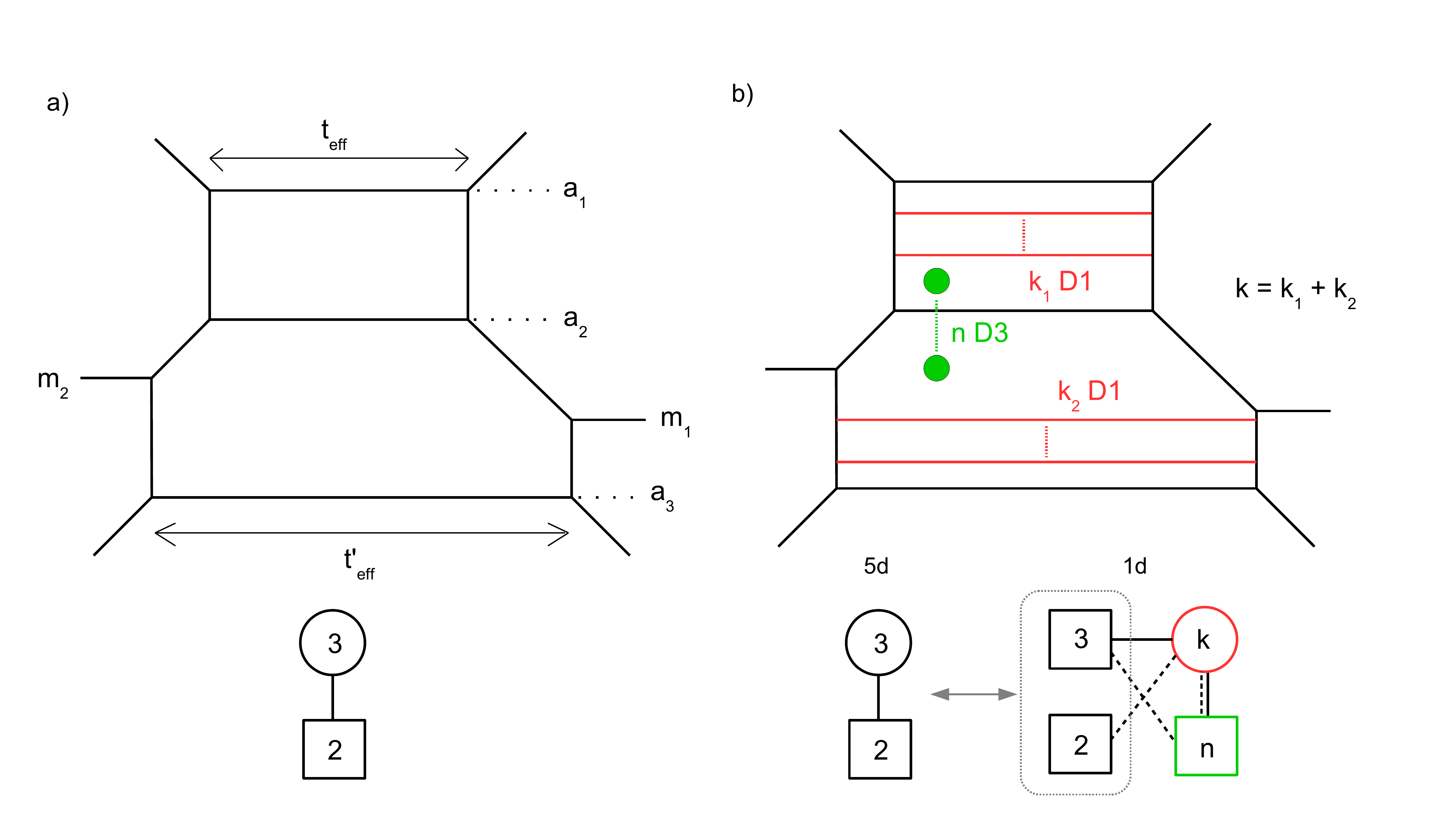}
\vspace{-0.5cm}
\caption{\footnotesize{a) Brane realization of the $SU(3)$ theory with $N_f=2$. b) Brane setup ADHM quiver SQM for the $k$-instanton sector in the presence of an $n$ D3 branes SQM loop.}}
\label{SU3Nf2}
\end{figure}

To start with we would like to compute the VEVs of Wilson loops on $S^1\times \bR^4_{\epsilon_1,\epsilon_2}$ in the $SU(3)$ theory. In particular, in analogy with the $SU(2)$ case, we will focus on Wilson loops in tensor product representations $\cR_{n_1,n_2} = \mathbf{3}^{\otimes n_1} \otimes \overline{\mathbf{3}}{}^{\otimes n_2}$. Loops in other representations can be obtained as linear combinations of those.

The Wilson loop VEVs will arise from various residues of SQM loops realized with $n = n_1+n_2$ D3 branes placed in the central regions of the brane web. The associated (0,4) SQM has $3n$ Fermi multiplets transforming in the $(\mathbf{3} , \mathbf{n})$ of $SU(3)\times U(n)_f$, where $U(n)_f$ is the flavor symmetry of the SQM and $SU(3)$ is gauged with 5d fields. 
\medskip

The string configurations contributing to a Wilson loop in the $\mathbf{3}$ are those with a D3 brane above the brane web and with a single string stretched from the D3 to any D5 segment. Upon moving the D3 brane towards the middle regions, taking into account Hanany-Witten effects, we reach configurations with a D3 brane placed between the top and middle D5s, with zero net number of strings attached. Such configurations are associated to states with a non-vanishing charge under the $U(1)_f$ flavor symmetry associated to the D3. Indeed the classical factor arising from D5 and D3 actions with this positioning is $\sqrt{\frac{\alpha_1}{x}}\sqrt{\frac{x}{\alpha_2}}\sqrt{\frac{x}{\alpha_3}} = \sqrt{\frac{\alpha_1x}{\alpha_2\alpha_3}}$ (as discussed in Section \ref{ssec:Loops}), thus in the $U(1)_f$ sector of charge $\frac 12$ (sector $x^{1/2}$). Additional strings do not change the $U(1)_f$ charge since their net number on the D3 is zero. Similarly a Wilson loop in the $\overline{\mathbf{3}}$ representation is realized from configurations with a D3 below the brane web and with a single string stretched from the D3 to any D5 segment. After Hanany-Witten moves they become configurations with the D3 placed between the middle and bottom D5s and with zero net number of strings ending in it. They carry a classical contribution $\sqrt{\frac{\alpha_1}{x}}\sqrt{\frac{\alpha_2}{x}}\sqrt{\frac{x}{\alpha_3}} = \sqrt{\frac{\alpha_1\alpha_2}{\alpha_3x}}$ and correspond to the SQM sector of $U(1)_f$ charge $-\frac 12$.

Similarly, for a Wilson loop in $\cR_{n_1,n_2}$ the string configurations contributing are those with $n_1$ D3s between the top and middle D5s and $n_2$ D3s between the middle and bottom D5s, and with zero net number of strings attached.
These configurations match the SQM sector of charge $(\underbrace{\frac 12, \cdots, \frac 12}_{n_1},\underbrace{-\frac 12, \cdots, -\frac 12}_{n_2})$ under $U(1)^{n_1} \times U(1)^{n_2} \subset U(n)$, with $n=n_1+n_2$. Here a charge $\frac 12$ or $-\frac 12$ is a flavor $U(1)$ charge associated to a single D3 brane in the upper or lower central region of the web.

We thus arrive at the following proposal for the residue relation between the SQM loop and the Wilson loops, isolating the relevant charge sector:
\be
\vev{W_{\cR_{n_1,n_2}}} = (-1)^{n}\oint_{\cC} \prod_{i=1}^{n_1}\frac{dx_i}{2\pi i x_i^{3/2}} \prod_{j=1}^{n_2}\frac{dx_j}{2\pi i x_j^{1/2}} \, \vev{L_{\rm SQM}^{n}}(x) \,,
\label{residueformulaSU3}
\ee
where $n=n_1+n_2$ , $x_i$ are the $U(n)$ fugacities, and the contour $\cC$ needs to be fixed to avoid spurious residues. As before, we will take $\cC$ to be unit circles with residues at $x_i = (q_1q_2)^{\pm 1} x_j$, $i<j$, excluded. The sign in \eqref{residueformulaSU3} is fixed a posteriori from the explicit computations.

\medskip

The evaluation of the SQM loop VEV proceeds with the $k$-instanton ADHM quiver of Figure \ref{SU3Nf2}-b, derived from the brane picture. We start from 
the computation for the $U(3)$  theory with $N_f = 2$ flavors and then impose the traceless condition $a_1 + a_2 + a_3 = 0$ on the Coulomb branch parameters.

We denote $m_1,m_2$ the flavor masses and work in the chamber $a_1 > a_2 > m_i > a_3$ as in the figure.  We define $a_{12} = a_1 - a_2$, $a_{23} = a_2 - a_3$ and the fugacities
\be 
\alpha_{12} = e^{a_{12}} \,, \quad \alpha_{23} = e^{a_{23}} \,,\quad \mu_1 = e^{m_1} \,,\quad \mu_2 = e^{m_2} \,.
\ee
The formulas that we find in terms of these parameters are too long to be reported here (we provide some explicit results in terms of other variables below). Still we find the expected structure,
for $n=1,2$,
\bea
\vev{L_{\text{SQM}}^{n = 1}} &= x_1^{3/2} - x_1^{1/2} \vev{W_{\mathbf{3}}} + x_1^{-1/2} \vev{W_{\overline{\mathbf{3}}}} - x_1^{-3/2} \,, \\[6 pt]
\vev{L_{\text{SQM}}^{n = 2}} & = x_1^{3/2} x_2^{3/2} + x_1^{-3/2} x_2^{-3/2} - x_1^{3/2} x_2^{-3/2} - x_1^{-3/2} x_2^{3/2} \cr
& - \Big( x_1^{3/2} x_2^{1/2} + x_1^{1/2} x_2^{3/2} - x_1^{-3/2} x_2^{1/2} - x_1^{1/2} x_2^{-3/2} \Big) \vev{W_{\mathbf{3}}} \cr
& - \Big( x_1^{-3/2} x_2^{-1/2} + x_1^{-1/2} x_2^{-3/2} - x_1^{3/2} x_2^{-1/2} - x_1^{-1/2} x_2^{3/2} \Big) \vev{W_{\overline{\mathbf{3}}}} \cr
& + x_1^{1/2} x_2^{1/2} \vev{W_{\mathbf{3} \, \otimes \mathbf{3}}} 
+ x_1^{-1/2} x_2^{-1/2} \vev{W_{\overline{\mathbf{3}} \, \otimes \overline{\mathbf{3}}}}
- \Big( x_1^{1/2} x_2^{-1/2} + x_1^{-1/2} x_2^{1/2} \Big) \vev{W_{\mathbf{3} \, \otimes \overline{\mathbf{3}}}} \cr
& + Q \dfrac{(1-q_1)(1-q_2)\sqrt{x_1 x_2}}{\sqrt{\mu_1 \mu_2}}
\dfrac{\sqrt{q_1 q_2} (x_1 + x_2)(\mu_1 + \mu_2) - (1 + q_1 q_2)(x_1 x_2 + \mu_1 \mu_2)}{(x_1 - q_1 q_2 x_2)(x_2 - q_1 q_2 x_1)} \,.
\label{ZSQMSU3}
\eea
The appearance of Wilson loops VEVs in \eqref{ZSQMSU3}, with the correct classical part (zero instanton sector), is in agreement and confirms the residue formula \eqref{residueformulaSU3}.
Here again we see spurious terms at one-instanton level in $\vev{L_{\rm SQM}^{n=2}}$ (last line in \eqref{ZSQMSU3}), whose poles are avoided by the contour prescription in \ref{residueformulaSU3}. 

\medskip

In order to check S-duality we introduce a new set of variables corresponding to (exponentiated) distances between D5 branes ($Q_{F_i}$) and between NS5 branes ($Q_{B_i}$),\footnote{The distances between NS5 branes are, in this case, the lengths of D5 segments and correspond to the effective abelian couplings on the Coulomb branch. They can be computed as the second derivative of the prepotential as in previous sections. Here $\cF = \frac t2 \sum_{i} a_i^2 + \frac{1}{6} \sum_{i<j}|a_i-a_j|^3 -\frac{1}{12} \sum_{i}\big[ |a_i-m_1|^3 + |a_i-m_2|^3  \big]$.}
\be
\begin{split}
& \hspace{2.4 cm} Q_{F_1} = e^{-a_{12}} = \alpha_{12}^{-1} \,,\quad Q_{F_2} = e^{-a_{23}} = \alpha_{23}^{-1} \,, \quad Q_m = e^{\frac{m_1-m_2}{2}} = \sqrt{\frac{\mu_1}{\mu_2}} \,, \\ 
& Q_{B_1} = e^{- t - \frac{4}{3}a_{12} - \frac{2}{3}a_{23} - \frac{m_1+m_2}{2}} = \frac{Q}{\alpha_{12}^{4/3} \alpha_{23}^{2/3} \sqrt{\mu_1 \mu_2}} \,, \quad
Q_{B_2} = e^{- t - \frac{2}{3}a_{12} - \frac{4}{3}a_{23} + \frac{m_1+m_2}{2}} = \frac{Q  \sqrt{\mu_1 \mu_2}}{\alpha_{12}^{2/3} \alpha_{23}^{4/3}} \,.
\end{split}
\ee
S-duality exchanges D5 and NS5 branes in  the brane web, therefore it will map $Q_B$ parameters of the $SU(3)$ theory to $Q_F$ parameters of the $SU(2)^2$ theory and vice-versa. To compare the vevs we will need a double expansion in $Q_B$ and $Q_F$ parameters. Thus we want to express the Wilson loop VEVs in terms of the new parameters and expand further in small $Q_F$. 
We show here the results at order two in $Q_F,Q_B$, and at order three in Appendix \ref{sapp:SU3Nf2},
\bea
Q_{F_1}^{2/3} Q_{F_2}^{1/3} \vev{W_{\mathbf{3}}} & = 1 + Q_{F_1} + Q_{B_1} + Q_{F_1} Q_{F_2} + Q_{F_1} Q_{B_2} + \chi^{A_1}_{\mathbf{3}}(q_+) Q_{F_1} Q_{B_1} \cr
& - \chi^{A_1}_{\mathbf{2}}(q_+) \chi^{A_1}_{\mathbf{2}}(Q_m) Q_{F_1} \sqrt{Q_{B_1}Q_{B_2}}  + \ldots \,, \cr
Q_{F_1}^{4/3} Q_{F_2}^{2/3} \vev{W_{\mathbf{3} \, \otimes \mathbf{3}}} & = 
1 + 2(Q_{F_1} + Q_{B_1}) + Q_{F_1}^2 + Q_{B_1}^2 + 2 Q_{F_1} Q_{F_2} + 2 Q_{F_1}  Q_{B_2} \cr
& + \Big( \chi^{A_1}_{\mathbf{3}}(q_+) + \chi^{A_1}_{\mathbf{2}}(q_+) \chi^{A_1}_{\mathbf{2}}(q_-) + 1 \Big) Q_{F_1} Q_{B_1} \cr
& - \Big( \chi^{A_1}_{\mathbf{2}}(q_+) + \chi^{A_1}_{\mathbf{2}}(q_-) \Big) \chi^{A_1}_{\mathbf{2}}(Q_m) Q_{F_1} \sqrt{Q_{B_1} Q_{B_2}}  + \ldots \,, \cr
Q_{F_1} Q_{F_2} \vev{W_{\mathbf{3} \, \otimes \overline{\mathbf{3}}}} & = 
1 + Q_{F_1} + Q_{F_2} + Q_{B_1} + Q_{B_2}  + 3 Q_{F_1} Q_{F_2}  \cr
&+ 2 Q_{F_1} Q_{B_2} + 2 Q_{F_2} Q_{B_1} + Q_{B_1} Q_{B_2} 
+ \chi^{A_1}_{\mathbf{3}}(q_+) (Q_{F_1} Q_{B_1} + Q_{F_2} Q_{B_2}) \cr
& - \chi^{A_1}_{\mathbf{2}}(q_+) \chi^{A_1}_{\mathbf{2}}(Q_m) (Q_{F_1} + Q_{F_2})\sqrt{Q_{B_1} Q_{B_2}} + \ldots \,.
\eea
The VEV of the Wilson loop $\vev{W_{\overline{\cR_{n_1,n_2}}}} := \vev{W_{\cR_{n_2,n_1}}} $ is obtained from $\vev{W_{\cR_{n_1,n_2}}}$ by exchanging $Q_{F_1} \leftrightarrow Q_{F_2}$, $Q_{B_1} \leftrightarrow Q_{B_2}$ and inverting $Q_m \to (Q_m)^{-1}$ (reflection about the $x^5$ axis in the brane picture).

We have multiplied the VEVs by appropriate factors $Q_{F_1}^{\frac{2n_1 + n_2}{3}}Q_{F_2}^{\frac{n_1 + 2n_2}{3}}$ to facilitate the comparison under S-duality. This normalization always corresponds to having expansions starting with a term $1$. This indicates that they are normalized indices counting some BPS states. It would be interesting to understand what these states are in detail in a future work.

\subsubsection{$SU(2)_{\pi} \times SU(2)_\pi$ loops}
\label{sssec:SU2xSU2}

\begin{figure}[th]
\centering
\includegraphics[scale=0.35]{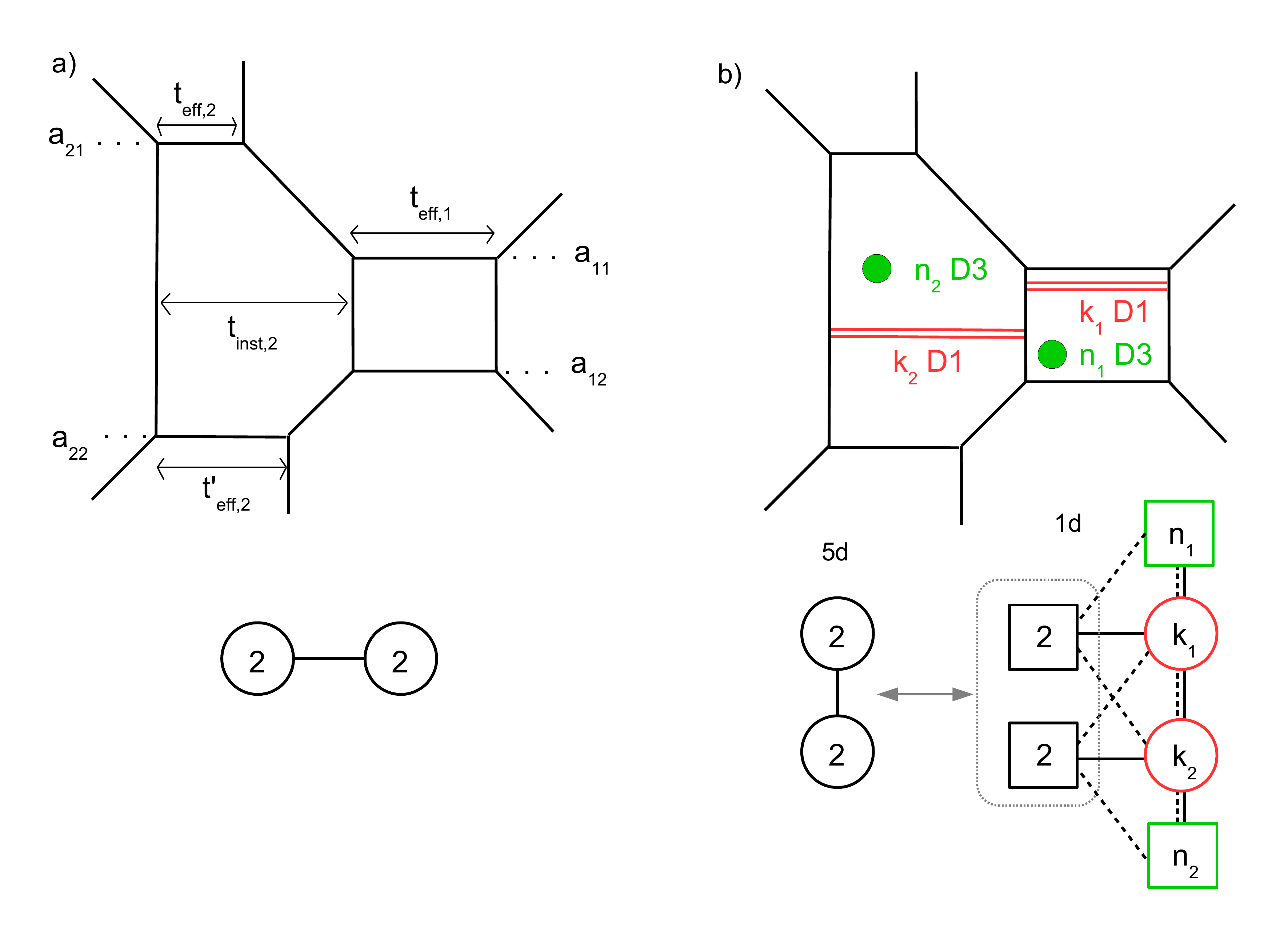}
\vspace{-0.5cm}
\caption{\footnotesize{a) Brane realization of the $SU(2)_{\pi} \times SU(2)_{\pi}$ theory. b) Brane setup ADHM quiver SQM for the $(k_1,k_2)$-instanton sector in the presence of an $(n_1,n_2)$ D3 branes SQM loop.}}
\label{SU2xSU2}
\end{figure}

In the $SU(2)_{\pi}\times SU(2)_{\pi}$ theory we consider Wilson loops in the tensor product representations $\ti\cR_{n_1,n_2} = (\mathbf{2}^{\otimes n_1} , \mathbf{2}^{\otimes n_2})$. Again other Wilson loops can be obtained as linear combination of those. 
These Wilson loops are related to the natural SQM loop that is engineered with $n_1$ D3 branes in the right-central region (between the middle and the rigth NS5 segment) and $n_2$ D3 branes in the left-central region (between the left and middle NS5 segments), as shown in Figure \ref{SU2xSU2}-b.  This SQM loop corresponds to a (0,4) SQM theory with $2n_1+2n_2$ Fermi multiplets transforming in the $(\mathbf{2},\mathbf{1},\mathbf{n_1}, \mathbf{1}) \oplus (\mathbf{1},\mathbf{2}, \mathbf{1} ,\mathbf{n_2})$ of  $SU(2)\times SU(2)\times U(n_1)_{f1}\times U(n_2)_{f2}$ with $U(n_1)_{f1}\times U(n_2)_{f2}$ the flavor symmetries and $SU(2)\times SU(2)$ gauged with 5d fields (this is the SQM theory in Figure \ref{SU2xSU2}-b when $k_1=k_2=0$).

\medskip

Following the usual heuristic argument, we say that the string configurations contributing to the Wilson loop VEV $\vev{W_{\ti\cR_{n_1,n_2}}}$ are those with $n_1$ D3s in the central right-region, $n_2$ D3s in the left-central region, and with zero net-number of strings attached. These contributions are extracted from the SQM loop VEV by selecting the $U(1)^{n_1}\times U(1)^{n_2} \subset U(n_1)_{f1}\times U(n_2)_{f2}$ neutral sector, namely by performing the residue computation
\be
\vev{W_{\ti\cR_{n_1,n_2}}} = (-1)^{n_1+n_2} \oint_{\cC} \prod_{i=1}^{n_1}\frac{dx_i}{2\pi i x_i} \prod_{j=1}^{n_2}\frac{dz_j}{2\pi i z_j} \, \vev{L_{\rm SQM}^{(n_1,n_2)}}(x,z) \,,
\label{residueformulaSU2xSU2}
\ee
 where $x_i$ and $z_j$ are the $U(n_1)_{f1}$ and $U(n_2)_{f2}$ fugacities, respectively, and the contour $\cC$ is chosen as unit circles with residues at $x_i = (q_1q_2)^{\pm 1}x_j$ and $z_i = (q_1q_2)^{\pm 1}z_j$ excluded.
\smallskip

The computation of $\vev{L_{\rm SQM}^{(n_1,n_2)}}$ is performed using the $(k_1,k_2)$-instanton ADHM quiver of figure \ref{SU2xSU2}-b, read from the brane setup with $k_1+k_2$ D1 segments.
In the presence of a non-zero theta angle for the $SU(2)$ gauge factors the computation of the half-index must be modified. We follow the prescription of \cite{Bergman:2013aca}, appendix A (see also Appendix \ref{app:ADHMcomputations}).

We start from the $U(2) \times U(2)$ theory (without Chern-Simons terms) with Coulomb parameters $a_{ij}$, $i=1,2$, $j=1,2$, and impose the trace condition $a_{11} + a_{12} = - ( a_{21} + a_{22} ) = m_{\rm bif}$ the mass of the bifundamental hypermultiplet. We then define the $SU(2)\times SU(2)$ Coulomb parameters $\ti a_1=\frac 12 (a_{11} - a_{12})$, $\ti a_2=\frac 12 (a_{21} - a_{22})$ and the fugacities\footnote{To be precise the $a_{ij}$ parameters corresponds to the $x^6$ positions of the D5 segments in the brane picture. They are related to the $a^{(I)}_j$ of Appendix \ref{app:ADHMcomputations} as $a_{1j} = a^{(1)}_j + m_{\rm bif}/2$ and $a_{2j} = a^{(2)}_j - m_{\rm bif}/2$.}
\be 
\ti\alpha_1 = e^{\ti a_1} \,, \quad \ti\alpha_2 = e^{\ti a_2} \,, \quad \ti\mu = e^{m_{\rm bif}} \,.
\ee
Here again the formulas are too long to be reported in terms of the gauge theory parameters. The result that we find from the residue formula \eqref{residueformulaSU2xSU2} reproduce the known classical parts of the Wilson loop VEVs.
 \medskip

To compare with the dual $SU(3)$ Wilson loops we introduce the new set of variables $\ti Q_{F_i}, \ti Q_{B_j}$ corresponding to (exponentiated) distances between D5 segments and between NS5 segments respectively.
\be 
\begin{split}
\widetilde{Q}_{F_1} = e^{-2 \ti a_1} = \ti\alpha_1^{-2} \,, \quad \widetilde{Q}_{F_2} = e^{-2 \ti a_2} = \ti \alpha_2^{-2} \,, \quad \widetilde{Q}_m = e^{\ti m} = \ti\mu \,, \\
\widetilde{Q}_{B_1} = e^{- \ti t_1 -2\ti a_1 + \ti a_2} = \ti Q_1 \ti \alpha_1^{-2} \ti \alpha_2 \,, \quad \widetilde{Q}_{B_2} = e^{- \ti t_2 + \ti a_1 - 2 \ti a_2} = \ti Q_2 \ti \alpha_1 \ti \alpha_2^{-2} \,.
\end{split}
\ee
We then express the results in terms a double expansion in $\ti Q_{F_i}, \ti Q_{B_j}$. We show here the expansions up to order two, and in Appendix \ref{sapp:SU2xSU2} up to order three, with appropriate multiplicative factors $\ti Q_{F_1}^{n_1/2}\ti Q_{F_2}^{n_2/2}$,
\bea
\widetilde{Q}_{F_1}^{1/2} \vev{W_{(\mathbf{2},\mathbf{1})}}  &= 1 + \widetilde{Q}_{B_1} + \widetilde{Q}_{F_1} + \widetilde{Q}_{B_1} \widetilde{Q}_{B_2} + \widetilde{Q}_{B_1} \widetilde{Q}_{F_2} + \chi^{A_1}_{\mathbf{3}}(q_+) \widetilde{Q}_{B_1} \widetilde{Q}_{F_1} \cr
& - \chi^{A_1}_{\mathbf{2}}(q_+) \chi^{A_1}_{\mathbf{2}}(\widetilde{Q}_m) \widetilde{Q}_{B_1} \sqrt{\widetilde{Q}_{F_1}\widetilde{Q}_{F_2}} + \ldots \,, \cr
\widetilde{Q}_{F_1} \vev{W_{(\mathbf{2} \, \otimes \mathbf{2},\mathbf{1})}}  &=  
1 + 2(\widetilde{Q}_{B_1} + \widetilde{Q}_{F_1}) + \widetilde{Q}_{B_1}^2 + \widetilde{Q}_{F_1}^2 + 2 \widetilde{Q}_{B_1} \widetilde{Q}_{B_2} + 2 \widetilde{Q}_{B_1}  \widetilde{Q}_{F_2} \cr
& + \Big( \chi^{A_1}_{\mathbf{3}}(q_+) + \chi^{A_1}_{\mathbf{2}}(q_+) \chi^{A_1}_{\mathbf{2}}(q_-) + 1 \Big) \widetilde{Q}_{B_1} \widetilde{Q}_{F_1} \cr
& - \Big( \chi^{A_1}_{\mathbf{2}}(q_+) + \chi^{A_1}_{\mathbf{2}}(q_-) \Big) \chi^{A_1}_{\mathbf{2}}(\widetilde{Q}_m) \widetilde{Q}_{B_1} \sqrt{\widetilde{Q}_{F_1} \widetilde{Q}_{F_2}}  + \ldots \,, \cr
\widetilde{Q}_{F_1}^{1/2} \widetilde{Q}_{F_2}^{1/2} \vev{W_{(\mathbf{2}, \mathbf{2})}} & = 
1 + \widetilde{Q}_{B_1} + \widetilde{Q}_{B_2} + \widetilde{Q}_{F_1} + \widetilde{Q}_{F_2} + 3 \widetilde{Q}_{B_1} \widetilde{Q}_{B_2} \cr
&  + 2 \widetilde{Q}_{B_1} \widetilde{Q}_{F_2} + 2 \widetilde{Q}_{B_2} \widetilde{Q}_{F_1} + \widetilde{Q}_{F_1} \widetilde{Q}_{F_2} 
+ \chi^{A_1}_{\mathbf{3}}(q_+) (\widetilde{Q}_{B_1} \widetilde{Q}_{F_1} + \widetilde{Q}_{B_2} \widetilde{Q}_{F_2}) \cr
& - \chi^{A_1}_{\mathbf{2}}(q_+) \chi^{A_1}_{\mathbf{2}}(\widetilde{Q}_m) (\widetilde{Q}_{B_1} + \widetilde{Q}_{B_2})\sqrt{\widetilde{Q}_{F_1} \widetilde{Q}_{F_2}}  + \ldots \,.
\eea
The Wilson loops $\vev{W_{\cR_{n_2,n_1}}}$ are obtained from $\vev{W_{\cR_{n_1,n_2}}}$ by the exchange $\ti Q_{F_1} \leftrightarrow \ti Q_{F_2}$, $\ti Q_{B_1} \leftrightarrow \ti Q_{B_2}$ and the inversion $\ti Q_m \to (\ti Q_m)^{-1}$, corresponding to a reflection about the $x^6$ axis in the brane picture.

\subsubsection{S-duality}
\label{sssec:SdualitySU3Nf2}

We are now ready to compare Wilson loops across S-duality and find the exact map.
The map of parameters is simply the exchange of the $Q_{F_i},Q_{B_j}$ with the $\ti Q_{B_i},\ti Q_{F_j}$:
\be
\underline{\text{S-duality map}}: \quad (Q_{F_i},Q_{B_j},Q_m) \leftrightarrow (\ti Q_{B_i}, \ti Q_{F_j}, \ti Q_m) \,.
\ee
From the brane realization of the loops we can already predict the map up to multiplicative factors. The Wilson loops realized with $n_1$ and $n_2$ D3 branes in the two central regions of the brane web, with zero net number of strings attached, are related across S-duality. We thus expect the duality to map the $SU(3)$ loop $W_{\cR_{n_1,n_2}}$ to the $SU(2)\times SU(2)$ loop $W_{\ti\cR_{n_1,n_2}}$ (we chose the notations purposefully).  From the low $n_1,n_2$ exact computations above we find the exact relation
\be
Q_{F_1}^{\frac{2n_1 + n_2}{3}}Q_{F_2}^{\frac{n_1 + 2n_2}{3}}\vev{W_{\cR_{n_1,n_2}}} = \ti Q_{F_1}^{\frac{n_1}{2}}\ti Q_{F_2}^{\frac{n_2}{2}} \vev{W_{\ti\cR_{n_1,n_2}}} \,,
\ee
which, expressed in terms of gauge theory parameters, yields
\be
\vev{W_{\cR_{n_1,n_2}}} =  Y_1^{-n_1} Y_2^{-n_2} \vev{W_{\ti\cR_{n_1,n_2}}} \,,
\ee
with $Y_1 = e^{-\frac{2\ti t_1+ \ti t_2}{3}} = e^{\frac t2 + \frac{m_1+m_2}{4}}$ and $Y_2= e^{-\frac{\ti t_1+ 2\ti t_2}{3}}= e^{\frac t2 - \frac{m_1+m_2}{4}}$. Therefore the S-duality action can be expressed as
\bea
& S.W_{\cR_{n_1,n_2}} = Y_1^{-n_1} Y_2^{-n_2} W_{\ti\cR_{n_1,n_2}} \,, \cr
& S.W_{\ti \cR_{n_1,n_2}} = Y_1^{n_1} Y_2^{n_2} W_{\cR_{n_1,n_2}} \,.
\label{SmapSU3Nf2}
\eea
The parameters $Y_1,Y_2$ can be understood as background Wilson loops of charge one for $U(1)$ subgroups of the global symmetry. For instance $Y_1$ is a charge one Wilson loop in $U(1)_{\rm diag} \subset U(1)_{\rm inst}\times U(2)_{\rm flavor}$ in the $SU(3)$ theory.

The fact that explicit computations are in agreement with the above simple formula is remarkable and provides a strong validation of the procedure we devised for extracting the Wilson loops VEVs.
\medskip

Importantly we focused on Wilson loops in the tensor product of (anti)fundamental representations $\cR_{n_1,n_2}, \ti\cR_{n_1,n_2}$. From this results one can deduce the S-duality map involving any chosen representation, however the map will be more complicated, in the sense that a given $SU(3)$ Wilson loop in representation $\cR$ will be mapped to a linear combination of $SU(2)\times SU(2)$ Wilson loops and vice-versa.


\subsection{$SU(3)$ $N_f = 6$ and $SU(2)\times SU(2)$ $N_f=2+2$}
\label{ssec:SU3Nf6duality}

As a second example we consider the $SU(3)$ theory with $N_f=6$ fundamental hypermultiplets (without Chern-Simons term). Its brane realization is shown in Figure \ref{SU3Nf6}-a. The S-dual brane configuration is that of Figure \ref{SU2xSU2Nf4}-a, which realizes the quiver theory  $SU(2) \times SU(2)$, with two fundamental hypermultiplets in each gauge node. We will call it the $SU(2)^2_{N_f=2+2}$ theory.

\subsubsection{$SU(3)$ $N_f = 6$}
\label{sssec:SU3Nf6}

\begin{figure}[th]
\centering
\includegraphics[scale=0.3]{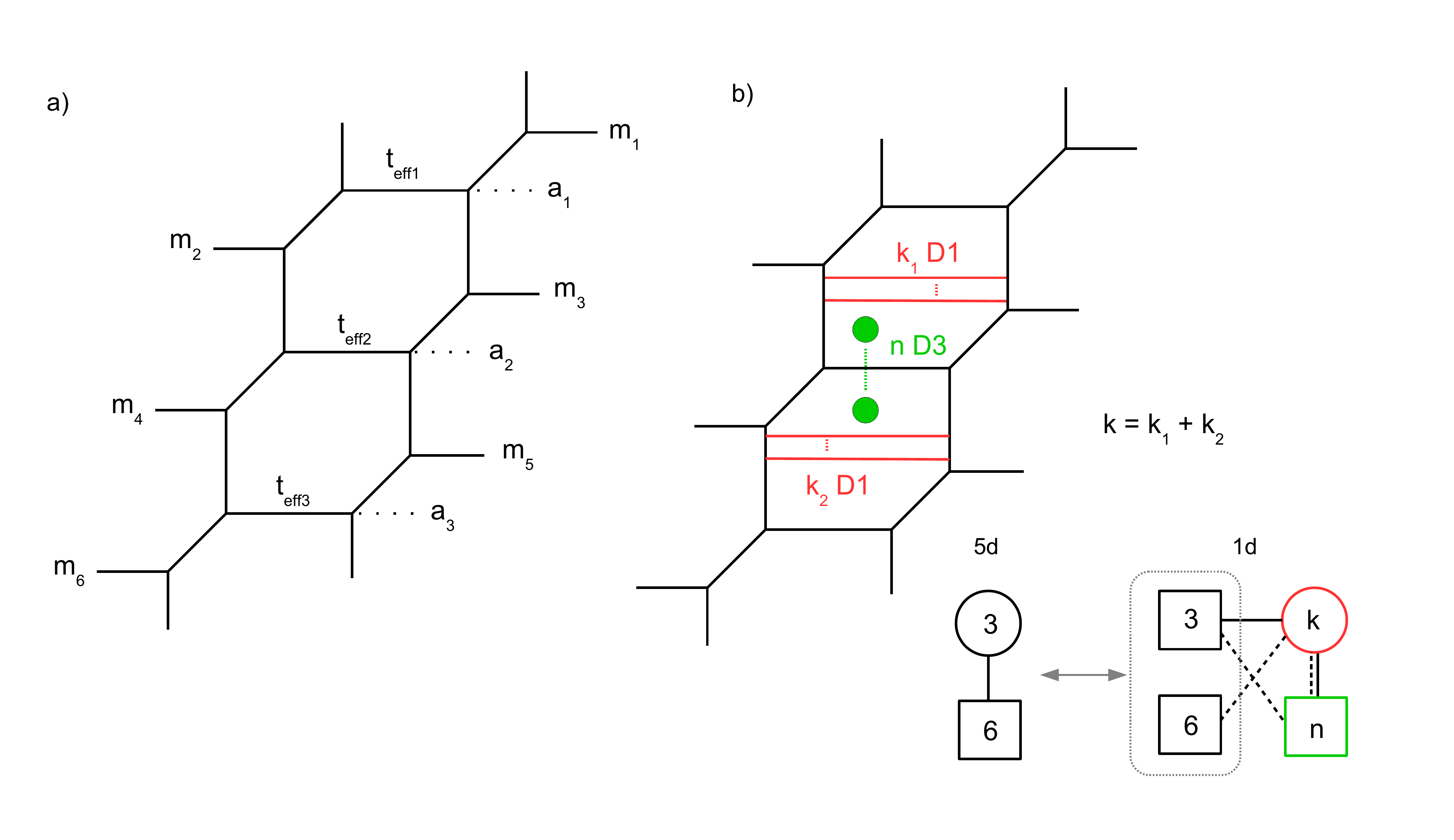}
\vspace{-0.5cm}
\caption{\footnotesize{a) Brane realization of the $SU(3)$ theory with $N_f=6$. b) Brane setup ADHM quiver SQM for the $k$-instanton sector in the presence of an $n$ D3 branes SQM loop.}}
\label{SU3Nf6}
\end{figure}

We first compute the VEVs of Wilson loops on $S^1\times \bR^4_{\epsilon_1,\epsilon_2}$ in the $SU(3)$ theory, and we focus on Wilson loops in tensor product representations $\cR_{n_1,n_2} = \mathbf{3}^{\otimes n_1} \otimes \overline{\mathbf{3}}{}^{\otimes n_2}$.

The computation is essentially the same as for the $SU(3)$ $N_f=2$ theory.
The Wilson loop VEVs will arise from residues of the SQM loops realized with $n = n_1+n_2$ D3 branes placed in the central regions of the brane web. The SQM loop 1d theory is the same as for the $SU(3)$ $N_f=2$ theory, but the $k$-instanton ADHM quiver is modified. It is given by the $(0,4)$ quiver of Figure \ref{SU3Nf6}-b.

The relation between the SQM loop and the Wilson loops is still given by \eqref{residueformulaSU3}.

\medskip

We start from the computation for the $U(3)$  theory with $N_f = 6$ flavors and then impose the traceless condition $a_1 + a_2 + a_3 = 0$ on the Coulomb branch parameters.
We denote $m_{i=1,\cdots,6}$ the flavor masses and work in the chamber $m_1 > a_1 > (m_2, m_3)  > a_2 > (m_4,m_5) > a_3 > m_6$ as depicted in the figure.  We define $a_{12} = a_1 - a_2$, $a_{23} = a_2 - a_3$ and the fugacities
\be 
\alpha_{12} = e^{a_{12}} \,, \quad \alpha_{23} = e^{a_{23}} \,,\quad Q = e^{-t} \,, \quad \mu_i = e^{m_i} \,.
\ee
The formulas that we find in terms of these parameters are again too long to be reported here. 

To check conveniently S-duality we express the results in terms of the new variables $A_1,A_2,w,z$ and $y_i$, satisfying $\prod_{i=1}^6 y_i =1$, defined as
\bea 
\alpha_{12} = \frac{1}{A_1 w} \,,\quad \alpha_{23} = \dfrac{1}{A_2 z} \,,\quad
Q = \frac{1}{w z} \,,\quad \mu_i = \left( \frac{w}{z} \right)^{1/3} y_i \quad  (i = 1, \ldots, 6) \,.
\eea
It is believed that the global symmetry group at the SCFT point is enhanced from $U(6)_{\rm flavor}\times U(1)_{\rm inst}$ to $SU(2)\times SU(2) \times SU(6)$ \cite{Bergman:2013aca,Mitev:2014jza} (see also \cite{Tachikawa:2015mha}). Our choice of parameters is such that $w$ and $z$ will be the fugacities of the two $SU(2)$ factors, while the $y_i$ will be the fugacities of the $SU(6)$. 

The new "Coulomb branch" parameters are $A_1,A_2$ and in order to check S-duality we need to expand further the results in small $A_1,A_2$. Using the ADHM quivers described in Figure \ref{SU3Nf6}-b and the residue relations, we obtain for $0 \le n_1,n_2 \le 1$,
\bea
A_1^{2/3} A_2^{1/3} w^{2/3} z^{1/3} \vev{W_{\mathbf{3}}} & = 
1 + A_1 \chi^{A_1}_{\mathbf{2}}(w) 
- A_1^{5/3} A_2^{1/3} \chi^{A_1}_{\mathbf{2}}(q_+) \chi^{A_5}_{\mathbf{6}}(\vec{y}) 
+ A_1^{4/3} A_2^{2/3} \chi^{A_5}_{\mathbf{15}}(\vec{y}) \cr
& + A_1 A_2 \chi^{A_1}_{\mathbf{2}}(w) \chi^{A_1}_{\mathbf{2}}(z) + A_1^2 \chi^{A_1}_{\mathbf{3}}(q_+) + A_1 A_2^2 \chi^{A_1}_{\mathbf{3}}(q_+)\chi^{A_1}_{\mathbf{2}}(w)\cr
& + A_1^2 A_2 \left(1 + \chi^{A_1}_{\mathbf{3}}(q_+) \right) \chi^{A_1}_{\mathbf{2}}(z) 
- A_1^2 A_2 \chi^{A_1}_{\mathbf{2}}(q_+) \chi^{A_1}_{\mathbf{2}}(w) \chi^{A_5}_{\mathbf{20}}(\vec{y}) \cr
& - A_1^{4/3} A_2^{5/3} \chi^{A_1}_{\mathbf{2}}(q_+) \chi^{A_1}_{\mathbf{2}}(w) \chi^{A_5}_{\overline{\mathbf{6}}}(\vec{y})
+ A_1^{5/3} A_2^{4/3} \chi^{A_1}_{\mathbf{2}}(w) \chi^{A_5}_{\overline{\mathbf{15}}}(\vec{y}) \cr
& - A_1^{5/3} A_2^{4/3} \chi^{A_1}_{\mathbf{2}}(q_+) \chi^{A_1}_{\mathbf{2}}(z) \chi^{A_5}_{\mathbf{6}}(\vec{y}) + A_1^3 \chi^{A_1}_{\mathbf{5}}(q_+) \chi^{A_1}_{\mathbf{2}}(w) \cr
 - A_1^{8/3} A_2^{1/3} & \chi^{A_1}_{\mathbf{4}}(q_+) \chi^{A_1}_{\mathbf{2}}(w) \chi^{A_5}_{\mathbf{6}}(\vec{y}) + A_1^{7/3} A_2^{2/3} \chi^{A_1}_{\mathbf{3}}(q_+) \chi^{A_1}_{\mathbf{2}}(w) \chi^{A_5}_{\mathbf{15}}(\vec{y}) + \ldots ,
 \label{W3}
\eea
\bea
A_1^{4/3} A_2^{2/3} w^{4/3} z^{2/3} \vev{W_{\mathbf{3} \otimes \mathbf{3}}} & = 
1 + 2 A_1 \chi^{A_1}_{\mathbf{2}}(w) + 2 A_1^{4/3} A_2^{2/3} \chi^{A_5}_{\mathbf{15}}(\vec{y}) + 2 A_1 A_2 \chi^{A_1}_{\mathbf{2}}(w) \chi^{A_1}_{\mathbf{2}}(z) \cr
& - A_1^{5/3} A_2^{1/3} \left( \chi^{A_1}_{\mathbf{2}}(q_+) + \chi^{A_1}_{\mathbf{2}}(q_-) \right) \chi^{A_5}_{\mathbf{6}}(\vec{y}) + A_1^2 \chi^{A_1}_{\mathbf{3}}(w) \cr
& + A_1^2 \left( \chi^{A_1}_{\mathbf{3}}(q_+) + \chi^{A_1}_{\mathbf{2}}(q_+) \chi^{A_1}_{\mathbf{2}}(q_-) \right) + 2 A_1 A_2^2 \chi^{A_1}_{\mathbf{3}}(q_+) \chi^{A_1}_{\mathbf{2}}(w) \cr
& + A_1^3 \chi^{A_1}_{\mathbf{4}}(q_+) \left( \chi^{A_1}_{\mathbf{2}}(q_+) + \chi^{A_1}_{\mathbf{2}}(q_-) \right) \chi^{A_1}_{\mathbf{2}}(w) \cr
& - A_1^{8/3} A_2^{1/3} \chi^{A_1}_{\mathbf{3}}(q_+) \left( \chi^{A_1}_{\mathbf{2}}(q_+) + \chi^{A_1}_{\mathbf{2}}(q_-) \right) \chi^{A_1}_{\mathbf{2}}(w) \chi^{A_5}_{\mathbf{6}}(\vec{y}) \cr
& + A_1^{7/3} A_2^{2/3} \chi^{A_1}_{\mathbf{2}}(q_+) \left( \chi^{A_1}_{\mathbf{2}}(q_+) + \chi^{A_1}_{\mathbf{2}}(q_-) \right) \chi^{A_1}_{\mathbf{2}}(w) \chi^{A_5}_{\mathbf{15}}(\vec{y}) \cr
 - 2 A_1^{4/3} A_2^{5/3} \chi^{A_1}_{\mathbf{2}} & (q_+) \chi^{A_1}_{\mathbf{2}}(w) \chi^{A_5}_{\overline{\mathbf{6}}}(\vec{y}) - A_1^2 A_2 \left( \chi^{A_1}_{\mathbf{2}}(q_+) + \chi^{A_1}_{\mathbf{2}}(q_-) \right) \chi^{A_1}_{\mathbf{2}}(w) \chi^{A_5}_{\mathbf{20}}(\vec{y}) \cr
 + A_1^2 A_2  \chi^{A_1}_{\mathbf{2}}(q_+) & \left( \chi^{A_1}_{\mathbf{2}}(q_+) + \chi^{A_1}_{\mathbf{2}}(q_-) \right) \chi^{A_1}_{\mathbf{2}}(z) + 2 A_1^2 A_2 \left( 1 + \chi^{A_1}_{\mathbf{3}}(w) \right) \chi^{A_1}_{\mathbf{2}}(z) \cr
 + 2 A_1^{5/3} A_2^{4/3} \chi^{A_1}_{\mathbf{2}}& (w) \chi^{A_5}_{\overline{\mathbf{15}}}(\vec{y}) - A_1^{5/3} A_2^{4/3} \left( \chi^{A_1}_{\mathbf{2}}(q_+) + \chi^{A_1}_{\mathbf{2}}(q_-) \right) \chi^{A_1}_{\mathbf{2}}(z) \chi^{A_5}_{\mathbf{6}}(\vec{y}) + \ldots ,
 \label{W33}
\eea
\bea
A_1 A_2 w z & \vev{W_{\mathbf{3} \otimes \overline{\mathbf{3}}}}  = 1 + A_1 \chi^{A_1}_{\mathbf{2}}(w) + A_2 \chi^{A_1}_{\mathbf{2}}(z) + A_1^2 \chi^{A_1}_{\mathbf{3}}(q_+) 
+ A_2^2 \chi^{A_1}_{\mathbf{3}}(q_+) \cr
& + 3 A_1 A_2 \chi^{A_1}_{\mathbf{2}}(w) \chi^{A_1}_{\mathbf{2}}(z) 
- A_1^{5/3} A_2^{1/3} \chi^{A_1}_{\mathbf{2}}(q_+) \chi^{A_5}_{\mathbf{6}}(\vec{y})
- A_1^{1/3} A_2^{5/3} \chi^{A_1}_{\mathbf{2}}(q_+) \chi^{A_5}_{\overline{\mathbf{6}}}(\vec{y}) \cr
& + A_1^{4/3} A_2^{2/3} \chi^{A_5}_{\mathbf{15}}(\vec{y}) + A_1^{2/3} A_2^{4/3} \chi^{A_5}_{\overline{\mathbf{15}}}(\vec{y}) + A_1^3 \chi^{A_1}_{\mathbf{5}}(q_+) \chi^{A_1}_{\mathbf{2}}(w) \cr
& + A_2^3 \chi^{A_1}_{\mathbf{5}}(q_+) \chi^{A_1}_{\mathbf{2}}(z) 
- A_1^{8/3} A_2^{1/3} \chi^{A_1}_{\mathbf{4}}(q_+) \chi^{A_1}_{\mathbf{2}}(w) \chi^{A_5}_{\mathbf{6}}(\vec{y}) \cr
& - A_1^{1/3} A_2^{8/3} \chi^{A_1}_{\mathbf{4}}(q_+) \chi^{A_1}_{\mathbf{2}}(z) \chi^{A_5}_{\overline{\mathbf{6}}}(\vec{y}) + A_1^{7/3} A_2^{2/3} \chi^{A_1}_{\mathbf{3}}(q_+) \chi^{A_1}_{\mathbf{2}}(w) \chi^{A_5}_{\mathbf{15}}(\vec{y}) \cr
& + A_1^{2/3} A_2^{7/3} \chi^{A_1}_{\mathbf{3}}(q_+) \chi^{A_1}_{\mathbf{2}}(z) \chi^{A_5}_{\overline{\mathbf{15}}}(\vec{y}) - A_1 A_2^2 \chi^{A_1}_{\mathbf{2}}(q_+) \chi^{A_1}_{\mathbf{2}}(z) \chi^{A_5}_{\mathbf{20}}(\vec{y}) \cr
& + A_1 A_2^2 \left( 2 \chi^{A_1}_{\mathbf{3}}(q_+) + \chi^{A_1}_{\mathbf{2}}(q_+) \chi^{A_1}_{\mathbf{2}}(q_-) + 1 \right) \chi^{A_1}_{\mathbf{2}}(w) + A_1 A_2^2 \chi^{A_1}_{\mathbf{2}}(w) \chi^{A_1}_{\mathbf{3}}(z) \cr
& - A_1^2 A_2 \chi^{A_1}_{\mathbf{2}}(q_+) \chi^{A_1}_{\mathbf{2}}(w) \chi^{A_5}_{\mathbf{20}}(\vec{y})  + A_1^2 A_2 \left( 2 \chi^{A_1}_{\mathbf{3}}(q_+) + \chi^{A_1}_{\mathbf{2}}(q_+) \chi^{A_1}_{\mathbf{2}}(q_-) + 1 \right) \chi^{A_1}_{\mathbf{2}}(z) \cr
& + A_1^2 A_2 \chi^{A_1}_{\mathbf{3}}(w) \chi^{A_1}_{\mathbf{2}}(z) 
- A_1^{5/3} A_2^{4/3} \left( 2\chi^{A_1}_{\mathbf{2}}(q_+) + \chi^{A_1}_{\mathbf{2}}(q_-) \right) \chi^{A_1}_{\mathbf{2}}(z) \chi^{A_5}_{\mathbf{6}}(\vec{y}) \\
& + 2 A_1^{5/3} A_2^{4/3} \chi^{A_1}_{\mathbf{2}}(w) \chi^{A_5}_{\overline{\mathbf{15}}}(\vec{y}) - A_1^{4/3} A_2^{5/3} \left( 2\chi^{A_1}_{\mathbf{2}}(q_+) + \chi^{A_1}_{\mathbf{2}}(q_-) \right) \chi^{A_1}_{\mathbf{2}}(w) \chi^{A_5}_{\overline{\mathbf{6}}}(\vec{y}) \cr
& + 2 A_1^{4/3} A_2^{5/3} \chi^{A_1}_{\mathbf{2}}(z) \chi^{A_5}_{\mathbf{15}}(\vec{y}) + \ldots .
\label{W33b}
\eea
As expected the coefficients in the expansion are characters of $SU(2)^2\times SU(6)$, providing a strong support to the symmetry enhancement proposal.

\subsubsection{$SU(2) \times SU(2)$, $N_f = 2+2$}
\label{sssec:SU2xSU2Nf22}

\begin{figure}[th]
\centering
\includegraphics[scale=0.35]{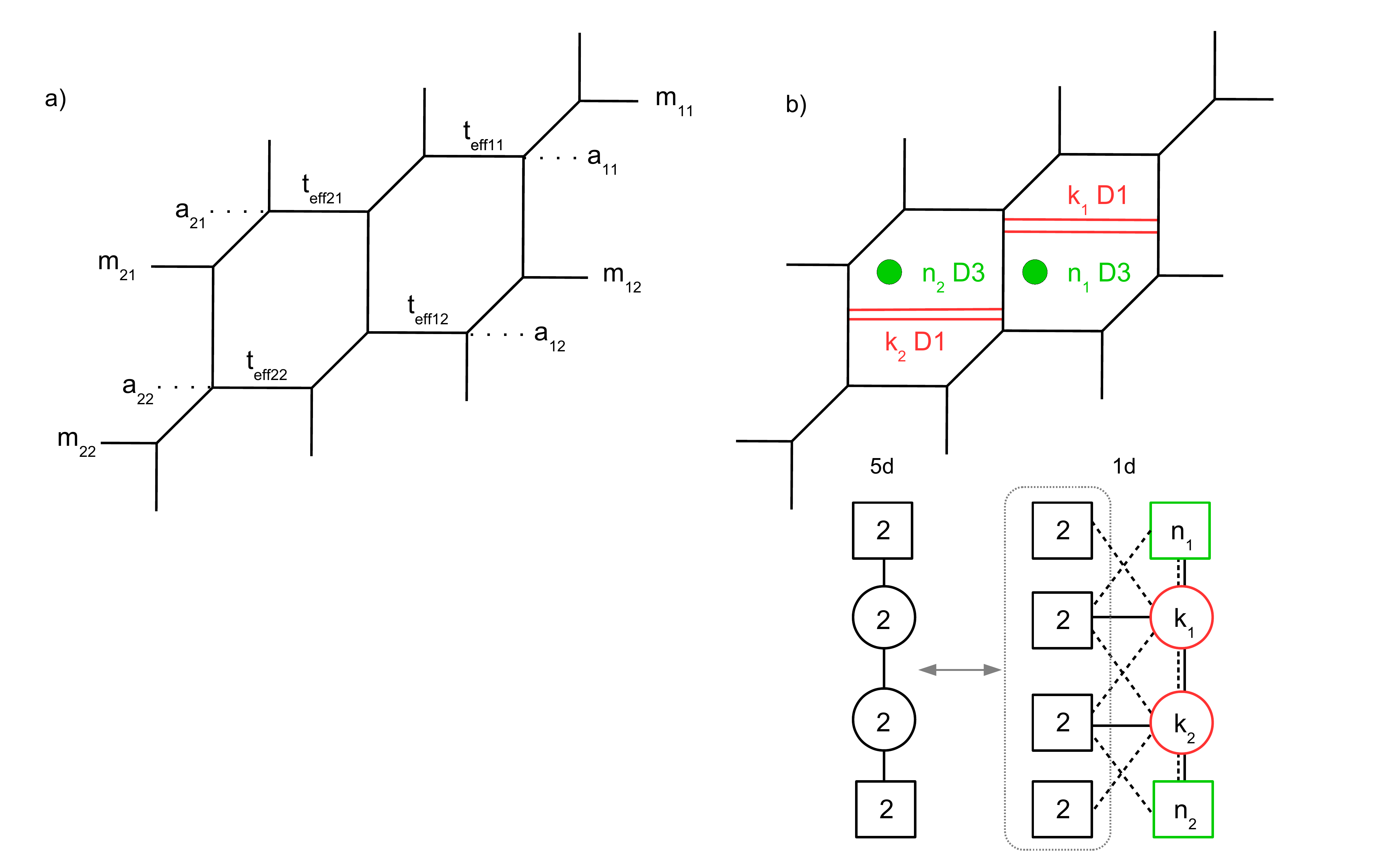}
\vspace{-0.5cm}
\caption{\footnotesize{a) Brane realization of the $SU(2)^2_{N_f=2+2}$ theory. b) Brane setup ADHM quiver SQM for the $(k_1,k_2)$-instanton sector in the presence of an $(n_1,n_2)$ D3 branes SQM loop.}}
\label{SU2xSU2Nf4}
\end{figure}

In the $SU(2)^2_{N_f=2+2}$ theory we consider Wilson loops in the tensor product representations $\ti\cR_{n_1,n_2} = (\mathbf{2}^{\otimes n_1} , \mathbf{2}^{\otimes n_2})$.
Their VEVs are computed from the residue formula \ref{residueformulaSU2xSU2} as before, from the same SQM loop VEVs, but with the $(k_1,k_2)$-instanton ADHM quiver of Figure \ref{SU2xSU2Nf4}-b. 
\medskip

The relevant fugacities are the same $\ti\alpha_1,\ti\alpha_2,\ti Q_1,\ti Q_2, \ti \mu$ as before, together with the flavor fugacities
$\ti\mu_{ij} = e^{m_{ij} - \frac 12 (a_{i1} + a_{i2})} \equiv e^{m'_{ij}}$, with $i,j=1,2$, for the 2+2 fundamental hypermultiplets. $m'_{ij}$ are the masses of the fundamental hypermultiplets in the $SU(2)^2$ theory.

We reorganize the fugacities to check S-duality and make the $SU(2)^2\times SU(6)$ symmetry manifest (enhanced from $SU(2)^2_{\rm fund}\times SU(2)_{\rm bif} \times U(1)^2_{\rm inst}$) as follows: 
\bea 
& \ti\alpha_{1} = \frac{(y_3 y_4 y_5 y_6)^{1/4}}{A_1^{2/3} A_2^{1/3} (y_1 y_2)^{1/4}} \,,\quad \ti\alpha_{2} = \frac{(y_5 y_6)^{1/4}}{A_1^{1/3} A_2^{2/3} (y_1 y_2 y_3 y_4)^{1/4}} \,, \quad
\ti Q_1 = \sqrt{\frac{y_3 y_4}{y_1 y_2}} \,,\quad  \ti Q_2 = \sqrt{\frac{y_5 y_6}{y_3 y_4}} \,, \cr
& \ti\mu = \sqrt{\frac{y_3}{y_4}} \,,\quad
\ti\mu_{11} = w \sqrt{\frac{y_1}{y_2}} \,,\quad
\ti\mu_{12} = \dfrac{1}{w} \sqrt{\frac{y_1}{y_2}}  \,, \quad
 \ti\mu_{21} = z \sqrt{\frac{y_6}{y_5}} \,,\quad
\ti\mu_{22} = \frac{1}{z} \sqrt{\dfrac{y_6}{y_5}}  \,,
\eea
where we used the same notations $A_1,A_2,w,z,y_i$ as in the previous section, providing implicitly the map of parameters under S-duality.

Expanding at small $A_1,A_2$ we find
\bea
A_1^{2/3} A_2^{1/3} \lp\frac{ y_1 y_2}{y_3 y_4 y_5 y_6}\rp^{1/4} \vev{W_{(\mathbf{2},\mathbf{1})}} & = 
\quad \text{r.h.s. of} \quad  \eqref{W3} \,, \cr
A_1^{4/3} A_2^{2/3} \lp\frac{ y_1 y_2}{y_3 y_4 y_5 y_6}\rp^{1/2} \vev{W_{(\mathbf{2} \, \otimes \mathbf{2},\mathbf{1})}} & = 
\quad \text{r.h.s. of} \quad  \eqref{W33} \,, \cr
A_1 A_2 \lp\frac{y_1 y_2}{y_5 y_6}\rp^{1/2} \vev{W_{(\mathbf{2}, \mathbf{2})}} & = 
\quad \text{r.h.s. of} \quad  \eqref{W33b} \,.
\eea

\subsubsection{S-duality}
\label{sssec:SdualitySU3Nf6}

Acting with S-duality on the brane setups realizing the loop insertions we can predict the same map between Wilson loop operators as in the previous section, up to the dressing with a background Wilson loop: the $SU(3)$ Wilson loop $W_{\cR_{n_1,n_2}}$ is mapped to the $SU(2)^2$ Wilson loop $W_{\ti\cR_{n_1,n_2}}$. The exact computations above support the precise relation
\be
\vev{W_{\cR_{n_1,n_2}}} =  Y_1^{-n_1} Y_2^{-n_2} \vev{W_{\ti\cR_{n_1,n_2}}} \,,
\ee
with 
\bea
& Y_1 = e^{-\frac{2\ti t_1+ \ti t_2}{3} + \frac 13(m'_{11}-m'_{12})+ \frac 16 (m'_{21}-m'_{22})} = e^{\frac t2 - \frac{m_1+m_2-m_3-m_4-m_5-m_6}{4}} \,, \cr
& Y_2= e^{-\frac{\ti t_1+ 2\ti t_2}{3}+ \frac 16 (m'_{11}-m'_{12})+ \frac 13 (m'_{21}-m'_{22})}= e^{\frac t2 - \frac{m_1+m_2 + m_3+m_4-m_5-m_6}{4}} \,.
\eea
The S-duality action can be expressed as
\bea
& S.W_{\cR_{n_1,n_2}} = Y_1^{-n_1} Y_2^{-n_2} W_{\ti\cR_{n_1,n_2}}  \,.
\label{SmapSU3Nf6}
\eea
The parameters $Y_1,Y_2$ can be understood as background Wilson loops of the global symmetry group.

\section{Generalization}
\label{sec:Generalization}

From the heuristic brane reasoning and the above exact results it is straightforward to conjecture the S-duality map relating the two following theories:
the $SU(M)_{(N_{f1}-N_{f2})/2}$ theory (the subscript indicates the Chern-Simons level) with $2N-4+N_{f1}+N_{f2}$ fundamental hypermultiplets, for $0\le N_{f1} \le 2$ and $0\le N_{f2} \le 2$, and  the $SU(2)^{M-1}$ linear quiver theory with $N_{f1}$ and $N_{f2}$ fundamental hypermultiplets in the left-most and right-most $SU(2)$ nodes respectively.\footnote{The conjectured duality actually extends to the ranges $0\le N_{f1} \le 4$, $0\le N_{f1} \le 4$, although this relates to more involved brane configurations compared to what we have been considering in this paper. 
The case $N_{f1}= N_{f2}=4$ is actually conjectured to describe a 6d $\cN=(1,0)$ SCFT compactified on a circle. We thank Gabi Zafrir for pointing this to us.} 

The Wilson loops transforming covariantly under S-duality are the $SU(M)$ loops $W_{(n_1,\cdots,n_{M-1})}$ in tensor product of rank $i$ antisymmetric representations $\cA_{i}$ and their dual $SU(2)^{M-1}$ loops $\ti W_{(n_1,\cdots,n_{M-1})}$ in the representation $\mathbf{2}^{\otimes n_i}$ for each quiver node:
\be
\cA_1^{\otimes n_1} \otimes \cA_2^{\otimes n_2} \otimes \cdots \otimes \cA_{M-1}^{\otimes n_{M-1}} \leftrightarrow (\mathbf{2}^{\otimes n_1}, \mathbf{2}^{\otimes n_2}, \cdots, \mathbf{2}^{\otimes n_{M-1}}) \,.
\ee
The associated SQM loops are realized with stacks of $n_1,n_2,\cdots,n_{M-1}$ D3 branes placed in the $M-1$ central regions of the brane system. The results in this paper generalize to the S-duality map
\bea
& S.W_{(n_1,\cdots,n_{M-1})} = Y_1^{-n_1} \cdots Y_{M-1}^{-n_{M-1}} \ti W_{(n_1,\cdots,n_{M-1})} \,,
\label{SmapGen}
\eea
with parameters $Y_i$ which are background Wilson loops, for which we conjecture the expressions in terms of $SU(M)$ parameters
$Y_i = e^{\frac t2 - \frac 14(\sum_{k=1}^{N_{f1}} \hat m_k + \sum_{k=1}^{2i-2}m_k - \sum_{k=2i-1}^{2M-4} m_k - \sum_{k=1}^{N_{f2}} \check m_k)}$, where $\hat m_k$, $m_k$ and $\check m_k$ are the masses of the $N_{f1}$, $2M-4$ and $N_{k2}$ fundamental hypermultiplets respectively.

Further generalization to the $SU(M)^{N-1} - SU(N)^{M-1}$ duality can also be worked out along the same lines.

\section*{Acknowledgements}

We would like to thank Joonho Kim and Chiung Hwang for discussions and technical support, as well as Hee-Cheol Kim, Diego Rodriguez-Gomez and Gabi Zafrir for valuable comments on the draft.


\appendix

\section{ADHM formulae}
\label{app:ADHMcomputations}

In this appendix we collect the formulae we used in the main text to compute the partition function on the Omega-deformed flat space $\mathbb{R}^4_{\epsilon_{1,2}} \times S^1$, or ``half-index", of 5d $\mathcal{N} = 1$ $SU(N)$ theories with $N_f$ flavors, engineered via 5-brane webs, in the presence of the SQM loop operator generated by the insertion of $n$ D3 branes. 
More precisely what we introduce below are the formulae for $U(N)_{\kappa}$ theories, with Chern-Simons level $\kappa$.\footnote{The specific choices of $\kappa$ is dictated by the number of flavors $N_f$ in the theory and the sign of their mass parameters (and related to the angles of the external 5-branes of the web).}
 $SU(N)_{\kappa}$ results can be recovered after imposing the traceless condition and performing some other minor changes, as discussed for example in \cite{Bergman:2013ala,Bergman:2013aca} and reviewed below.
Most of this short review is based on \cite{Kim:2016qqs,Hwang:2014uwa}.

\subsection{Single gauge node case}

Let us start by considering the half-index $Z_{\text{5d}}$ for a 5d $\mathcal{N} = 1$ $U(N)_{\kappa}$ theory with $N_f$ fundamental matter (without loop operators), which is just the partition function of the 5d theory on the Omega-deformed background $\mathbb{R}^4_{\epsilon_{1,2}} \times S^1$.
This is known to factorize as
\begin{equation}
Z_{\text{5d}} = Z_{\text{5d}}^{\text{pert}} \, Z_{\text{5d}}^{\text{inst}} \,; \label{factor}
\end{equation} 
here $Z_{\text{5d}}^{\text{pert}}$ contains the perturbative (classical + 1-loop) contribution to the partition function whose explicit form will not be needed in the following, while $Z_{\text{5d}}^{\text{inst}}$ contains non-perturbative corrections due to instantons. The instanton part of the partition function takes the form of a series expansion in the instanton fugacity $Q = e^{-t}$:
\begin{equation}
Z_{\text{5d}}^{\text{inst}} = 1+ 
\sum_{k \geqslant 1} Q^k Z_{\text{5d}}^{\text{inst},(k)} \,,
\end{equation}
where $Z_{\text{5d}}^{\text{inst},(k)}$ is the partition function of the $\cN=(0,4)$ ADHM quantum mechanics for $k$ instantons. This reduces to the contour integral
\begin{equation}
Z_{\text{5d}}^{\text{inst},(k)} = \dfrac{1}{k!} \oint \left[ \prod_{s=1}^k \dfrac{d \phi_s}{2\pi i} \right]
Z_{\text{vec}}^{(k)} Z_{\text{fund}}^{(k)} Z_{\text{CS}}^{(k)} \,, \label{ADHM}
\end{equation}
with (defining $\text{sh}(x) = 2\sinh(\frac{x}{2})$)
\begin{equation}
\begin{split}
Z_{\text{vec}}^{(k)} \,=\, & \left( -\dfrac{\text{sh}(2\epsilon_+)}{\text{sh}(\epsilon_1) \text{sh}(\epsilon_2)} \right) ^k
\prod_{s\neq t}^k \dfrac{\text{sh}(\phi_s - \phi_t)\text{sh}(\phi_s - \phi_t + 2\epsilon_+)}{\text{sh}(\phi_s - \phi_t + \epsilon_1) \text{sh}(\phi_s - \phi_t+ \epsilon_2)} \\
& \prod_{s=1}^k \prod_{r=1}^N \dfrac{1}{\text{sh}(\phi_s - a_r + \epsilon_+) \text{sh}(-\phi_s + a_r + \epsilon_+)} \, , \\
Z_{\text{fund}}^{(k)} \, = \, & \prod_{s=1}^k \prod_{b=1}^{N_f} \text{sh}(-\phi_s + m_b) \,, \;\;\;\;\;\;\;\; Z_{\text{CS}}^{(k)} = \prod_{s=1}^k e^{-\kappa \phi_s} \,.
\end{split}
\end{equation}
In this expression $\epsilon_{1,2}$ are the Omega background deformation parameters of $\mathbb{R}^4_{\epsilon_{1,2}} \times S^1$, and we define $\epsilon_{\pm} = \frac{\epsilon_1 \pm \epsilon_2}{2}$, while diag$(a_1,a_2,\cdots,a_N)$ correspond to the Cartan VEV of the real scalar in the 5d vector multiplet, and $m_b$ are the masses of the 5d fundamental matter multiplets. In terms of SQM symmetries, $\epsilon_+$ is the $SU(2) =$ diag$(SU(2)_2\times SU(2)_R)$ R-symmetry equivariant parameter, while $\epsilon_-$ is a flavor symmetry parameter.
The above factors combine contributions from various 1d $\cN=(0,4)$ multiplets of the ADHM SQM. $Z_{\text{vec}}^{(k)}$ contains the contributions of a $U(k)$ vector multiplet with an adjoint hypermultiplet, and $N$ fundamental hypermultiplets, $Z_{\text{fund}}^{(k)}$ matches the contribution of $N_f$ fundamental Fermi multiplets with single complex fermion, and $Z_{\text{CS}}^{(k)}$ is only the classical contribution from a 1d supersymmetric Chern-Simons term at level $-\kappa$.\footnote{For references on 1d $\cN=(0,4)$ multiplets see footnote \ref{foot:1d04}.}

The integral \eqref{ADHM} is evaluated by residues, and the contour choice is dictated by the Jeffrey-Kirwan prescription. For the case at hand, the poles selected by this prescription are classified by $N$-tuples of Young tableaux $\vec{Y} = \{Y_1, \ldots, Y_N\}$ with total number of boxes $\vert \vec{Y} \vert = \sum_{r=1}^N \vert Y_r \vert = k$. Practically, this corresponds to taking residues at
\begin{equation}
\phi_s = a_r + \epsilon_+ + (i-1)\epsilon_1 + (j-1)\epsilon_2 \,, \label{poles}
\end{equation}
with $(i,j)$ box in the tableau $Y_r$. As explained in \citep{Bergman:2013ala,Bergman:2013aca}, the $SU(N)_{\kappa}$ partition function (or $SU(2)$ partition function with some discrete $\theta$-angle for $N = 2$) is obtained from the $U(N)_{\kappa}$ one after we impose the traceless condition $\sum_{r=1}^N a_r = 0$, redefine $Q \rightarrow (-1)^{\kappa + N_f/2} Q$ and remove (by hand) additional $U(1)$ factors if parallel external NS5 branes are present \cite{Bao:2013pwa,Hayashi:2013qwa}. 
\medskip

These results are modified by the presence of the SQM loop realized by the addition of $n$ D3 branes in the brane setup. The half-index $Z_{\text{5d-1d}}$ computes the partition function of a 5d $\mathcal{N} = 1$ $U(N)_{\kappa}$ theory with $N_f$ flavors coupled to the 1d $\cN=(0,4)$ SQM by gauging 1d flavor symmetries with 5d fields.
It factorizes as
\begin{equation}
Z_{\text{5d-1d}} = Z_{\text{5d}}^{\text{pert}}  \, Z_{\text{5d-1d}}^{\text{inst}} \,; 
\end{equation}
here $Z_{\text{5d}}^{\text{pert}}$ is as in \eqref{factor}, 
while $Z_{\text{5d-1d}}^{\text{inst}}$ contains the non-perturbative instanton corrections to the 5d-1d system. The instanton part can again be written as a series expansion in $Q$,
\begin{equation}
Z_{\text{5d-1d}}^{\text{inst}} = \sum_{k \geqslant 0} Q^k Z_{\text{5d-1d}}^{\text{inst},(k)} \,,
\end{equation}
where this time $Z_{\text{5d-1d}}^{\text{inst},(k)}$ is the partition function of a modified (0,4) ADHM quantum mechanics for $k$ instantons, the modifications being due to additional matter multiplets arising from strings stretched between D3 and D1 or D5 branes. This takes the contour integral form
\begin{equation}
Z_{\text{5d-1d}}^{\text{inst},(k)} = \frac{1}{k!} \oint \left[ \prod_{s=1}^k \dfrac{d \phi_s}{2\pi i} \right]
Z_{\text{vec}}^{(k)} Z_{\text{fund}}^{(k)} Z_{\text{CS}}^{(k)} Z_{\text{SQM}}^{(k)} \,, \label{ADHMmod}
\end{equation}
which is the same as \eqref{ADHM} apart from the additional contribution 
\begin{equation}
Z_{\text{SQM}}^{(k)} = \prod_{r=1}^{N} \prod_{l=1}^{n} \text{sh}(M_l - a_r)  
\prod_{s=1}^k \prod_{l=1}^{n} \frac{\text{sh}(\phi_s - M_l + \epsilon_-)\text{sh}(-\phi_s + M_l + \epsilon_-)}{\text{sh}(\phi_s - M_l + \epsilon_+)\text{sh}(-\phi_s + M_l + \epsilon_+)} \,,
\end{equation}
corresponding to $n N$ Fermi multiplets with a single fermion ($n$ multiplets in the fundamental representation of the $SU(N)$ global symmetry, gauged with 5d fields), $n$ fundamental twisted hypermultiplets and $n$ fundamental Fermi multiplets with two fermions. 
Here $M_l$ are the masses of the Fermi multiplets, associated to the $U(n)$ global symmetry of the ADHM SQM. The contour choice for \eqref{ADHMmod} is still dictated by the Jeffrey-Kirwan prescription:
in this case, apart from the $N$-tuples of Young tableaux \eqref{poles}, additional poles of the form
\begin{equation}
\phi_s = M_l - \epsilon_+ \label{newpoles}
\end{equation}
contribute, where however at most only one pole of the form \eqref{newpoles} can be selected for each $l = 1, \ldots, n$; for example, in the $N=n = k = 2$ case there are five new poles given by 
\begin{equation}
\left\{
\begin{array}{rl}
\phi_{1} = & a_{1,2} + \epsilon_+ , \\
\phi_{2} = & M_{1,2} - \epsilon_+ ,
\end{array} \right.
\;\;\;\;\;\; \text{or} \;\;\;\;\;\;
\left\{
\begin{array}{rl}
\phi_{1} = & M_{1} - \epsilon_+, \\
\phi_{2} = & M_{2} - \epsilon_+. 
\end{array} \right.
\end{equation}
As before, $SU(N)_{\kappa}$ results can be obtained from $U(N)_{\kappa}$ ones after imposing the traceless condition $\sum_{r=1}^N a_r = 0$, redefining $Q \rightarrow (-1)^{\kappa + N_f/2} Q$ and removing additional $U(1)$ factors if parallel external NS5 branes are present. 
However, in the main text we always work with the normalized VEV of the SQM loop observable 
\begin{equation}
\vev{L_{\rm SQM}} 
= \dfrac{Z_{\text{5d-1d}}}{Z_{\text{5d}}} 
= \dfrac{Z_{\text{5d-1d}}^{\text{inst}}}{Z_{\text{5d}}^{\text{inst}}}, \label{norm}
\end{equation}
where we divided by the partition function of the 5d theory.
In addition to removing the $Z_{\text{5d}}^{\text{pert}}$ factor, this procedure also eliminates extra $U(1)$ factors due to parallel external NS5 branes; the normalized $SU(N)_{\kappa}$ observable is therefore obtained from the $U(N)_{\kappa}$ one \eqref{norm} simply by imposing the traceless conditions and redefining $Q \rightarrow (-1)^{\kappa + N_f/2} Q$.

\subsection{Linear quiver case}

We can now move to the half-index $Z_{\text{5d}}$ for a 5d $\mathcal{N} = 1$ $\prod_{I=1}^p U(N_I)_{\kappa_I}$ linear quiver gauge theory with $p$ nodes and bifundamental fields (without SQM loop). This is simply the partition function of the 5d quiver theory on $\mathbb{R}^4_{\epsilon_{1, 2}} \times S^1$, which factorizes as
\begin{equation}
Z_{\text{5d}} = Z_{\text{5d}}^{\text{pert}} \, Z_{\text{5d}}^{\text{inst}} \,.
\end{equation}
We will only be interested in the instanton part, which takes the form of a series expansion in the instanton fugacities $Q_i = e^{-t_i}$, $i=1,\dots, p$, of the $p$ gauge groups:
\begin{equation}
Z_{\text{5d}}^{\text{inst}} =
\sum_{k_1, \ldots, k_p \geqslant 0} Q_1^{k_1} \ldots Q_p^{k_p}  Z_{\text{5d}}^{\text{inst},(\vec{k})} \,,
\end{equation}
where we denoted $\vec{k} = \{ k_1, \ldots, k_p \}$. Here $Z_{\text{5d}}^{\text{inst},(\vec{k})}$ is the partition function for a generalized quiver ADHM quantum mechanics of $\vec{k}$ instantons, which reduces to the contour integral
\begin{equation}
Z_{\text{5d}}^{\text{inst},(\vec{k})}
= \dfrac{1}{k_1! \ldots k_p!} \oint \left[ \prod_{I=1}^p\prod_{s=1}^{k_I} \dfrac{d \phi_s^{(I)}}{2\pi i} \right]
Z_{\text{vec}}^{(\vec{k})} Z_{\text{bif}}^{(\vec{k})} Z_{\text{CS}}^{(\vec{k})} \,, \label{ADHMquiver}
\end{equation}
with the various factors entering the integrand given by
\begin{equation}
\begin{split}
Z_{\text{vec}}^{(\vec{k})} = & \prod_{I=1}^p \left( -\dfrac{\text{sh}(2\epsilon_+)}{\text{sh}(\epsilon_1) \text{sh}(\epsilon_2)} \right)^{k_I}
\prod_{I=1}^p \prod_{s\neq t}^{k_I} \dfrac{\text{sh}(\phi_s^{(I)} - \phi_t^{(I)})\text{sh}(\phi_s^{(I)} - \phi_t^{(I)} + 2\epsilon_+)}{\text{sh}(\phi_s^{(I)} - \phi_t^{(I)} + \epsilon_1) \text{sh}(\phi_s^{(I)} - \phi_t^{(I)} + \epsilon_2)} \\
& \prod_{I=1}^p \prod_{s=1}^{k_I} \prod_{r = 1}^{N_I}
\dfrac{1}{\text{sh}(\phi_s^{(I)} - a_r^{(I)} + \epsilon_+) \text{sh}(-\phi_s^{(I)} + a_r^{(I)} + \epsilon_+)} \,,
\end{split}
\end{equation}
\begin{equation}
\begin{split}
Z_{\text{bif}}^{(\vec{k})} = & \prod_{I=1}^{p-1} \prod_{s=1}^{k_I} \prod_{r=1}^{N_{I+1}}
\text{sh}(-\phi_s^{(I)} + a_r^{(I+1)} + m^{(I)})
\prod_{I=1}^{p-1} \prod_{s=1}^{k_{I+1}} \prod_{r=1}^{N_{I}}
\text{sh}(\phi_s^{(I+1)} - a_r^{(I)} - m^{(I)}) \\
& \prod_{I = 1}^{p-1} \prod_{s=1}^{k_I} \prod_{t=1}^{k_{I+1}} 
\dfrac{\text{sh}(-\phi_s^{(I)} + \phi_t^{(I+1)} + m^{(I)} + \epsilon_-)
\text{sh}(\phi_s^{(I)} - \phi_t^{(I+1)} - m^{(I)} + \epsilon_-)}{\text{sh}(-\phi_s^{(I)} + \phi_t^{(I+1)} + m^{(I)} + \epsilon_+) \text{sh}(\phi_s^{(I)} - \phi_t^{(I+1)} - m^{(I)} + \epsilon_+)} \,,
\end{split}
\end{equation}
\begin{equation}
Z_{\text{CS}}^{(\vec{k})} = 
\prod_{I=1}^p \prod_{s=1}^{k_I} e^{-\kappa_I \phi_s^{(I)}} \,.
\end{equation}
$Z_{\text{bif}}^{(\vec{k})}$ is also a combination of factors arising from (0,4) SQM multiplets in the ADHM quiver.
In this expression $a_r^{(I)}$ are the VEV of the real scalar field in the 5d $U(N_I)_{\kappa_I}$ vector multiplet, while $m^{(I)}$ can be identified with the mass of the $I$-th 5d bifundamental matter multiplet, because of constraints from potentials coupling ADHM and 5d matter fields. The integration contour for \eqref{ADHMquiver} is determined by the Jeffrey-Kirwan prescription; similarly to the single gauge node case, the relevant poles to be considered for the integration variables $\phi_s^{(I)}$ are classified in terms of $N_I$-tuples of Young tableaux $\vec{Y}^{(I)} = \{Y_1^{(I)}, \ldots, Y_{N_I}^{(I)} \}$ with total number of boxes $\vert \vec{Y}^{(I)} \vert = \sum_{r=1}^{N_I} \vert Y_r^{(I)} \vert = k_I$, that is
\begin{equation}
\phi_s^{(I)} = a_r^{(I)} + \epsilon_+ + (i-1)\epsilon_1 + (j-1)\epsilon_2 \label{poles2}
\end{equation}
with $(i,j)$ box in the tableau $Y_r^{(I)}$. As before, the partition function for the $\prod_{I=1}^p SU(N_I)_{\kappa_I}$ quiver theory is obtained from the $\prod_{I=1}^p U(N_I)_{\kappa_I}$ one by imposing the traceless condition $\sum_{r=1}^{N_I} a_r^{(I)} = 0$ for all gauge nodes $I = 1, \ldots, p$, redefining $Q_I \rightarrow (-1)^{\kappa_I + N_{I+1}/2 + N_{I-1}/2} Q_I$ (here $N_0 = N_{p+1} = 1$) and removing extra $U(1)$ factors if the 5-brane web involves parallel external NS5 branes.  \\

\noindent The computation is modified by the insertion of the SQM loop, associated with the presence of $n_I$ D3 branes for each set of $N_I$ D5 branes. The half-index $Z_{\text{5d-1d}}$ computes the partition function of a 5d $\mathcal{N} = 1$ $\prod_{I=1}^p U(N_I)_{\kappa_I}$ linear quiver gauge theory with bifundamental matter fields coupled to a 1d SQM, which lives at the intersection of the D3 and D5-branes, in the usual way.
The half-index factorizes as
\begin{equation}
Z_{\text{5d-1d}} = Z_{\text{5d}}^{\text{pert}} Z_{\text{5d-1d}}^{\text{inst}} \,.
\end{equation} 
$Z_{\text{5d-1d}}^{\text{inst}} $ is expressed as a series expansion in the instanton fugacities $Q_1$, \ldots, $Q_p$ of the 5d gauge groups,
\begin{equation}
Z_{\text{5d-1d}}^{\text{inst}} = 
\sum_{k_1, \ldots, k_p \geqslant 0} Q_1^{k_1} \ldots Q_p^{k_p} Z_{\text{5d-1d}}^{\text{inst},(\vec{k})} \,,
\end{equation}
where $Z_{\text{5d-1d}}^{\text{inst},(\vec{k})}$ is the partition function for a quiver (0,4) ADHM quantum mechanics of $\vec{k}$ instantons modified by additional matter multiplets sourced by strings between D3 and D5 or D1 branes; this can be written as the contour integral
\begin{equation}
Z_{\text{5d-1d}}^{\text{inst},(\vec{k})} = \dfrac{1}{k_1! \ldots k_p!} \oint \left[ \prod_{I = 1}^p \prod_{s=1}^{k_I} \dfrac{d \phi_s^{(I)}}{2\pi i} \right]
Z_{\text{vec}}^{(\vec{k})} Z_{\text{bif}}^{(\vec{k})} Z_{\text{CS}}^{(\vec{k})} 
Z_{\text{SQM}}^{(\vec{k})} \,, \label{ADHMquivermod}
\end{equation}
which is similar to \eqref{ADHMquiver} modulo the additional contribution to the integrand
\begin{equation}
Z_{\text{SQM}}^{(\vec{k})} = \prod_{I=1}^p \prod_{r=1}^{N_I} \prod_{l=1}^{n_I} \text{sh}(M_l^{(I)} - a_r^{(I)})  
\prod_{I=1}^p \prod_{s=1}^{k_I} \prod_{l=1}^{n_I} \frac{\text{sh}(\phi_s^{(I)} - M_l^{(I)} + \epsilon_-)\text{sh}(-\phi_s^{(I)} + M_l^{(I)} + \epsilon_-)}{\text{sh}(\phi_s^{(I)} - M_l^{(I)} + \epsilon_+)\text{sh}(-\phi_s^{(I)} + M_l^{(I)} + \epsilon_+)} \,,
\end{equation}
with $M_l^{(I)}$ the masses of Fermi multiplets and twisted hypermultiplets, associated to the $\prod_{I=1}^p U(n_I)$ global symmetry of the ADHM SQM.

The contour of integration is dictated once more by the Jeffrey-Kirwan prescription: apart from the 
$N_I$-tuples of Young tableaux \eqref{poles2} there are additional poles contributing, some of the form \eqref{newpoles}, i.e.
\begin{equation}
\phi_s^{(I)} = M_l^{(I)} - \epsilon_+ \,, \label{poles2quiver}
\end{equation} 
and some other of the form 
\begin{equation}
\phi^{(I + 1)}_t = \phi^{(I)}_s - m^{(I)} - \epsilon_+ \,, \quad \quad
\phi^{(I - 1)}_t = \phi^{(I)}_s + m^{(I-1)} - \epsilon_+ \,,
\end{equation}
for $\phi_s^{(I)}$ as in \eqref{poles2quiver}; we are however not able to write them in full generality, therefore we find these additional poles with a case by case analysis.
Results for $\prod_{I=1}^p SU(N_I)_{\kappa_I}$ quiver theories are as usual recovered by imposing the traceless condition $\sum_{r=1}^{N_I} a_r^{(I)} = 0$ for all $I = 1, \ldots, p$, redefining $Q_I \rightarrow (-1)^{\kappa_I + N_{I+1}/2 + N_{I-1}/2} Q_I$ and normalizing by the 5d partition function $Z_{\text{5d}}$ (this removes additional $U(1)$ factors arising if parallel external NS5 branes are present).

\section{$SU(2)$ theory with $N_f=3,4$}
\label{app:Nf34}

In this appendix we extend the computations of Wilson loop VEVs and we express the S-duality map for the $SU(2)$ theory with $N_f =3$ and $N_f =4$ flavors.

\subsection{$N_f = 3$}
\label{sapp:Nf3}

\begin{figure}[th]
\centering
\includegraphics[scale=0.3]{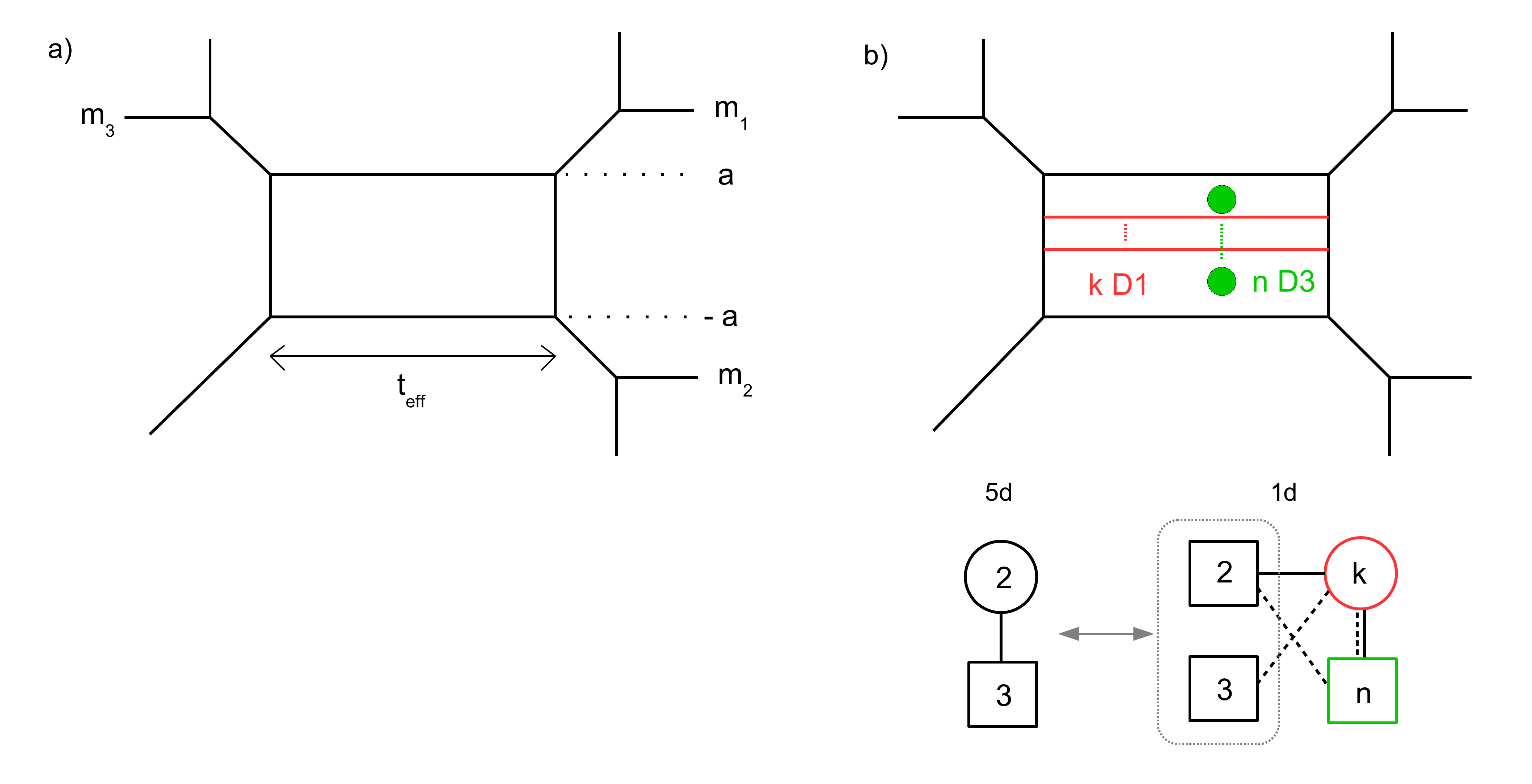}
\vspace{-0.5cm}
\caption{\footnotesize{Brane realization of the $SU(2)$ theory with $N_f=3$ and brane setup ADHM quiver SQM for the $k$-instanton sector.}}
\label{SU2Nf3}
\end{figure}

The brane realization of the $SU(2)$ theory with $N_f=3$ flavors is shown in Figure \ref{SU2Nf3}.
We can regard the theory as arising from the $U(2)$ theory with $N_f=3$ with Chern-Simons term $\kappa = -\frac 12$, by ungauging the diagonal $U(1)$. 
We denote $m_1,m_2,m_3$ the masses of the flavor hypermultiplets.
With $a_1 = -a_2 = a$, we define the fugacities
\be
\alpha = e^a \,, \quad \mu_i = e^{m_i} \,.
\ee
The Wilson loops $W_{\mathbf{2}^{\otimes n}}$ are evaluated from the residue formula \eqref{residueformula} with the SQM loop $L^{n}_{\rm SQM}$ associated to the $k$-instanton ADHM SQM shown in Figure \ref{SU2Nf3}-b. $\vev{L^{n}_{\rm SQM}}$ is evaluated following the recipe of Appendix \ref{app:ADHMcomputations}, taking into account the corrections due to the $U(2) \to SU(2)$ projection and the presence of parallel external NS5 branes.
We find
\bea
\vev{W_{\mathbf{2}}} &= \alpha + \alpha^{-1} 
+ Q \sqrt{\dfrac{q_1 q_2}{\mu_1 \mu_2 \mu_3}} \dfrac{(1 + q_1 q_2)(\mu_1 + \mu_2 + \mu_3 + \mu_1 \mu_2 \mu_3)}{(1-\alpha^2 q_1 q_2)(1 - \alpha^{-2} q_1 q_2)}  \cr
& - Q \dfrac{q_1 q_2}{\sqrt{\mu_1 \mu_2 \mu_3}} \dfrac{(\alpha + \alpha^{-1})(1 + \mu_1 \mu_2 + \mu_1 \mu_3 + \mu_2 \mu_3)}{(1-\alpha^2 q_1 q_2)(1 - \alpha^{-2} q_1 q_2)} 
+ O(Q^2) \,, \cr
\vev{W_{\mathbf{2} \,\otimes \mathbf{2}}}  &= \alpha^2 + 2 + \alpha^{-2}  \cr
& + Q \dfrac{1 + \mu_1 \mu_2 + \mu_1 \mu_3 + \mu_2 \mu_3}{\sqrt{\mu_1 \mu_2 \mu_3}} \dfrac{(1 - q_1)(1 - q_2)(1 + q_1 q_2) - 2 q_1 q_2 (\alpha^2 + 2 + \alpha^{-2})}{(1-\alpha^2 q_1 q_2)(1-\alpha^{-2} q_1 q_2)} \cr
& + Q \dfrac{\mu_1 \mu_2 \mu_3 + \mu_1 + \mu_2 + \mu_3}{\sqrt{\mu_1 \mu_2 \mu_3}} \dfrac{\sqrt{q_1 q_2}(1 + q_1)(1 + q_2) (\alpha + \alpha^{-1})}{(1-\alpha^2 q_1 q_2)(1-\alpha^{-2} q_1 q_2)} + O(Q^2) \,.
\eea
To exhibit the $E_{4} = SU(5)$ enhanced flavor symmetry we introduce the fugacities
\bea
& A = e^{-\frac{2t}{5} - a} \,,  \quad y_1 = e^{\frac{4t}{5}} \,, \quad y_2 = e^{-\frac{t}{5} +\frac{m_1-m_2+m_3}{2}} \,, \cr
& y_3 = e^{-\frac{t}{5} - \frac{m_1-m_2-m_3}{2}} \,, \quad  y_4 = e^{-\frac{t}{5} - \frac{m_1+m_2+m_3}{2}}   \,, \quad y_5 = e^{-\frac{t}{5} + \frac{m_1+m_2-m_3}{2}}   \,,
\eea
where $y_i$ are the $SU(5)$ fugacities satisfying $\prod_{i} y_i =1$. In terms of the new parameters, the S action is 
\be
\underline{\text{S-duality}}: \quad  y_1 \leftrightarrow y_2 \,, \quad y_3 \leftrightarrow y_4 \,.
\ee
It is a Weyl transformation in $SU(5)$. It does not commute with the Weyl flavor symmetries given by permutations of $m_1,m_2,m_3$, which correspond to the permutations of $y_2,y_3,y_5$ in the $SU(5)$ Weyl group.

We then expand further the Wilson loop VEVs at small $A$,
\bea
A y_1^{1/2} \vev{W_{\textbf{2}}} & = 1 + \chi^{A_4}_{\textbf{5}}(\vec{y}) A^2 
- \chi^{A_1}_{\mathbf{2}}(q_+) \chi^{A_4}_{\overline{\mathbf{5}}}(\vec{y}) A^3
+ \chi^{A_1}_{\mathbf{3}}(q_+) \chi^{A_4}_{\mathbf{10}}(\vec{y}) A^4 \cr
& - \chi^{A_1}_{\mathbf{4}}(q_+) \chi^{A_4}_{\mathbf{24}}(\vec{y}) A^5
- \Big( \chi^{A_1}_{\mathbf{4}}(q_+) + \chi^{A_1}_{\mathbf{2}}(q_+) 
+ \chi^{A_1}_{\mathbf{5}}(q_+) \chi^{A_1}_{\mathbf{2}}(q_-) \Big) A^5 \cr
& + \Big( \chi^{A_1}_{\mathbf{5}}(q_+) + \chi^{A_1}_{\mathbf{3}}(q_+) 
+ \chi^{A_1}_{\mathbf{6}}(q_+) \chi^{A_1}_{\mathbf{2}}(q_-) \Big) \chi^{A_4}_{\overline{\mathbf{10}}}(\vec{y}) A^6 \cr
& + \chi^{A_1}_{\mathbf{5}}(q_+) \chi^{A_4}_{\overline{\mathbf{40}}}(\vec{y}) A^6 
+ \chi^{A_1}_{\mathbf{5}}(q_+) \chi^{A_4}_{\overline{\mathbf{15}}}(\vec{y}) A^6 + \ldots \,, 
\eea
\bea
A^2 y_1 \vev{W_{\mathbf{2} \,\otimes \mathbf{2}}} & =
1 + 2 \chi^{A_4}_{\mathbf{5}}(\vec{y}) A^2 - \Big( \chi^{A_1}_{\mathbf{2}}(q_+) + \chi^{A_1}_{\mathbf{2}}(q_-) \Big) \chi^{A_4}_{\overline{\mathbf{5}}}(\vec{y}) A^3 \cr
& + \chi^{A_4}_{\mathbf{15}}(\vec{y}) A^4  
+ \Big( \chi^{A_1}_{\mathbf{3}}(q_+) + \chi^{A_1}_{\mathbf{2}}(q_+) \chi^{A_1}_{\mathbf{2}}(q_-) \Big) \chi^{A_4}_{\mathbf{10}}(\vec{y}) A^4 \cr
& - \Big( \chi^{A_1}_{\mathbf{4}}(q_+) + \chi^{A_1}_{\mathbf{2}}(q_+) + \chi^{A_1}_{\mathbf{3}}(q_+) \chi^{A_1}_{\mathbf{2}}(q_-) \Big) \chi^{A_4}_{\mathbf{24}}(\vec{y}) A^5 \cr
& - \Big( \chi^{A_1}_{\mathbf{4}}(q_+) + \chi^{A_1}_{\mathbf{2}}(q_+) + \chi^{A_1}_{\mathbf{3}}(q_+) \chi^{A_1}_{\mathbf{2}}(q_-) \cr
& \;\;\;\;\;\; + \chi^{A_1}_{\mathbf{5}}(q_+) \chi^{A_1}_{\mathbf{2}}(q_-) + \chi^{A_1}_{\mathbf{4}}(q_+) \chi^{A_1}_{\mathbf{3}}(q_-) + \chi^{A_1}_{\mathbf{2}}(q_-)  \Big) A^5 
\cr
& + \Big( \chi^{A_1}_{\mathbf{5}}(q_+) + 2\chi^{A_1}_{\mathbf{3}}(q_+) + \chi^{A_1}_{\mathbf{6}}(q_+) \chi^{A_1}_{\mathbf{2}}(q_-) + \chi^{A_1}_{\mathbf{5}}(q_+) \chi^{A_1}_{\mathbf{3}}(q_-) \cr
& \;\;\;\;\;\; + \chi^{A_1}_{\mathbf{4}}(q_+) \chi^{A_1}_{\mathbf{2}}(q_-) + \chi^{A_1}_{\mathbf{2}}(q_+) \chi^{A_1}_{\mathbf{2}}(q_-) \Big) \chi^{A_4}_{\overline{\mathbf{10}}}(\vec{y}) A^6 \cr
& + \Big( \chi^{A_1}_{\mathbf{5}}(q_+) + \chi^{A_1}_{\mathbf{4}}(q_+) \chi^{A_1}_{\mathbf{2}}(q_-) + 1 \Big) \chi^{A_4}_{\overline{\mathbf{15}}}(\vec{y}) A^6 \cr
& + \Big( \chi^{A_1}_{\mathbf{5}}(q_+) + \chi^{A_1}_{\mathbf{3}}(q_+) + \chi^{A_1}_{\mathbf{4}}(q_+) \chi^{A_1}_{\mathbf{2}}(q_-) \Big) \chi^{A_4}_{\overline{\mathbf{40}}}(\vec{y}) A^6 + \ldots \,.
\eea
The expansions exhibit the expected $SU(5)$($\sim A_4$) characters. 

Under S-duality, we have 
\be
S. W_{\mathbf{2}^{\otimes n}}(A,y_1,y_2,y_3,y_4,y_5) =  \lp\frac{y_1}{y_2}\rp^{-\frac n2} \, W_{\mathbf{2}^{\otimes n}}(A,y_2,y_1,y_4,y_3,y_5) \,,
\label{StransfoNf3}
\ee
with the multiplicative parameter $(y_1/y_2)^{\frac 12} = e^{\frac t2 - \frac{m_1-m_2+m_3}{4}}$.

\subsection{$N_f = 4$}
\label{sapp:Nf4}

\begin{figure}[th]
\centering
\includegraphics[scale=0.3]{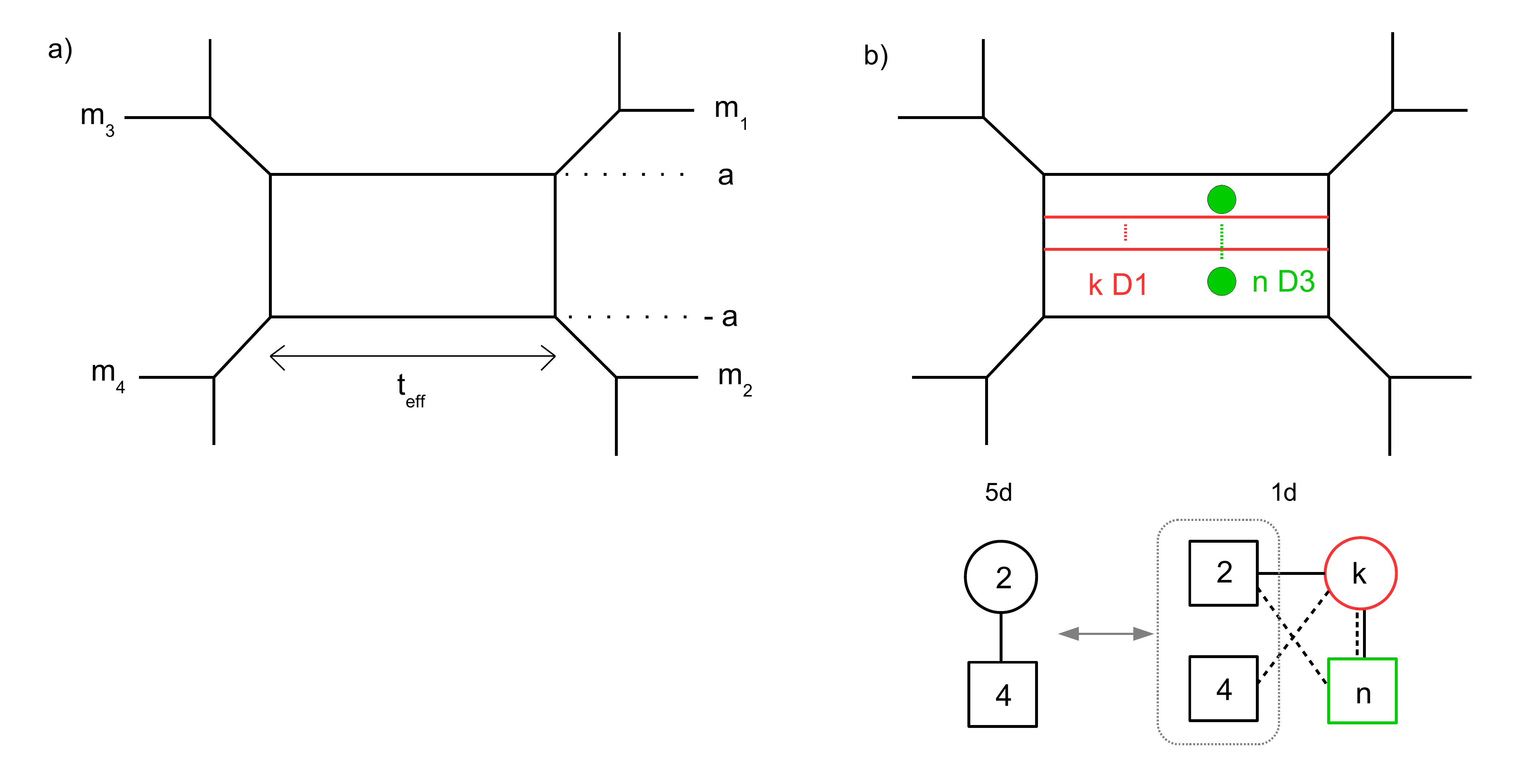}
\vspace{-0.5cm}
\caption{\footnotesize{Brane realization of the $SU(2)$ theory with $N_f=4$ and brane setup ADHM quiver SQM for the $k$-instanton sector.}}
\label{SU2Nf4}
\end{figure}

The brane realization of the $SU(2)$ theory with $N_f=4$ flavors is shown in Figure \ref{SU2Nf4}.
We can regard the theory as arising from the $U(2)$ theory with $N_f=4$ (without Chern-Simons term), by ungauging the diagonal $U(1)$. 
We denote $m_i$ the masses of the flavor hypermultiplets, with fugacities $\mu_i = e^{m_i}$.

The Wilson loops $W_{\mathbf{2}^{\otimes n}}$ are evaluated from the residue formula \eqref{residueformula} with the SQM loop $L^{n}_{\rm SQM}$ associated to the $k$-instanton ADHM SQM shown in Figure \ref{SU2Nf4}-b. $\vev{L^{n}_{\rm SQM}}$ is evaluated following the recipe of Appendix \ref{app:ADHMcomputations}, taking into account the corrections due to the $U(2) \to SU(2)$ projection and the presence of parallel external NS5 branes.
We find
\bea
\vev{W_{\mathbf{2}}}  &= \alpha + \alpha^{-1} 
+ Q \sqrt{\dfrac{q_1 q_2}{\prod_i\mu_i}} \dfrac{(1 + q_1 q_2)(\sum_i \mu_i + \sum_{i<j<k}\mu_i\mu_j\mu_k)}{(1-\alpha^2 q_1 q_2)(1 - \alpha^{-2} q_1 q_2)} \cr
& - Q \dfrac{q_1 q_2}{\sqrt{\prod_i\mu_i}} \dfrac{(\alpha + \alpha^{-1})(1 + \sum_{i<j} \mu_i \mu_j + \prod_i\mu_i)}{(1-\alpha^2 q_1 q_2)(1 - \alpha^{-2} q_1 q_2)} 
+ O(Q^2) \,, \cr
\vev{W_{\mathbf{2} \,\otimes \mathbf{2}}} &= \alpha^2 + 2 + \alpha^{-2} \cr
& + Q \dfrac{1 + \sum_{i<j} \mu_i \mu_j + \prod_i\mu_i}{\sqrt{\prod_i\mu_i}} \dfrac{(1 - q_1)(1 - q_2)(1 + q_1 q_2) - 2 q_1 q_2 (\alpha^2 + 2 + \alpha^{-2})}{(1-\alpha^2 q_1 q_2)(1-\alpha^{-2} q_1 q_2)} \cr
& + Q \dfrac{\sum_i \mu_i + \sum_{i<j<k}\mu_i\mu_j\mu_k}{\sqrt{\prod_i\mu_i}} \dfrac{\sqrt{q_1 q_2}(1 + q_1)(1 + q_2) (\alpha + \alpha^{-1})}{(1-\alpha^2 q_1 q_2)(1-\alpha^{-2} q_1 q_2)} + O(Q^2) \,.
\eea
To exhibit the $E_{5} = $Spin$(10)$ enhanced flavor symmetry we introduce the fugacities
\bea
& A = e^{-\frac{t}{2} - a} \,,  \quad y_1 = e^{t} \,, \quad y_2 = e^{\frac{m_1-m_2+m_3-m_4}{2}} \,, \cr
& y_3 = e^{\frac{-m_1+m_2+m_3-m_4}{2}} \,, \quad  y_4 =  e^{-\frac{m_1+m_2+m_3+m_4}{2}}  \,, \quad y_5 =  e^{\frac{m_1+m_2-m_3-m_4}{2}}  \,,
\eea
where $y_i$ are the $\mathfrak{so}(10)$ fugacities.
In terms of the new parameters, the S action is 
\be
\underline{\text{S-duality}}: \quad  y_1 \leftrightarrow y_2 \,, \quad  y_3 \leftrightarrow y_4 \,.
\ee
We then expand further the Wilson loop VEVs at small $A$,
\bea
A y_1^{1/2} \vev{W_{\mathbf{2}}} & = 1 + \chi^{E_5}_{\mathbf{10}}(\vec{y}) A^2 - \chi^{A_1}_{\mathbf{2}}(q_+) \chi^{E_5}_{\overline{\mathbf{16}}}(\vec{y}) A^3 \cr
& + \chi^{A_1}_{\mathbf{3}}(q_+) \chi^{E_5}_{\mathbf{45}}(\vec{y}) A^4 + \Big( \chi^{A_1}_{\mathbf{3}}(q_+) + \chi^{A_1}_{\mathbf{4}}(q_+)\chi^{A_1}_{\mathbf{2}}(q_-) + 1 \Big) A^4 \cr
& - \Big( \chi^{A_1}_{\mathbf{4}}(q_+) + \chi^{A_1}_{\mathbf{2}}(q_+) 
+ \chi^{A_1}_{\mathbf{5}}(q_+) \chi^{A_1}_{\mathbf{2}}(q_-) \Big) \chi^{E_5}_{\mathbf{16}}(\vec{y}) A^5 - \chi^{A_1}_{\mathbf{4}}(q_+) \chi^{E_5}_{\mathbf{144}}(\vec{y}) A^5 + \ldots \,, 
\eea
\bea
A^2 y_1 \vev{W_{\mathbf{2} \,\otimes \mathbf{2}}} & = 1 + 2 \chi^{E_5}_{\mathbf{10}}(\vec{y}) A^2 - \Big( \chi^{A_1}_{\mathbf{2}}(q_+) + \chi^{A_1}_{\mathbf{2}}(q_-) \Big) \chi^{E_5}_{\overline{\mathbf{16}}}(\vec{y}) A^3 \cr
& + \chi^{E_5}_{\mathbf{54}}(\vec{y}) A^4 + \Big( \chi^{A_1}_{\mathbf{3}}(q_+) + \chi^{A_1}_{\mathbf{2}}(q_+) \chi^{A_1}_{\mathbf{2}}(q_-) \Big) \chi^{E_5}_{\mathbf{45}}(\vec{y}) A^4 \cr
& + \Big( \chi^{A_1}_{\mathbf{3}}(q_+) + \chi^{A_1}_{\mathbf{4}}(q_+) \chi^{A_1}_{\mathbf{2}}(q_-) + \chi^{A_1}_{\mathbf{3}}(q_+) \chi^{A_1}_{\mathbf{3}}(q_-) + \chi^{A_1}_{\mathbf{2}}(q_+) \chi^{A_1}_{\mathbf{2}}(q_-) + 1 \Big) A^4 \cr
& - \Big( \chi^{A_1}_{\mathbf{2}}(q_+) + \chi^{A_1}_{\mathbf{2}}(q_-) \Big) 
\Big( \chi^{A_1}_{\mathbf{4}}(q_+) \chi^{A_1}_{\mathbf{2}}(q_-) + 1 \Big) 
\chi^{E_5}_{\mathbf{16}}(\vec{y}) A^5 \cr
& - \Big( \chi^{A_1}_{\mathbf{2}}(q_+) + \chi^{A_1}_{\mathbf{2}}(q_-) \Big) \chi^{A_1}_{\mathbf{3}}(q_+) \chi^{E_5}_{\mathbf{144}}(\vec{y}) A^5 + \ldots \,.
\eea
The expansions exhibit the expected $Spin(10)$ characters. 

Under S-duality, we have 
\be
S. W_{\mathbf{2}^{\otimes n}}(A,y_1,y_2,y_3,y_4,y_5) =  \lp\frac{y_1}{y_2}\rp^{-\frac n2} \, W_{\mathbf{2}^{\otimes n}}(A,y_2,y_1,y_4,y_3,y_5) \,,
\label{StransfoNf4}
\ee
with the multiplicative parameter $(y_1/y_2)^{\frac 12} = e^{\frac t2 - \frac{m_1-m_2+m_3-m_4}{4}}$.

\section{Results}
\label{app:longresults}

In this appendix we collect various results which are too long to be presented in the main text.

\subsection{Wilson loops in $SU(3)$, $N_f=2$ theory}
\label{sapp:SU3Nf2}

Using the notations of the main text, we have
\bea
Q_{F_1}^{2/3} & Q_{F_2}^{1/3}  \vev{W_{\mathbf{3}}} \cr
&= 1 + Q_{F_1} + Q_{B_1} + Q_{F_1} Q_{F_2} + Q_{F_1} Q_{B_2} + \chi^{A_1}_{\mathbf{3}}(q_+) Q_{F_1} Q_{B_1} \cr
& - \chi^{A_1}_{\mathbf{2}}(q_+) \chi^{A_1}_{\mathbf{2}}(Q_m) Q_{F_1} \sqrt{Q_{B_1}Q_{B_2}} - \chi^{A_1}_{\mathbf{2}}(q_+) \chi^{A_1}_{\mathbf{2}}(Q_m) Q_{F_1} Q_{F_2} \sqrt{Q_{B_1}Q_{B_2}} \cr
& - \chi^{A_1}_{\mathbf{4}}(q_+) \chi^{A_1}_{\mathbf{2}}(Q_m) Q_{F_1} (Q_{F_1} + Q_{B_1}) \sqrt{Q_{B_1}Q_{B_2}} \cr
& + Q_{F_1} Q_{F_2} Q_{B_1} + \chi^{A_1}_{\mathbf{5}}(q_+) Q_{F_1} Q_{B_1} (Q_{F_1} + Q_{B_1}) \cr
& + \chi^{A_1}_{\mathbf{3}}(q_+) Q_{F_1} (Q_{B_1} Q_{B_2} + Q_{F_1} Q_{B_2} + Q_{F_2} Q_{B_1} + Q_{F_2} Q_{B_2})  + \ldots \,, 
\eea
\bea
Q_{F_1}^{4/3} & Q_{F_2}^{2/3}  \vev{W_{\mathbf{3} \, \otimes \mathbf{3}}} \cr
&= 
1 + 2(Q_{F_1} + Q_{B_1}) + Q_{F_1}^2 + Q_{B_1}^2 + 2 Q_{F_1} Q_{F_2} + 2 Q_{F_1}  Q_{B_2} \cr
& + \Big( \chi^{A_1}_{\mathbf{3}}(q_+) + \chi^{A_1}_{\mathbf{2}}(q_+) \chi^{A_1}_{\mathbf{2}}(q_-) + 1 \Big) Q_{F_1} Q_{B_1} \cr
& - \Big( \chi^{A_1}_{\mathbf{2}}(q_+) + \chi^{A_1}_{\mathbf{2}}(q_-) \Big) \chi^{A_1}_{\mathbf{2}}(Q_m) Q_{F_1} \sqrt{Q_{B_1} Q_{B_2}} \cr
& + 2 Q_{F_1}^2 Q_{F_2} + \Big( \chi^{A_1}_{\mathbf{5}}(q_+) + \chi^{A_1}_{\mathbf{3}}(q_+) + \chi^{A_1}_{\mathbf{4}}(q_+) \chi^{A_1}_{\mathbf{2}}(q_-) \Big) Q_{F_1} Q_{B_1}(Q_{F_1} + Q_{B_1}) \cr
& + \Big( \chi^{A_1}_{\mathbf{3}}(q_+) + \chi^{A_1}_{\mathbf{2}}(q_+) \chi^{A_1}_{\mathbf{2}}(q_-) + 1 \Big) Q_{F_1} Q_{B_2}(Q_{F_1} + Q_{B_1}) \cr
& + \Big(\chi^{A_1}_{\mathbf{3}}(q_+) + \chi^{A_1}_{\mathbf{2}}(q_+) \chi^{A_1}_{\mathbf{2}}(q_-) + 3 \Big) Q_{F_1} Q_{F_2} Q_{B_1} + 2 \chi^{A_1}_{\mathbf{3}}(q_+) Q_{F_1} Q_{F_2} Q_{B_2} \cr
& - \Big( \chi^{A_1}_{\mathbf{4}}(q_+) + \chi^{A_1}_{\mathbf{2}}(q_+) + \chi^{A_1}_{\mathbf{3}}(q_+) \chi^{A_1}_{\mathbf{2}}(q_-) \Big) \chi^{A_1}_{\mathbf{2}}(Q_m) Q_{F_1} (Q_{F_1} + Q_{B_1}) \sqrt{Q_{B_1} Q_{B_2}} \cr
& - 2 \chi^{A_1}_{\mathbf{2}}(q_+) \chi^{A_1}_{\mathbf{2}}(Q_m) Q_{F_1} Q_{F_2} \sqrt{Q_{B_1} Q_{B_2}} + \ldots \,, 
\eea
\bea
Q_{F_1} & Q_{F_2}  \vev{W_{\mathbf{3} \, \otimes \overline{\mathbf{3}}}} \cr
&= 
1 + Q_{F_1} + Q_{F_2} + Q_{B_1} + Q_{B_2} \cr
& + 3 Q_{F_1} Q_{F_2} + 2 Q_{F_1} Q_{B_2} + 2 Q_{F_2} Q_{B_1} + Q_{B_1} Q_{B_2} 
+ \chi^{A_1}_{\mathbf{3}}(q_+) (Q_{F_1} Q_{B_1} + Q_{F_2} Q_{B_2}) \cr
& - \chi^{A_1}_{\mathbf{2}}(q_+) \chi^{A_1}_{\mathbf{2}}(Q_m) (Q_{F_1} + Q_{F_2})\sqrt{Q_{B_1} Q_{B_2}} + Q_{F_1}^2 Q_{F_2} + Q_{F_1} Q_{F_2}^2 + Q_{F_1} Q_{B_2}^2 + Q_{F_2} Q_{B_1}^2 \cr
& + \chi^{A_1}_{\mathbf{5}}(q_+)(Q_{F_1} Q_{B_1}^2 + Q_{F_1}^2 Q_{B_1} + Q_{F_2} Q_{B_2}^2 + Q_{F_2}^2 Q_{B_2}) \cr
& + \chi^{A_1}_{\mathbf{3}}(q_+) (Q_{F_1}^2 Q_{B_2} + Q_{F_2}^2 Q_{B_1} + 2 Q_{F_1} Q_{B_1} Q_{B_2} + 2 Q_{F_2} Q_{B_1} Q_{B_2}) \cr
& + \Big( 2 \chi^{A_1}_{\mathbf{3}}(q_+) + \chi^{A_1}_{\mathbf{2}}(q_+) \chi^{A_1}_{\mathbf{2}}(q_-) + 2 \Big) Q_{F_1} Q_{F_2} (Q_{B_1} + Q_{B_2}) \cr
& - \chi^{A_1}_{\mathbf{4}}(q_+) \chi^{A_1}_{\mathbf{2}}(Q_m) (Q_{F_1} Q_{B_1} + Q_{F_2} Q_{B_2} + Q_{F_1}^2 + Q_{F_2}^2) \sqrt{Q_{B_1} Q_{B_2}} \cr
& - \chi^{A_1}_{\mathbf{2}}(q_+) \chi^{A_1}_{\mathbf{2}}(Q_m) (Q_{F_1} Q_{B_2} + Q_{F_2} Q_{B_1}) \sqrt{Q_{B_1} Q_{B_2}} \cr
& - \Big( 3 \chi^{A_1}_{\mathbf{2}}(q_+) + \chi^{A_1}_{\mathbf{2}}(q_-) \Big) \chi^{A_1}_{\mathbf{2}}(Q_m) Q_{F_1} Q_{F_2} \sqrt{Q_{B_1} Q_{B_2}} + \ldots \,.
\eea

\subsection{Wilson loops in $SU(2)_\pi\times SU(2)_\pi$ theory}
\label{sapp:SU2xSU2}

Using the notations of the main text, we have
\bea
\widetilde{Q}_{F_1}^{1/2} & \vev{W_{(\mathbf{2},\mathbf{1})}}   \cr
&= 1 + \widetilde{Q}_{B_1} + \widetilde{Q}_{F_1} + \widetilde{Q}_{B_1} \widetilde{Q}_{B_2} + \widetilde{Q}_{B_1} \widetilde{Q}_{F_2} + \chi^{A_1}_{\mathbf{3}}(q_+) \widetilde{Q}_{B_1} \widetilde{Q}_{F_1} \cr
& - \chi^{A_1}_{\mathbf{2}}(q_+) \chi^{A_1}_{\mathbf{2}}(\widetilde{Q}_m) \widetilde{Q}_{B_1} \sqrt{\widetilde{Q}_{F_1}\widetilde{Q}_{F_2}} - \chi^{A_1}_{\mathbf{2}}(q_+) \chi^{A_1}_{\mathbf{2}}(\widetilde{Q}_m) \widetilde{Q}_{B_1} \widetilde{Q}_{B_2} \sqrt{\widetilde{Q}_{F_1}\widetilde{Q}_{F_2}} \cr
& - \chi^{A_1}_{\mathbf{4}}(q_+) \chi^{A_1}_{\mathbf{2}}(\widetilde{Q}_m) \widetilde{Q}_{B_1} (\widetilde{Q}_{B_1} + \widetilde{Q}_{F_1}) \sqrt{\widetilde{Q}_{F_1}\widetilde{Q}_{F_2}} \cr
& + \widetilde{Q}_{B_1} \widetilde{Q}_{B_2} \widetilde{Q}_{F_1} + \chi^{A_1}_{\mathbf{5}}(q_+) \widetilde{Q}_{B_1} \widetilde{Q}_{F_1} (\widetilde{Q}_{B_1} + \widetilde{Q}_{F_1}) \cr
& + \chi^{A_1}_{\mathbf{3}}(q_+) \widetilde{Q}_{B_1} (\widetilde{Q}_{F_1} \widetilde{Q}_{F_2} + \widetilde{Q}_{B_1} \widetilde{Q}_{F_2} + \widetilde{Q}_{B_2} \widetilde{Q}_{F_1} + \widetilde{Q}_{B_2} \widetilde{Q}_{F_2})  + \ldots \,, 
\eea
\bea
\widetilde{Q}_{F_1} & \vev{W_{(\mathbf{2} \, \otimes \mathbf{2},\mathbf{1})}}  \cr
&=  
1 + 2(\widetilde{Q}_{B_1} + \widetilde{Q}_{F_1}) + \widetilde{Q}_{B_1}^2 + \widetilde{Q}_{F_1}^2 + 2 \widetilde{Q}_{B_1} \widetilde{Q}_{B_2} + 2 \widetilde{Q}_{B_1}  \widetilde{Q}_{F_2} \cr
& + \Big( \chi^{A_1}_{\mathbf{3}}(q_+) + \chi^{A_1}_{\mathbf{2}}(q_+) \chi^{A_1}_{\mathbf{2}}(q_-) + 1 \Big) \widetilde{Q}_{B_1} \widetilde{Q}_{F_1} \cr
& - \Big( \chi^{A_1}_{\mathbf{2}}(q_+) + \chi^{A_1}_{\mathbf{2}}(q_-) \Big) \chi^{A_1}_{\mathbf{2}}(\widetilde{Q}_m) \widetilde{Q}_{B_1} \sqrt{\widetilde{Q}_{F_1} \widetilde{Q}_{F_2}} \cr
& + 2 \widetilde{Q}_{B_1}^2 \widetilde{Q}_{B_2} + \Big( \chi^{A_1}_{\mathbf{5}}(q_+) + \chi^{A_1}_{\mathbf{3}}(q_+) + \chi^{A_1}_{\mathbf{4}}(q_+) \chi^{A_1}_{\mathbf{2}}(q_-) \Big) \widetilde{Q}_{B_1} \widetilde{Q}_{F_1}(\widetilde{Q}_{B_1} + \widetilde{Q}_{F_1}) \cr
& + \Big( \chi^{A_1}_{\mathbf{3}}(q_+) + \chi^{A_1}_{\mathbf{2}}(q_+) \chi^{A_1}_{\mathbf{2}}(q_-) + 1 \Big) \widetilde{Q}_{B_1} \widetilde{Q}_{F_2}(\widetilde{Q}_{B_1} + \widetilde{Q}_{F_1}) \cr
& + \Big(\chi^{A_1}_{\mathbf{3}}(q_+) + \chi^{A_1}_{\mathbf{2}}(q_+) \chi^{A_1}_{\mathbf{2}}(q_-) + 3 \Big) \widetilde{Q}_{B_1} \widetilde{Q}_{B_2} \widetilde{Q}_{F_1} + 2 \chi^{A_1}_{\mathbf{3}}(q_+) \widetilde{Q}_{B_1} \widetilde{Q}_{B_2} \widetilde{Q}_{F_2} \cr
& - \Big( \chi^{A_1}_{\mathbf{4}}(q_+) + \chi^{A_1}_{\mathbf{2}}(q_+) + \chi^{A_1}_{\mathbf{3}}(q_+) \chi^{A_1}_{\mathbf{2}}(q_-) \Big) \chi^{A_1}_{\mathbf{2}}(\widetilde{Q}_m) \widetilde{Q}_{B_1} (\widetilde{Q}_{B_1} + \widetilde{Q}_{F_1}) \sqrt{\widetilde{Q}_{F_1} \widetilde{Q}_{F_2}} \cr
& - 2 \chi^{A_1}_{\mathbf{2}}(q_+) \chi^{A_1}_{\mathbf{2}}(\widetilde{Q}_m) \widetilde{Q}_{B_1} \widetilde{Q}_{B_2} \sqrt{\widetilde{Q}_{F_1} \widetilde{Q}_{F_2}} + \ldots \,, 
\eea
\bea
\widetilde{Q}_{F_1}^{1/2} & \widetilde{Q}_{F_2}^{1/2} \vev{W_{(\mathbf{2}, \mathbf{2})}} \cr 
&= 
1 + \widetilde{Q}_{B_1} + \widetilde{Q}_{B_2} + \widetilde{Q}_{F_1} + \widetilde{Q}_{F_2} \cr
& + 3 \widetilde{Q}_{B_1} \widetilde{Q}_{B_2} + 2 \widetilde{Q}_{B_1} \widetilde{Q}_{F_2} + 2 \widetilde{Q}_{B_2} \widetilde{Q}_{F_1} + \widetilde{Q}_{F_1} \widetilde{Q}_{F_2} 
+ \chi^{A_1}_{\mathbf{3}}(q_+) (\widetilde{Q}_{B_1} \widetilde{Q}_{F_1} + \widetilde{Q}_{B_2} \widetilde{Q}_{F_2}) \cr
& - \chi^{A_1}_{\mathbf{2}}(q_+) \chi^{A_1}_{\mathbf{2}}(\widetilde{Q}_m) (\widetilde{Q}_{B_1} + \widetilde{Q}_{B_2})\sqrt{\widetilde{Q}_{F_1} \widetilde{Q}_{F_2}} + \widetilde{Q}_{B_1}^2 \widetilde{Q}_{B_2} + \widetilde{Q}_{B_1} \widetilde{Q}_{B_2}^2 + \widetilde{Q}_{B_1} \widetilde{Q}_{F_2}^2 + \widetilde{Q}_{B_2} \widetilde{Q}_{F_1}^2 \cr
& + \chi^{A_1}_{\mathbf{5}}(q_+)(\widetilde{Q}_{B_1} \widetilde{Q}_{F_1}^2 + \widetilde{Q}_{B_1}^2 \widetilde{Q}_{F_1} + \widetilde{Q}_{B_2} \widetilde{Q}_{F_2}^2 + \widetilde{Q}_{B_2}^2 \widetilde{Q}_{F_2}) \cr
& + \chi^{A_1}_{\mathbf{3}}(q_+) (\widetilde{Q}_{B_1}^2 \widetilde{Q}_{F_2} + \widetilde{Q}_{B_2}^2 \widetilde{Q}_{F_1} + 2 \widetilde{Q}_{B_1} \widetilde{Q}_{F_1} \widetilde{Q}_{F_2} + 2 \widetilde{Q}_{B_2} \widetilde{Q}_{F_1} \widetilde{Q}_{F_2}) \cr
& + \Big( 2 \chi^{A_1}_{\mathbf{3}}(q_+) + \chi^{A_1}_{\mathbf{2}}(q_+) \chi^{A_1}_{\mathbf{2}}(q_-) + 2 \Big) \widetilde{Q}_{B_1} \widetilde{Q}_{B_2} (\widetilde{Q}_{F_1} + \widetilde{Q}_{F_2}) \cr
& - \chi^{A_1}_{\mathbf{4}}(q_+) \chi^{A_1}_{\mathbf{2}}(\widetilde{Q}_m) (\widetilde{Q}_{B_1} \widetilde{Q}_{F_1} + \widetilde{Q}_{B_2} \widetilde{Q}_{F_2} + \widetilde{Q}_{B_1}^2 + \widetilde{Q}_{B_2}^2) \sqrt{\widetilde{Q}_{F_1} \widetilde{Q}_{F_2}} \cr
& - \chi^{A_1}_{\mathbf{2}}(q_+) \chi^{A_1}_{\mathbf{2}}(\widetilde{Q}_m) (\widetilde{Q}_{B_1} \widetilde{Q}_{F_2} + \widetilde{Q}_{B_2} \widetilde{Q}_{F_1}) \sqrt{\widetilde{Q}_{F_1} \widetilde{Q}_{F_2}} \cr
& - \Big( 3 \chi^{A_1}_{\mathbf{2}}(q_+) + \chi^{A_1}_{\mathbf{2}}(q_-) \Big) \chi^{A_1}_{\mathbf{2}}(\widetilde{Q}_m) \widetilde{Q}_{B_1} \widetilde{Q}_{B_2} \sqrt{\widetilde{Q}_{F_1} \widetilde{Q}_{F_2}} + \ldots \,.
\eea

\bibliography{5dbib}
\bibliographystyle{JHEP}

\end{document}